\documentclass[twocolumn,superscriptaddress,
	showpacs,amsfonts,amssymb,amsmath,floats,prb]{revtex4-2}
\usepackage{graphicx}
\usepackage{dcolumn}
\usepackage{multirow}

\usepackage[normalem]{ulem}
\def\expect#1{\left\langle#1\right\rangle}
\usepackage{braket}
\def\ie{i.e.\ }

\def\e{\varepsilon}

\def\w{\omega}
\def\e{\epsilon}

\def\k{\vec{k}}

\def\mat#1{\bm{#1}}
\usepackage{upgreek}
\usepackage{bbold}
\usepackage{bm}

\begin{document}

\title{Kondo holes in strongly correlated impurity arrays:\\
RKKY driven Kondo screening and hole-hole interactions}

\author{Fabian Eickhoff}
\affiliation{Fakult\"at Physik, Technische Universit\"at Dortmund, 44221 Dortmund, Germany}
\author{Frithjof B. Anders}
\affiliation{Fakult\"at Physik, Technische Universit\"at Dortmund, 44221 Dortmund, Germany}

\date{\today}

\begin{abstract}

The emerging and screening of local magnetic moments in solids has been investigated for more than 60 years.
Local vacancies as in graphene or in Heavy Fermions can induce decoupled bound states that lead to the formation
of local moments. In this paper, we address the puzzling question how these local moments 
can be screened and what determines the additionally emerging low temperature scale.
We review the initial problem for half-filled conduction bands from two complementary perspectives: By a single-particle supercell analysis in the uncorrelated limit and by the Lieb-Mathis theorem for systems with a large Coulomb interaction $U$.
Applying Wilson's numerical renormalization group approach to a recently developed mapping of the problem onto an effective low-energy description of a Kondo hole with up to $N_f=7$ correlated impurities as background, 
we proof that the stable local moments are subject to screening by three different mechanisms.  
Firstly the local moments are delocalized  by a finite $U$ beyond the single-particle bound state. We find a Kosterlitz-Thouless type transition governed by an exponentially suppressed low energy scale of a counterintuitive 
Kondo form with $J_{\rm eff} \propto U^n$ for small $U$, where $n>1$ depends on the precise model. 
Secondly, we show that 
away from half-filling the local moment phase becomes unstable and is replaced by two types of singlet phases
 that are adiabatically connected. At a critical value for the band center, the physics is governed by an exponentially suppressed Kondo scale approaching the strong coupling phase that is replaced by an singlet formation via antiferromagnetic RKKY interaction for large deviation from the critical values.
Thirdly, we show that the local magnetic moment can be screened by a  Kondo hole orbital at finite energy, even though the orbital occupation is negligible: An additional low energy scale emerges below which the localized moment is quenched. Similarities to the experimental findings in Ce$_{1-x}$La$_x$Pd$_3$ are pointed out.

\end{abstract}

\maketitle

\section{Introduction}

A dilute concentration of magnetic impurities in a metal gives rise to the Kondo effect \cite{DeHaas36}.
The narrow resonance in the impurity spectral function, right at the Fermi energy, as well as the minimum
in the temperature dependent electrical resistivity are a manifestation
of strong (incoherent) magnetic scattering of the conduction electrons at these local moments \cite{Kondo1964,Hewson1993}.
If such local moments are regularly placed in each unit cell, as a consequence of translational symmetry,
the scattering needs to become coherent at low temperature
and a hybridization gap opens \cite{Grewe1984-SSC,Grewe91} . Depending on the 
electron filling of the hybridized bands 
the material is either insulating or becomes a metal with heavy quasi-particles.

The replacement of single magnetic atoms by their nonmagnetic counterparts
in a charge neutral substitution is called creation of Kondo hole which
gradually destroys the coherence of the heavy Fermion (HF) ground state and, consequently,
results in new properties of the highly correlated material.
In recent decades, the physics of Kondo holes has been of great interest for several experiments on heavy fermions
\cite{Lawrence85,Lawrence96,Hamidian2011,Rosa2016,shimozawa2012,Pietrus2008,Rotundu2007,lebraski2010}.
Kondo holes in the metallic phase of the PAM in general leads to a continuous crossover from the coherent heavy Fermi liquid (FL) to the single impurity behavior
\cite{ONUKI1987,Grewe91}. In contrast to this, the influence of a very low concentration of single Kondo holes on the ground state in the insulating phase induce bound states
 \cite{Malik1995,Adroja1996,Borstel1998} leading to exotic transport properties \cite{Rotundu2007}.

The effect of the Kondo holes on Kondo insulators has been studied perturbatively and with various numerical techniques, such as the DMRG in 1d, and a combination of DMFT and self-consistent mean field theory
\cite{Schlottmann91I,Schlottmann91II,Schlottmann92,Schlottmann96,Clare96,Figgins2011,Vojta2014,Neng2017,Sen2015,Kumar2014,Zhu2012,Maruyama2002,Wermbter96}, where the basic properties of the clean system 
are described by
the periodic Anderson model (PAM) or the Kondo lattice (KL). 
Sollie and Schlottmann \cite{Schlottmann91I,Schlottmann91II} employed the DMFT solution for the PAM \cite{SchweitzerCzycholl90b} and investigated the change of the single
particle properties in the vicinity of the hole site via second order perturbation theory 
in the Coulomb repulsion $U$. By examining the local $f$-electron density of states in the insulating phase of the PAM they found mid-gap states and demonstrated that these states have magnetic properties which
result in a Curie susceptibility and a Schottky anomaly in the specific heat \cite{Schlottmann92,Schlottmann96}.
They further showed that these bound states are solely localized on the nearest neighbors of the hole site in the presence of particle hole symmetry. 
 
Clare C. Yu \cite{Clare96} 
studied the physics of a missing local moment in the strongly interacting case of the one dimensional Kondo insulator 
via DMRG, which includes spatial fluctuations in contrary to the DMFT.
She confirmed the emergence of a stable magnetic bound state, however, in contrast to the weakly interacting DMFT solution, she found that the 
induced spin-density extends over the adjacent sites and falls off exponentially with some localization length that increases with decreasing strength of the Kondo coupling $J_K$.
In addition, the bound state was found to have pure $f$-character in the weak coupling limit but gradually localizes at the $c$-orbital of the hole site when increasing $J_K$.

However, there are still some open questions:
What is the fate of the Kondo effect of the unscreened local moments which contribute to the spin-density and
the Curie susceptibility?
Can the spin-density induced by a Kondo hole act as a magnetic impurity in a metal? 
For example, $\text{CePd}_3$ is a heavy fermion metal that is considered to be close to a Kondo insulator but still maintains Fermi liquid properties at low temperatures.
However, when Ce ions are substituted by nonmagnetic La ions in $\text{Ce}_{1-x}\text{La}_x\text{Pd}_3$,
the resistivity below 50 Kelvin increases with decreasing temperature as with a magnetic impurity in a metal \cite{Lawrence85,Lawrence96}
which has been attributed to a secondary Kondo effect even though the previous theories
\cite{Schlottmann91I,Schlottmann91II,Clare96}  predict localized bound states which do not interact with the itinerant states.

In this paper we (i) review the effect of Kondo holes in lattice and impurity models from two complementary perspectives using a supercell analysis in the uncorrelated limit and the Lieb-Mattis theorem in the strongly interacting regime with well defined local moments at PH symmetry.
Further we (ii) demonstrate that the emergence of decoupled localized states in a Kondo insulator due to Kondo hole substitution can be understood from a local perspective and does not rely on the periodicity and translational invariance of the lattice model.
Using a combination of the NRG and a wide band approximation \cite{MIAM2020} we (iii) study the effect of Kondo holes in finite impurity clusters as function of the local hole orbital energy and the band center of the conduction electrons and show that breaking PH symmetry can lead to Kondo screening of the hole induced magnetic bound states on low energy scales.

The onset of magnetic scattering with the remaining quasi-particles of the Fermi liquid potentially explains the logarithmic increase of the resistivity in $\text{Ce}_{1-x}\text{La}_x\text{Pd}_3$.

The paper is organized as follows.
In Sec.\ \ref{sec:theory} we introduce the Hamiltonian of the MIAM which includes the PAM and the SIAM as two limiting cases.
In Sec.\ \ref{sec:supercell} we use a supercell analysis for the non-interacting limit to study the effect of Kondo-holes in lattice and impurity models, which is compared with the Lieb-Mattis theorem for the subset of PH symmetric models on a bipartite lattice in the strongly interacting limit in Sec.\ \ref{sec:LiebMattis}.
For both limits we predict the existence of hole induced decoupled bound states. 
The combination of the two comprehensive perspectives in Sec.\ \ref{sec:Classification} allows us to differentiate between conventional MIAMs and three different types of unconventional MIAMs when Kondo-holes are introduced.
Further in Sec.\ \ref{sec:RealSpaceInterpretation} we provide a real space interpretation of the decoupling in lattice and impurity models  in terms of local pseudo gap physics from a local impurity point of view.
In order to solve the MIAM in the strongly interacting limit we use the NRG in combination with a wide band approximation which is reviewed in Sec.\ \ref{sec:low-energy-model}.
In Sec.\ \ref{sec:SreeningMechanism} we analyze the potential screening of the Kondo hole induced magnetic bound states.
We study the interaction between magnetic bound states originating from two different holes as function of the distance between the hole sites in  Sec.\ \ref{sec:HHInteraction}.
We apply the results of our NRG analysis and propose a microscopic mechanism that can explain
the unusual transport properties of $\text{Ce}_{1-x}\text{La}_x\text{Pd}_3$ in Sec.\ \ref{sec:experiment},
before we close  with a short summary and discussion in Sec.\ \ref{sec:summary}.

\section{Modeling of Kondo holes}
\label{sec:theory}

\subsection{Hamiltonian}

In order to include a wide range of different cases for Kondo holes, 
with periodic lattices (absence of Kondo holes) and the single impurity (all but one correlated site removed)
as the two extreme limits
but keep the complexity and the number of parameters manageable, 
we consider a Anderson type model which contain two type of orbitals: the uncorrelated
conduction electrons that are accounted for in a tight-binding model
\begin{align}
 H_{\rm host}=\sum_{\substack{i,j, \sigma\\i\not=j}}\left(-t_{ij} c^\dagger_{i,\sigma}c_{j,\sigma}+\e^c_ic^\dagger_{i,\sigma}c_{i,\sigma}\right),
\label{eq:host}
\end{align}
where $t_{ij}$, $\e^c_i$ denote the transfer parameter and single particle energy and $i,j$ denote the lattice sites of
the underlying lattice with
the annihilation (creation) operator $c^{(\dagger)}_{i,\sigma}$ of an electron on the lattice site  $i$ and spin $\sigma=\pm$.
For a translational invariant system $ H_{\rm host}$ can be diagonalized in $k$ space.

The localized $f$-electrons on a subset of $N_f$ lattice sites $l\in i$ are modeled by the usual local part
of a Hubbard Hamiltonian
\begin{align}
\label{eqn:himp}
H_{\text{corr}} = & \sum_{l,\sigma}\epsilon^f_{l} f^{\dagger}_{l,\sigma}f_{l,\sigma}
+\frac{1}{2}\sum_{l,\sigma}U_l f^{\dagger}_{l,\sigma}f_{l, \sigma}f^\dagger_{l,\bar{\sigma}}f_{l,\bar{\sigma}},
\end{align}
where $f_{l}^{(\dagger)}$ destroys (creates) an electron on impurity $l$, whose on-site energy is labeled by $\e^f_l$, $\bar\sigma=-\sigma$, and $U$ denotes the on-site Coulomb repulsion.

The coupling between these correlated local orbitals and the itinerant band
are accounted for by the single particle hopping term 
\begin{align}
 H_{\rm hyb}=\sum_{l,\sigma}V_l c^\dagger_{l,\sigma}f_{l,\sigma}+\rm h.c.\,,
\label{eq:Hhyb}
\end{align}
where $V_l$ denotes the local hybridization of the impurity at lattice site $l$
with the corresponding local lattice orbital.
The strength of the coupling is typically discussed in terms of
$\Gamma_{0,l}=\pi V^2_l\rho(0)$, which describes the effective hybridization of a single impurity with
a conduction band density of states (DOS) $\rho(\e)$. 

The total Hamiltonian is given by
\begin{eqnarray}
H &=&  H_{\rm host} + H_{\text{corr}}  +  H_{\rm hyb}.
\label{eq:Hamilton}
\end{eqnarray}
This formulation includes two well studied limits.
If the orbital index $l$ exhausts all lattice sites, we recover the PAM. If
$l$ only accounts for a single site, the model is known as single-impurity Anderson
model that was accurately solved using the NRG  \cite{Krishna-murthy1980I,Krishna-murthy1980II} and the Bethe ansatz  \cite{AndreiFuruyaLowenstein83,Schlottmann89}
 almost 40 years ago.
If the number of sites $N_f =\# l>1$ is small and finite,
we refer to a multi impurity Anderson model (MIAM), $ H_{\rm MIAM}$,
whose simplest realization is the two-impurity Anderson model (TIAM)
\cite{Jones1987,Eickhoff2018}.

\subsection{Supercell analysis of Kondo holes in lattice and impurity models: Formation of localized orbitals}
\label{sec:supercell}

It is well established, that single Kondo holes in the half filled PAM and Kondo insulator induce stable local moments whose spatial location and extent depends on the hybridization strength \cite{Schlottmann91I,Schlottmann91II,Schlottmann92,Schlottmann96,Clare96,Figgins2011,Vojta2014}. 
Interestingly, the basic understanding of the formation of localized orbitals can already be obtained by investigating the exact solution of the non-interacting PAM with $U=0$.

\subsubsection{General analysis}

\begin{figure}[t]
\begin{center}
\includegraphics[width=0.485\textwidth,clip]{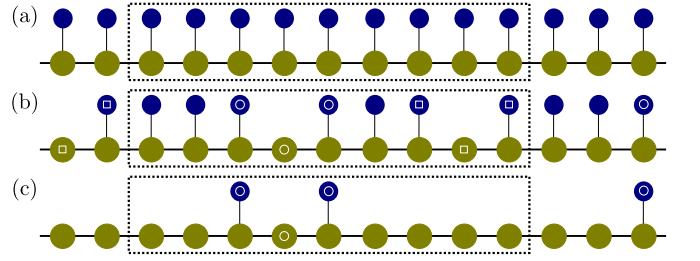}
\caption{Schematic of 1d PAM with $f$- (blue) and $c$-orbitals (green). The solid line denotes hopping elements between the corresponding orbitals. 
Panel (a) depicts an exemplary supercell with $n=9$ sites, indicated by the dashed rectangle.
In (b) $N_h=2$ $f$-orbitals per supercell have been removed leading to $N_h^d=N_h$ decoupled $d$-orbitals per supercell.
The white bordered symbols indicate the $f$- and $c$-orbitals that contribute to the corresponding decoupled state.
In Panel (c) $N_h=7$ $f$-orbitals per supercell have been removed leading to $N_h^d=1<N_h$ decoupled $d$-orbitals per supercell.
}
\label{fig:Supercell}
\end{center}
\end{figure}

In order to study the effect of Kondo holes but still maintain the useful translational invariance, we artificially define 
a supercell comprising $n$ sites ($n$ $f$- and $n$ $c$-orbitals) as schematically depicted in Fig.\ \ref{fig:Supercell}(a) for the 1d PAM with $n=9$, and remove $N_h$ of the $f$-orbitals in each supercell. This is exemplified in Fig.\ \ref{fig:Supercell}(b) for $N_h=2$ and in Fig.\ \ref{fig:Supercell}(c) for $N_h=7$.
This procedure allows to study two different scenarios in the limit of $n\to\infty$:
In case of $N_h\ll n$ (Fig.\ \ref{fig:Supercell}(b)) we can study the effect of single holes in a dense lattice, whereas 
$N_h \approx n$ corresponds to a periodic continued MIAM with $N_f=(n-N_h)$ $f$-orbitals (Fig.\ \ref{fig:Supercell}(c)).
For the supercell analysis, however, we keep all $n$ $f$-orbitals in the consideration but
use the parameters $V_i$ and $\e^f_i$ to decoupled or remove the Kondo hole orbitals.

After the Fourier formation of the periodic real space supercell structure into $k$-space
the Hamiltonian, Eq.~\eqref{eq:Hamilton}, becomes $k$-diagonal for $U=0$,
\begin{eqnarray}
H&=&\sum_{\vec{k}} H_{\vec{k}},
\end{eqnarray}
due to Bloch's theorem.
Let us label the $n$ f-orbitals, the $n$ c-orbitals by $\alpha,\beta$
and suppress the spin index $\sigma$ for better readability. 
By  defining the supercell vector operator
\begin{align}
\vec{\Psi}_{\k}=(c_{\k1},...,c_{\k n},f_{\k1},...,f_{\k n})^T
\end{align}
$H_{\vec{k}}$ reads
\begin{eqnarray}
H_{\vec{k}} &=&\vec{\Psi}^\dagger_{\k}\mat{M}_{\k}\vec{\Psi}_{\k},
\end{eqnarray}
with an appropriate matrix $\mat{M}_{\k}$. 
Since the single-particle dispersion is obtained from diagonalizing the Hermitian matrix $\mat{M}_{\k}$,
we analyze some of its fundamental properties for supercells with Kondo holes present.

From Eq.~\eqref{eq:Hamilton}, we obtain 
\begin{eqnarray}
H_{\vec{k}} &=&
\sum_{\alpha\beta} T_{\k}^{\alpha\beta} c^\dagger_{\k\alpha} c_{\k\beta} 
+ \sum_{\alpha} \e_{\alpha}^f f^\dagger_{\k\alpha} f_{\k\alpha}
\nonumber
\\
&&+ \sum_{\alpha} V_{\alpha}( f^\dagger_{\k\alpha} c_{\k\alpha} +c^\dagger_{\k\alpha}  f_{\k\alpha} )
\nonumber
\\
&=&
\sum_{\alpha} c^\dagger_{\k\alpha} C_{\k\alpha}  + \sum_{\alpha} f^\dagger_{\k\alpha} F_{\k\alpha}.
\label{eq:Hk}
\end{eqnarray}
The transfer matrix $T_{\k}^{\alpha\beta}$ denotes the 
tight-binding representation of real-space hopping parameters and orbital energies  in $ H_{\rm host}$,
with $ T_{\k}^{\alpha\beta} = [T_{\k}^{\beta\alpha}]^*$ and  $T_{\k}^{\alpha\alpha}= \e^c_\alpha$.
In Eq.\ \eqref{eq:Hk} we identified $2n$ new operators defined as
\begin{eqnarray}
\label{eq:def-Ca}
C_{\k\alpha} &=& \sum_{\beta} T_{\k}^{\alpha\beta} c_{\k\beta} 
+  V_{\alpha} f_{\k\alpha},
\\
\label{eq:def-Fa}
F_{\k\alpha} &=& \e_{\alpha}^f  f_{\k\alpha} +V_{\alpha} c_{\k\alpha} .
\end{eqnarray}

As long as the new operators $C_{\k\alpha}$ and $F_{\k\alpha}$ are linear independent, $H_{\vec{k}}$
operates on a $2n$-dimensional space.
Now let us introduce a number of $N_h$ Kondo holes placed on a subset of sites $\alpha=l_h\in L_h$, by decoupling some of the correlated orbitals from the host, \ie $V_{l_h}=0$.
By setting all 
other $\e^f_\beta=0,\beta\not\in L_h,$ and restricting to $\e^c_{l_h}=0$, we obtain from Eq.\ \eqref{eq:def-Fa}
\begin{align}
c_{\k\beta}=  F_{\k\beta}/V_{\beta}\quad \text{for}\quad \beta\not\in L_h.
\label{eq:decoupling_condition}
\end{align}
Substituting this expression back into Eq.\ \eqref{eq:def-Ca} for $\alpha=l_h$ with $V_{l_h}=0$ yields
\begin{align}
C_{\k l_h} = \sum_{\beta\not\in L_h} \frac{T_{\k}^{l_h\beta}}{ V_{\beta}} F_{\k\beta}+\sum_{\gamma\in L_h} T_{\k}^{l_h\gamma} c_{\k\gamma}.
\label{eq:Cklh}
\end{align}

Focusing on the case $T_{\k}^{l_h m_h}=0$ for a moment, which means that the $c$-orbitals at different hole sites are not directly coupled via the tight binding hopping elements $t_{ij}$ in Eq.\ \eqref{eq:host} (as is the case in Fig.\ \ref{fig:Supercell}(b) for example), the second sum on the right hand side of Eq.\ \eqref{eq:Cklh} vanishes.
This observation has a profound consequence onto the single-particle spectrum of $H_{\k}$. While the Kondo hole 
degrees of freedom $F_{\k l_h} \propto   f_{\k l_h}$ are eigenoperators
to the eigenvalues $\e^f_{l_h}$, we just showed that the operators
$C_{\k l_h}$ are linear dependent on the operators $ F_{\k\beta}$: The rank of $\mat{M}_{\k}$ 
is reduced to $2n-N_h$, and $N_h$ eigenvalues must be always zero and $\k$-independent. 
$N_h$ flat bands are formed which can be associated with $N_h$ localized states in each supercell. 
It is straight forward to prove that 
the annihilation operators  $d_{\k,l_h}$,
\begin{align}
d_{\k,l_h}=\zeta_{\k,l_h} \left(\sum_{\alpha}\frac{T_{\k}^{l_h\alpha}}{V_\alpha}f_{\k \alpha} - c_{\k l_h}\right),
\label{eq:d-k}
\end{align}
 are eigenoperators of the Hamiltonian $H_{\vec{k}}$, \ie,
$[H_{\vec{k}} ,d_{\k,l_h}]=\e_{\k} d_{\k,l_h}$ to the eigenvalue  $\e_{k}=0$, and
their normalization constants $\zeta_{\k,l_h}$ are given by
\begin{align}
\zeta_{\k,l_h}
=\left(\sum_{\alpha}\frac{|T_{\k}^{l_h\alpha}|^2}{V_\alpha^2}+1 \right)^{-1/2}.
\end{align}

Note that decoupled orbitals $d_{\k,l_h}$ not necessarily need to be orthogonal and eventually overlap if the spatial distance between the corresponding hole sites $l_h$ is small:
The spatial extend of the $d_{\k,l_h}$-orbitals is determined by most distant site $\alpha$ for which $T_{\k}^{l_h\alpha}\not=0$ holds, such that several $d_{\k,l_h}$-orbitals in Eq.\ \eqref{eq:d-k} might share some $f_{\k,\alpha}$-orbitals and, therefore, $\{d_{\k,1},d^\dagger_{\k,1}\} \not=0$.
For example, if we restrict ourselves to nearest neighbor hopping this happens if two hole sites are separated by exactly one $f$-orbital.
However, one can always define a new set of decoupled orbitals, $d_{\k,\tilde{l}_h}=\sum_{l_h}a_{\tilde{l}_h}d_{\k,l_h}$, to ensure orthogonality.
The corresponding localized Wannier orbitals in each unit cell $s$ are obtained by Fourier transformation
of $d_{\k,l_h}$ back into the real space, which will be a mixing of the $f_{s,\alpha}$-operators surrounding the hole sites and the conduction electron operators $c_{s,l_h}$ right at the hole sites. 
These $f$- and $c$-orbitals are indicated by the white bordered symbols in Fig.\ \ref{fig:Supercell}(b), where different symbols (circles and squares) denote different $d$-orbitals.

Up to here we assumed $T_{\k}^{l_h m_h}=0$.
If we allow for $T_{\k}^{l_h m_h}\not=0$ (as is the case in Fig.\ \ref{fig:Supercell}(c) for example), the second sum of the right hand side of Eq.\ \eqref{eq:Cklh} doesn't vanish in general and, consequently, the operators $C_{\k l_h}$ are not necessarily linear dependent on the operators $ F_{\k\beta}$ any longer.
However, rotating Eq.\ \eqref{eq:Cklh} into the eigenbase of the matrix $T_{\k}^{l_h m_h}$, $l_h\to \tilde{l}_h$, we obtain one linear dependent operator $C_{\k \tilde{l}_h}$ for each zero eigenvalue of $T_{\k}^{l_h m_h}$ and a corresponding decoupled orbital $d_{\k,\tilde{l}_h}$.

In general, $T_{\k}^{l_h m_h}$ is block diagonal and each block $i$ connects a subset of $n_{h,i}\leq N_h$ $c_{\k}$-orbitals. For example, if we restrict ourselves to nearest neighbor tunneling matrix elements $t_{ij}$, each subspace contains adjacent hole sites only.
Since the $c$-orbitals at the hole sites are PH symmetric, we obtain one zero eigenvalue and a corresponding decoupled orbital $d_{\k,\tilde{l}_h}$ for each odd dimensional subspace of $T_{\k}^{l_h m_h}$.

To this end,
we obtain $N_h^d\leq N_h$ decoupled $d_{\k}$ orbitals with eigenenergy $\e_{\k}=0$
by introducing
$N_h$ Kondo holes by decoupling the corresponding $f$-orbitals. $N_h^d= N_h$ holds in case of $T_{\k}^{l_h m_h}=0$ and the corresponding decoupled $d_{k}$-orbitals are than given by Eq.\ \eqref{eq:d-k}.

\subsubsection{Half-filled case}

Let us now focus on the half-field case, $\e_\alpha^f= \e_\alpha^c=0$, and ignore the decoupled $f$-orbitals. Reintroducing the spin and filling the bands with $(2n-N_h)$ electrons per unit cell yields $(n-[N_h+N_h^d]/2)$ fully filled bands, and $N_h^d$ half-filled non-dispersive bands where the electrons are mainly located at the $f$-orbitals for small couplings $V/D$.
We can divide the finite $U$ interaction term in Eq.\ \eqref{eqn:himp} into a Hartree term that is
 absorbed into $\e_l^f\to \tilde \e_l^f= \e_l^f +U_l/2$ and a charge fluctuation term $U(N^f_l-1)^2/2$  \cite{Krishna-murthy1980I} responsible for the generating of an effective magnetic moment. The zero-energy localized states
emerge as long as $\tilde \e_l^f=0$. The finite $U$ generates
an effective moment on the decoupled orbitals which might interact with each other if the spatial distance between different hole sites is not too large.
So far, the dimensionality of the model as well as the gemoetry of the underlying lattice has not entered: Therefore the bound state formation is generic in arbitrary spatial dimensions for any type of lattice.

\subsubsection{Embedding the supercell analysis into the literature}

More than thirty years ago, the
existence of hole induced bound states has already been proposed by Sollie and Schlottmann \cite{Schlottmann91I,Schlottmann91II,Schlottmann92,Schlottmann96} in the framework of the dynamical mean field theory approach to the PAM in the Kondo insulator limit.

Using a large-$N$ mean field decoupling to solve a 2d Kondo lattice and by assuming additional local potential scattering terms in the conduction electron band at the hole sites $l_h$, 
Figgins and Morr \cite{Figgins2011} found that the hole induced bound states also occur in models with asymmetric conduction bands away from half filling.

This perfectly fits to the supercell analysis since such local potential scattering terms just shift the local on site energies $\e^c_{l_h}$:
The decoupling of the $d$-orbitals in Eq.\ \eqref{eq:d-k} only requires local PH symmetry at the hole sites, $\e^c_{l_h}=0$, and is independent from the filling of the entire conduction band, \ie allows for $\e^c_\alpha\not=0$ for $\alpha\not\in L_h$.
Consequently, in case of a asymmetric conduction band, $\e_i^c=\e^c\not=0$, a local potential scattering $u_0$ at the hole site $l_h$, as introduced by Figgins and Morr in Ref.\ \cite{Figgins2011}, can lead to a local reduction of PH asymmetry, $\e^c_{l_h}=\e^c+u_0\approx0$, and, therefore, stabilize the hole induced bound states.
Whereas the decoupling of the effective orbitals $d_{\k,l_h}$ is unstable against a weak deviation from $\e^c_{l_h}=0$ in the non-interacting limit, the local moment that forms in case of a finite interaction $U$ is stable against small deviations from $\e^c_i\not=0$ and $\e^f_i\not=-U/2$, as we demonstrate later.

Moreover, the expression \eqref{eq:d-k}
of the dispersionless band orbitals $d_{\k,l_h}$ 
comprising different orbital contributions 
is already sufficient to
understand the spatial variation of the local moments in a 1d Kondo hole problem 
investigated by a DMRG calculation as function of the local Kondo interaction $J$ -- see Fig.\ 3 in Ref.\ \cite{Clare96}. In the Schrieffer-Wolff limit \cite{SchriefferWol66} a larger $J$ corresponds to larger on-site hybridization $V$. In the limit $V/t\to 0$, the $d_{\k,l_h}$ states
have mainly $f$-character and, therefore, the local moment induced by a finite $U$ 
is mainly located at the $f$-orbitals surrounding the hole sites.
With increasing $V/t$, the $c$-orbitals right at the hole sites are more and more mixed into $d_{\k,l_h}$,
 and the local moment moves from nearby $f$-orbitals to the $c$-orbitals at the hole sites. In the limit $V/t\to \infty$, the localized orbitals are localized at the disconnected $c$-orbitals on the hole sites.

Notably, the decoupling of the orbitals $d_{\k,l_h}$ does not depend on some spatial isotropy within the supercell since no restrictions concerning the hopping elements $t_{ij}$ were made and the individual couplings $V_i$ are completely independent.
These parameters solely enter the composition of $d_{\k,l_h}$ as can be seen in Eq.\ \eqref{eq:d-k}.
Due to $T_{\k}^{i\alpha} \propto t_{i\alpha}$ only $f$-orbitals for which $t_{ni}$ is finite are involved and the relative amount of these individual orbitals is controlled by the strength of the coupling $t_{ni}/V_i$.
If the hopping between the $c$-orbitals is restricted to nearest neighbors, the decoupled state is solely localized on the nearest  $f$-orbital neighbors of the hole site.

Whereas this result perfectly fits to the weakly interacting (small $U$) DMFT solution of Solli and Schlottmann \cite{Schlottmann91I,Schlottmann91II}, the DMRG calculations of Clare C. Yu \cite{Clare96} for the half filled 1d Kondo lattice demonstrate, that the spin density induced by a single Kondo hole, however, extends beyond the nearest neighbors.
As we demonstrate later on, this is a result of the restriction to singly occupied $f$-orbitals in the Kondo lattice which corresponds to a large interaction $U$ and $\e^f\ll 0$.

In accordance with the DMFT solution of Schlottmann \cite{Schlottmann1995} 
for a  weakly interacting system, hole induced bound states originating from different holes within the unit cell do not interact with each other in the non-interacting limit. The corresponding dispersionless bands are degenerate due to the lack of interaction.
The aforementioned delocalization of the induced spin density in case of large $U$, however,
leads to an overlap between magnetic bound states originating from different holes which results in a finite exchange interaction
as demonstrated below.

\subsection{Lieb-Mattis theorem applied to the strongly interacting MIAM: Prediction of stable local moments}
\label{sec:LiebMattis}

For the strongly interacting limit of the depleted, finite size Kondo lattice there is a modified version of the Lieb-Mattis theorem \cite{LiebMattis}
proven by Shen \cite{Shen1996}, which states that for a number of $N_f$ local moments coupled by a local antiferromagnetic exchange interaction to a half-filled system of 
conduction electrons on a bipartite d-dimensional lattice with $N_c\geq N_f$ sites, interacting via a finite Hubbard-type interaction,  the ground state has a total $S_z$ component of
\begin{align}
S^\text{tot}_z = \frac{1}{2}|N_{c,\text{A}} - N_{c,\text{B}} + N_{f,\text{B}} - N_{f,\text{A}}|
\label{eq:LiebMattis}
\end{align}
(see theorem VI in Ref.\ \cite{Shen1996}).
In case of non-interacting conduction electrons degeneracy of the ground state (apart from the trivial (2$S^\text{tot}_z$+1) fold degeneracy) can only be excluded for the dense case, $N_f=N_c$. However, even if there is degeneracy, one of the ground states is always in the sector of $S^\text{tot}_z$ given by Eq.\ \eqref{eq:LiebMattis}.
Moreover, Titvinidze et al. \cite{Potthoff2014} demonstrated the applicability of the theorem
to the 1d regularly depleted Kondo lattice by employing the DMRG, showing
that the ground state is unique even if the conduction electrons are non-interacting.

Nevertheless, the modified Lieb-Mattis theorem assumes a finite size system, whereas we are also interested in the MIAM with a finite number of $f$-orbitals coupled to an electron continuum. Consequently, we need to slightly modify the theorem in order to apply it to multi impurity models.

The predictions of  Eq.\ \eqref{eq:LiebMattis} are limited to a finite size system. They perfectly agree with the
well \cite{ Wilson1975,Bulla2008} established SC FP structure of the SIAM or the single impurity Kondo model (SIKM) where two FPs are found, one for even and one for odd chain length.
Nevertheless, the term Kondo singlet ground state  has been used numerously in the literature \cite{Wilson1975}
when a local spin 1/2 is
coupled antiferromagnetically to  fermionic continuum in the thermodynamic limit:
The precise state of the infinitely large
conduction band sea is considered to be irrelevant, and the Fermi sea is treated  as a singlet state,  regarding the
even-odd oscillations as trivial and irrelevant point. 
Wilson realized that the  Kondo singlet formation is better quantified by calculating local quantities defined as difference between the total system with and without impurity \cite{Wilson1975,Bulla2008}: a spatially extended singlet is formed which decouples from the remaining conduction electron band whose precise properties do not matter for $N\to\infty$.

\begin{figure}[tbp]
\begin{center}
\includegraphics[width=0.45\textwidth,clip]{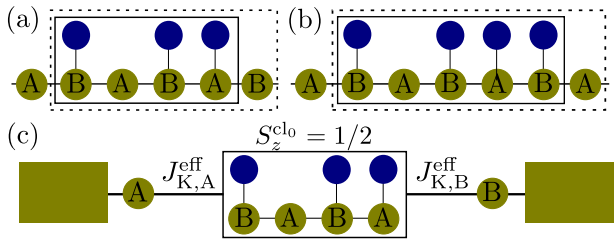}
\caption{(a) and (b): Schematic of two 1d MIAM's on a bipartite lattice with A and B sites.
The solid line rectangle defines the smallest finite size cluster that includes all sites with $f$-orbitals present.
Applying Eq.\ \eqref{eq:LiebMattis} to these finite size cluster results in $S^{\text{cl}_0}_z=1/2$, however, the cluster are connected to the remaining continuum via $J^\text{eff}_{\text{K,A/B}}$, as schematically depicted in (c) for the model in panel (a).
Here $J^\text{eff}_{\text{K,A/B}}$ indicates the coupling between the cluster and nearest A/B site of the continuum.
In order to determine the sign of these couplings, Eq.\ \eqref{eq:LiebMattis} is used to obtain $S^{\text{cl}_0^\prime}_z(A/B)$ of an enlarged cluster that includes one additional A/B site, as indicated by the dashed rectangle in panel (a) and (b).
$J^\text{eff}_{\text{K,A/B}}$ can be derived via $S^{\text{cl}_0}_z$ and $S^{\text{cl}_0^\prime}_z$ using Eq.\ \eqref{eq:JeffAB}.
}
\label{fig:Model_LiebMattis}
\end{center}
\end{figure}

In order to reveal the magnetic ground state properties of a MIAM 
in Wilson's spirit using the Lieb-Mattis theorem, we can proceed as follows:
In a first step we define the smallest finite size cluster $\text{CL}_0$ in such a way that it includes all sites with $f$-orbitals present, as indicated by the solid line rectangle in Fig.\ \ref{fig:Model_LiebMattis} (a) and (b) for two exemplary 1d models.
For this finite size subsystem we can apply Eq.\ \eqref{eq:LiebMattis} to predict the $S^{\text{cl}_0}_z$ component of the finite size cluster.
In case of $S^{\text{cl}_0}_z=0$, as one obtains for the SIAM and dense MIAM, we are already done and can conclude $S^\text{tot}_z=S^{\text{cl}_0}_z=0$, since the remaining infinite size lattice 
is treated as $S^\text{lattice}_z=0$ 
\footnote{Note that we cannot make a statement about the  nature of the FP: In the TIAM, for instance, two  adiabatically connected FP are found, one originating from a RKKY interaction the other driven by the Kondo effect \cite{Jones1987,Affleck1995}}.
However, when the magnetic moment of the cluster is finite, $S^{\text{cl}_0}_z\not=0$, it can be quenched via the Kondo effect by the remaining electron continuum if the coupling $J^\text{eff}_{\text{K},\text{A/B}}$ between the cluster and the nearest A/B sites of the continuum is antiferromagnetic. 
This situation is schematically depicted in Fig.\ \ref{fig:Model_LiebMattis}(c) for the 1d model in panel \ref{fig:Model_LiebMattis}(a) and can also be generalized to arbitrary spatial dimensions.

In order to determine the sign of $J^\text{eff}_{\text{K},A/B}$ we now enlarge the cluster $\text{CL}_0\to\text{CL}^\prime_0$ by including one additional A
(or B) site, as indicated by the dashed rectangle in Fig.\ \ref{fig:Model_LiebMattis} (a) and (b) for example, and use Eq.\ \eqref{eq:LiebMattis} again to calculate $S^{\text{cl}^\prime_0}_z(A/B)$ of the enlarged cluster.
At zero temperature a antiferromagnetic (ferromagnetic) $J^\text{eff}_{\text{K},\text{A/B}}$ would decrease (increase) the original $S^{\text{cl}_0}_z$ by $1/2$ and, consequently, the $S^{\text{cl}^\prime_0}_z(A/B)$ component of the enlarged cluster can be written as
\begin{align}
S^{\text{cl}^\prime_0}_z(A/B)=S^{\text{cl}_0}_z+\frac{J^\text{eff}_{\text{K},\text{A/B}}}{|J^\text{eff}_{\text{K},\text{A/B}}|}\frac{1}{2}.
\label{eq:JeffAB}
\end{align}
If the finite size cluster $\text{CL}_0$ with $S^{\text{cl}_0}_z\not=0$ is coupled ferromagnetically to all nearest neighbors of the remaining infinite lattice we can conclude $S^\text{tot}_z=S^{\text{cl}_0}_z\not=0$: The $S_z$ component of any other finite size cluster $\text{CL}$ that contains an arbitrary number of sites can never be smaller than that of $\text{CL}_0$: $S^{\text{cl}}_z\geq S^{\text{cl}_0}_z\not=0$.

An example for such a situation is depicted in Fig.~\ref{fig:Model_LiebMattis}(b):
Applying Eq.\ \eqref{eq:LiebMattis} to the cluster CL$_0$ (solid rectangle) results in $S^{\text{cl}_0}_z=1/2$ and including one additional A site (dashed rectangle) would increase the size of the local cluster magnetic moment, $S^{\text{cl}_0^\prime}_z(A)=1$.
Consequently, the cluster CL$_0$ is FM coupled to the left and right electron continuum and the LM FP is stable: $S^\text{tot}_z=S^{\text{cl}_0}_z=1/2$.

In contrast to that, in case of $K\not=0$ antiferromagnetic couplings one can always find a new finite size cluster $\text{CL}$ whose $S_z$ component is reduced compared to that of $\text{CL}_0$: $S^{\text{cl}}_z < S^{\text{cl}_0}_z\not=0$.
For $K\geq 2 S^{\text{cl}_0}_z$ the magnetic moment can be quenched completely, $S^{\text{cl}}_z\geq 0$, whereas for $K< 2 S^{\text{cl}_0}_z$ a finite local moment will always remain, $S^{\text{cl}}_z\geq (S^{\text{cl}_0}_z-K/2)$, and all the cluster for which $S^{\text{cl}}_z=(S^{\text{cl}_0}_z-K/2)$ holds are only ferromagnetically coupled to the remaining continuum: $K=0$.

An example for such a situation is depicted in Fig.\ \ref{fig:Model_LiebMattis}(a):
Applying Eq.\ \eqref{eq:LiebMattis} to the cluster CL$_0$ (solid rectangle) results in $S^{\text{cl}_0}_z=1/2$.
However, by including one additional B site (dashed rectangle) the size of the local cluster magnetic moment is reduced, $S^{\text{cl}_0^\prime}_z(B)=0$, and,
consequently, the cluster CL$_0$ is AF coupled to the right electron continuum and the LM FP is unstable: $S^\text{tot}_z=S^{\text{cl}_0}_z-1/2=0$.

To this end, using the modified Lieb-Mattis theorem of Eq.\ \eqref{eq:LiebMattis} to calculate $S^{\text{cl}_0}_z$, and Eq.\ \eqref{eq:JeffAB} to determine the number $K$ of AF couplings $J^\text{eff}_\text{K,A/B}$, the magnetic ground state properties of a MIAM in Wilson's spirit are given by
\begin{align}
S^\text{tot}_{z}=\left\{\begin{array}{ll} 
																				  0, & \,\text{if}\, S^{\text{cl}_0}_z\leq K/2\\
																					S^{\text{cl}_0}_z-K/2, & \,\text{if}\, S^{\text{cl}_0}_z> K/2
								 \end{array}\right. .
         \label{eq:55}
\end{align}

Note that this result is valid in arbitrary spatial dimensions. 
While the cluster $S^{\text{cl}_0}_z$ component is always given by Eq.\ \eqref{eq:LiebMattis}, the number of screening channels $K$ depends on the
geometric embedding of the cluster into the lattice and, consequently, on its spatial dimension.
Due to the left/right structure in 1d the number of screening channels can never be larger than two in this case, $K^\text{1d}\leq 2$, which provides an alternative interpretation of our result in Ref.\ \cite{MIAM2020}, where we demonstrated that the maximum number of possible screening channels for any MIAM in arbitrary spatial dimensions is limited to the number of Fermi surface states.

\subsection{Combination of the supercell analysis and the Lieb-Mattis theorem: Conventional and unconventional MIAM}
\label{sec:Classification}

Using the supercell analysis for the PAM in Sec.\ \ref{sec:supercell} we demonstrated that the removal of a number of $N_h$ $f$-orbitals per unit cell (uc) can lead to $N_h^d\leq N_h$ decoupled $d$-orbitals in the non-interacting limit and, consequently, to an impurity induced entropy of $S^\text{uc}_\text{imp}=N_h^d k_\text{B}\ln(4)$ per unit cell.
If we take the MIAM as a representation of the impurity physics with an arbitrary large real space supercell such that $S^\text{tot}_\text{imp}=S^\text{uc}_\text{imp}$ holds, we can strictly differentiate between different types of MIAMs by combining the predictions from the supercell analysis and the Lieb-Mattis theorem at half-filling on bi-partite lattice.

In order to separate our investigation from the conventional MIAM,
we define the unconventional  MIAM as a model where
$N_h^d$ localized orbitals 
decouple from the rest of the system at  $U=0$  leading to a finite 
ground state entropy of $S^\text{tot}_\text{imp}>0$. At finite $U$, usually a local moment arises which
might remain unscreened indicated by a finite ground state entropy.
Well studied examples are gaped or system with pseudo gap density of states \cite{Vojta2006}.
In graphene, for example, carbon vacancies  generated such single particle bound states \cite{Pareira2006,GrapheneRMP2009,Nanda2012} which are subject to Kondo screening \cite{MayGraphen2018,AndreiGraphen2018}.
In this paper, however, we focus on conventional metallic conduction band hosts, where such localized orbitals
are induced by vacancies in dense systems called Kondo holes.
The pseudo gap physics  only implicitly enters   
via the reduced rank of $\mat{\Gamma}$ as reviewed in Sec.\ \ref{sec:low-energy-model} below
based on the mapping presented in Ref.~\cite{MIAM2020}.

In the conventional  MIAM no such decoupled localized state exist, and one or several intermediate unstable LM FPs develop with increasing $U/\Gamma_0$. The emerging local moments are quenched on a low energy scale which results from mixture of the Kondo effect and the RKKY interaction in general. The SIKM, SIAM and dense MIAM are typical representatives of that category where we always find a vanishing residual entropy: $S^\text{tot}_\text{imp}=0$.

In the unconventional MIAM we distinguish between a non-interacting  ($U=0$) and an
interacting case ($U\not = 0$). For $U=0$ we find a residual entropy of $S^\text{tot}_\text{imp}=N_h^d k_\text{B} \ln(4)$: each hole induces a decoupled localized orbital with the single-particle energy $\e=0$. In the NRG
language we have  a free orbital (FO) fixed point of these orbitals while the rest of the system is represented by a ground state of a Fermi sea.

For the interacting problem, $U>0$, we find the hierarchy 
\begin{eqnarray}
0\leq S^\text{tot}_\text{imp} \leq S^{\text{cl}_0}_\text{imp} \leq N_h^d k_\text{B} \ln(2).
\end{eqnarray}
Without any coupling to the conduction band, free local moments are developing when $\beta U>1 $ which provide
an upper bound for the cluster and the impurity residual entropy. The hybridization mediated RKKY mechnism
leads to a reduction of the entropy by alignment of local moments that are subject to potentially 
incomplete Kondo screening.

The stable low-temperature FP entropy $S^\text{tot}_\text{imp}$ is discontinuous at $U=0$ for  $T\to 0$ defining a quantum phase transition (QPT). This transition is either 
of first order due to a level crossing of the ground state energies or of KT type \cite{Vojta2006}.:  An arbitrary small Coulomb interaction $U>0$ can already be sufficient to obtain a LM FP at intermediate temperature $T<U$, resulting in strong correlation effects.
We can further differentiate between three types of unconventional  MIAM.

\subsubsection{Unconventional MIAM of Type I}

In the type I model, the ground state magnetic moment of the local cluster couples via a FM  
$J^\text{eff}_\text{K}$ to the conduction band channels. Therefore the residual entropy remains finite, and we have
the hierarchy 
\begin{eqnarray}
0&<&S^\text{tot}_\text{imp}=S^{\text{cl}_0}_\text{imp}\leq N_h^d k_\text{B}\ln(2)
\end{eqnarray}

An example for this kind of model is depicted in Fig.\ \ref{fig:Model_LiebMattis}(b).
A finite local cluster magnetic moment $S^{\text{cl}_0}_z\geq 1/2$ is FM coupled to the remaining continuum such that the LM FP is stable.  

The QPT is of first order in case of a single Kondo hole, $N_h^d=N_h=1$ at $U=0$:
The unstable FO  FP with residual entropy $S^\text{FO}_\text{imp}=k_\text{B}\ln(4)$ crosses over to the LM FP with $S^\text{LM}_\text{imp}=k_\text{B}\ln(2)$ on the energy scale of the Coulomb interaction $U$, which, obviously, vanishes linear at $U_c=0$.

\subsubsection{Unconventional MIAM of Type II}

In the type II model, the ground state magnetic moment of the local cluster couples via an AF  
$J^\text{eff}_\text{K}$ to the effective conduction band channels. An example for this kind of model is depicted in Fig.\ \ref{fig:Model_LiebMattis}(a). In this case the cluster moment is reduced by the conduction band screening channels
and the hierarchy 
\begin{eqnarray}
0& \leq & S^\text{tot}_\text{imp}< S^{\text{cl}_0}_\text{imp}\leq N_h^d k_\text{B}\ln(2)
\end{eqnarray}
holds. It turns out that the low energy scales depend exponentially on the effective Kondo couplings $J^\text{eff}_\text{K,A/B}$ which vanishes at $U^c=0$.  Consequently, the QPT is of KT type as in SIKM at $J^c_\text{K}=0$.

The interesting question arises how $J^\text{eff}_\text{K,A/B}$ depends on the Coulomb interaction $U$ since $J^\text{eff}_\text{K,A/B}(U=0)=0$ must be fulfilled.
We will demonstrate that NRG calculations that are presented in Sec.\ \ref{sec:TypeII} result in $J^\text{eff}_\text{K,A/B}\propto U^n$, where $n>1$ depends on the model.

\subsubsection{Unconventional MIAM of Type III}

In this class, the residual entropy of the cluster as well as the total effective impurity
always vanishes for $U>0$: $S^\text{tot}_\text{imp}=S^{\text{cl}_0}_\text{imp}=0.$
This scenario requires an even number of decoupled orbitals $N^d_h=2n$.
The hole induced local moments are AF coupled such that they lock into an intra-cluster singlet state.
We study such a scenario in Sec.\ \ref{sec:HHInteraction}.

\subsection{Real space interpretation of the decoupling in lattice and impurity models: Local pseudo-gap physics}
\label{sec:RealSpaceInterpretation}

Using the supercell analysis in Sec.\ \ref{sec:supercell} we demonstrated that Kondo holes in non-interacting lattice and impurity models quite general lead to the occurrence of decoupled states, localized in the vicinity of the hole sites. Moreover, the modified Lieb-Mattis theorem for the strongly interacting limit, which we discussed in Sec.\ \ref{sec:LiebMattis}, predicts a stable LM FP for several MIAMs and a macroscopic magnetization for certain regularly depleted lattice models.
While these results for the two complementary perspectives already allowed us to distinguish between conventional MIAMs and three types of unconventional MIAMs, a detailed understanding of the microscopic mechanism, responsible for the decoupling and spatial redistribution of the localized orbitals and magnetic moments, from the local impurity point of view is still missing.  

In this section we demonstrate that the decoupling in both, lattice and impurity models, can be understood in terms of local pseudogap physics.

\subsubsection{Periodically depleted PAM in 1d}

In order to obtain a real space interpretation of the mechanism that leads to the decoupling of the states $d_{\k,l_h}$ in Eq.\ \eqref{eq:d-k}, we focus on the 1d PAM with nearest neighbor hopping between the $c$-orbitals, $t_{ij}\propto \delta_{i,j\pm 1}$, and consider the smallest supercell, $n=2$, in which one can insert a Kondo hole without removing all $f$-orbitals in the Hamiltonian.

\begin{figure}[t]
\begin{center}
\vspace{5mm}
\includegraphics[width=0.4\textwidth,clip]{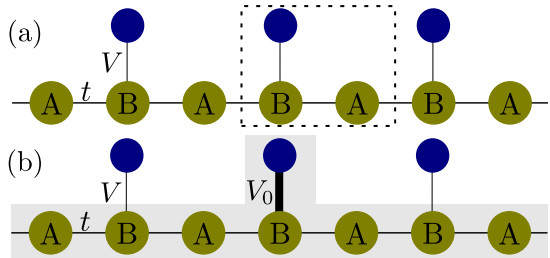}
\caption{(a) schematic of the depleted PAM in 1d with $f$-orbitals (blue) connected to the B-sublattice of the $c$-orbitals (green). The unit cell is indicated by the dashed rectangular.
(b) if a single coupling $V_0$ dominates, the other $f$-orbitals ''feel'' an effective medium indicated by the gray area in the background including the $f_0$-orbital.}
\label{fig:1dDPAM}
\end{center}
\end{figure}
This model, known as the depleted PAM, is schematically depicted in Fig.\ \ref{fig:1dDPAM}(a), where the $f$-orbitals (blue) are coupled to the B-sublattice of the $c$-orbitals (green) and the unit cell (dashed rectangular in Fig.\ \ref{fig:1dDPAM}(a)) contains two $c$- and one $f$-orbital.
For this special setup we obtain one decoupled $d_{\k}$-orbital and can evaluate Eq.\ \eqref{eq:d-k} to obtain
\begin{align}
d_{\k}=\zeta\left(\k\right)\left(\frac{2t\cos(ka)}{V} f_{B,\k} + c_{A,\k}\right),
\end{align}
where the index $A$ and $B$ labels the site of the unit cell.
In the wide band limit, $t/V\to\infty$, the dispersionless band has pure $f$-character and, consequently, the $f$-orbitals seem to decouple from the itinerant electrons.

\begin{figure}[t]
\begin{center}
\includegraphics[width=0.5\textwidth,clip]{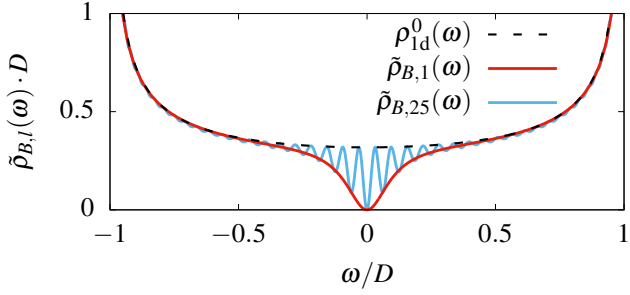}
\caption{Local conduction electron DOS of the $B$ sublattice: (i)  DOS $\rho^0_{\rm 1d}(\w)$ in the absence of an impurity, (ii) 
$\tilde \rho_{B,l}(\w)$ in the presence of an resonant level at $R=0$ for two different sites $l=1$ and $l=25$ corresponding to $\Delta R=2a,50a$ and $D/\Gamma_0=10$, $V_0/\Gamma_0=\sqrt{D/\Gamma_0}$}
\label{fig:EffDOS}
\end{center}
\end{figure}

This surprising finding can be very easily explained by studying a slightly modified version of the model, schematically depicted in Fig. \ref{fig:1dDPAM}(b), as we have done in a previous publication \cite{MIAM2020}.
If the coupling to the impurity site at the origin dominates over all others: $V_0\gg V_l$, $l\not=0$, the $f_{B,l}$-orbitals at the sites $l\not=0$ ''feel'' an effective medium (gray background in Fig.\ \ref{fig:1dDPAM}(b)), which includes the influence of the $f_{B,0}$-orbital on the conduction band electrons. 
If the $f_{B,0}$-orbital is assumed to be non-interacting, $U_0=0$, the effective $c$-density of states at the B-sublattice of site $l$ is approximately given by \cite{MIAM2020}
\begin{align}
\label{eq:rho-eom-disance-R}
&\rho_{B,l}(\omega)\approx\tilde{\rho}_{B,l}(\w)=\quad\rho_\text{1d}^0(\omega)\,\,-\nonumber\\
&\left[\rho_\text{1d}^0(\omega)\cos\left\{l\cos^{-1}\left[\frac{\omega}{D}\right]\right\}\right]^2
\frac{\pi^2V^4_0\rho_\text{1d}^0(0)}{\omega^2+\left[\pi V^2_0 \rho_\text{1d}^0(0)\right]^2},
\end{align}
where the approximation $\Re G_{c,ij}^0(\w-i0^+)\approx\Re G_{c,ij}^0(-i0^+)=0$ for the real part of the free conduction band electron propagator has entered \cite{MIAM2020}.
The comparison of the local conduction electron DOS without impurity $\rho^0_{\rm 1d}(\w)$ and the local DOS at site $l$ with the impurity present is shown in Fig.\ \ref{fig:EffDOS}. Focusing on lattice sites for 
 the second impurity that are on the same bi-partite sublattice as the first impurity reveals a pseudo-gap formation
 of the spectrum: the larger the distance the faster the DOS oscillations in energy space, the smaller the energy intervall
 of the pseudo-gap. Since the pseudo-gap always vanishes quadratically in this energy 
 window, $\rho_{B,l}(\w) \propto |\w|^2$, a local magnetic moment of a second impurity coupled to the lattice at site $l$ 
 decouples in the limit $T\to 0$ since $V_l$ is irrelevant in the sense of an RG treatment \cite{Vojta2004I}.  

This result not only enables a simple interpretation of the decoupling in terms of local pseudo-gap physics but also reveals another important property:
The decoupling of some localized orbitals due to Kondo holes does not rely on the periodicity and translational invariance of the Hamiltonian.
For example, in the 1d model discussed above it would be enough to consider only one additional $f_{B,l}$-orbital among the $f_{B,0}$-orbital in order to obtain a decoupled state. Indeed, the emergence of stable local moments in the 1d TIAM has already been studied in Ref.\ \cite{TIAM2017}.

\subsubsection{Kondo holes in finite impurity cluster}
\label{sec:finite_cluster}

Having demonstrated the equivalence of the local pseudo gap formation in the regularly depleted 1d lattice and 1d MIAM, we now extend the latter model to arbitrary dimensions.

\begin{center}
\textit{Local point group analysis: decomposition of the single particle subspace} 
\end{center}

In this section, we generalized the strategy applying to the two impurity problem \cite{Jones1987,Jones1988,Affleck1995}  which uses even and odd parity sectors. 
Parity conservation results in decomposing the Hilbert space in irreducible representation of the $C_2$ point group in the two impurity problem.

\begin{figure}[tbp]
\begin{center}
\includegraphics[width=0.42\textwidth,clip]{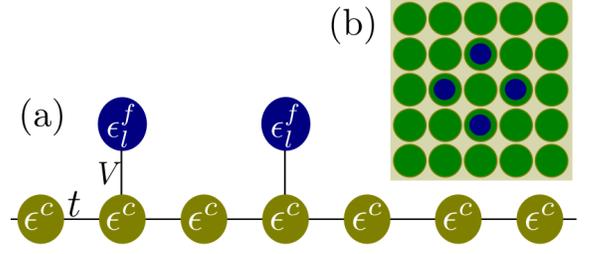}
\caption{Simplest realization of a hole in a finite impurity cluster in (a) 1d with two $f$-orbitals and (b) 2d with four $f$-orbitals.}
\label{fig:CN_1d2d}
\end{center}
\end{figure}

In order to avoid complications by more complex lattices, 
we only consider Bravais lattices for $H_{\rm host}$ in Eq.\ \eqref{eq:host} and restrict the correlated lattice sites to the nearest neighbors 
of the hole site at $\vec{R}=0$ for the moment, i.\ e.\
\begin{eqnarray}
\label{eqn:H_MIAM}
H_{\rm MIAM} &=& H_{\rm host} +  H_{\rm hyb}
\\
&& \nonumber
+
\sum_{<l,0>,\sigma}\epsilon^f_{l} f^{\dagger}_{l,\sigma}f_{l,\sigma}
+\frac{1}{2}\sum_{l,\sigma}U_l f^{\dagger}_{l,\sigma}f_{l, \sigma}f^\dagger_{l,\bar{\sigma}}f_{l,\bar{\sigma}},
\end{eqnarray}
with the same restriction of the index $l$ in $H_{\rm hyb}$. For a 1d and  a 2d simple cubic lattice,
the setup is depicted in Fig.\ \ref{fig:CN_1d2d}(a) and \ref{fig:CN_1d2d}(b), respectively. 
The correlated impurity sites
are invariant under the point group $P=C_2$ (1d) or $P=C_4$ (2d) symmetry. In 3d and in different geometries, we refer to the  appropriated point group $P$ of the lattice of interest.

After diagonalizing $H_{\rm host}$ in k-space, the hybridization term $H_{\rm hyb}$ takes the form
\begin{eqnarray}
 H_{\rm hyb}&=& \frac{V}{\sqrt{N_c}} \sum_{\vec{k}\sigma} c^\dagger_{\vec{k},\sigma}
 \sum_{l,\sigma} e^{-i\vec{k}\vec{R}_l} f_{l,\sigma}+\rm h.c.\,,
\end{eqnarray}
assuming an equal hybridization strength for all $V_l=V$
to ensure the point group symmetry.

There are $N_f$ different $f$-electron annihilation operators $f_{l,\sigma}$ that span 
the single particle vector space of the $f$-orbitals on which an 
reducible representation of the point group is operating on.
By applying the projector $P^{\Gamma}$
of each irreducible representation $\Gamma$ of $P$,
\begin{eqnarray}
\hat P^\Gamma &=& \frac{d_\Gamma}{|P|} \sum_{g\in P} \chi^{\Gamma}(g) \hat O(g)
\end{eqnarray}
onto the operators $f_{l,\sigma}$, all operators $f_{\Gamma \alpha,\sigma}$
of the irreducible representations can be constructed 
\begin{eqnarray}
\hat P^\Gamma f_{l,\sigma} &\to & f_{\Gamma \alpha,\sigma},
\end{eqnarray}
where $\alpha$ labels the different degrees of freedom in a possibly multidimensional irreducible representation (irrep) $\Gamma$.
Applying the mapping $\hat P^\Gamma f_{l,\sigma}$ creates an operator that requires
normalisation such that
\begin{eqnarray}
\{ f_{\Gamma \alpha,\sigma},f_{\Gamma' \alpha',\sigma'}^\dagger \} = \delta_{\Gamma\Gamma'} 
\delta_{\alpha\alpha'} 
\delta_{\sigma\sigma'} 
\label{eq:anti-commuator-21}
\end{eqnarray}
The number of group elements in P is given by $N_g=|P]$, $\chi^{\Gamma}(g)$ denotes
the character of $g$ in the representation ${\Gamma}$, and $\hat O(g)$ is the representation of the group element $g$ in the vector space spanned by the $f_{l,\sigma}$. This leads to expansion
\begin{eqnarray}
\label{eqn:f-expansion-irreps}
f_{l,\sigma} &=& \sum_{\Gamma\alpha} U_{l,\Gamma\alpha} f_{\Gamma \alpha,\sigma}
\end{eqnarray}
with the unitary transformation $\mat{U}$
that is substituted into  $H_{\rm hyb}$ reading
\begin{eqnarray}
 H_{\rm hyb}&=&  \sum_{\vec{k}\Gamma\alpha,\sigma} 
 V_{\vec{k}\Gamma\alpha} 
 c^\dagger_{\vec{k},\sigma}
 \sum_{l,\sigma} f_{\Gamma\alpha,\sigma}+\rm h.c.\,,
\end{eqnarray}
where
\begin{eqnarray}
\label{eq:H-hyp}
 V_{\vec{k} \Gamma\alpha}  &=& \frac{V}{\sqrt{N_c}} \sum_l U_{l,\Gamma\alpha} e^{-i\vec{k}\vec{R}_l}.
\end{eqnarray}
Since different orbital energies $\epsilon^f_{l}$ break the local point group symmetry \cite{AU-PTCDA-dimer}, leading to
single-particle transfer matrix elements between the new orbitals $f_{\Gamma\alpha,\sigma}$  we focus on $\epsilon^f_{l}=\epsilon^f=const$ in the following.

Note that the conduction
electron dispersion in case of nearest neighbor tight binding description with a uniform single particle on site energy, $\e^c_i=\e^c$, is given by
\begin{eqnarray}
\e^c_{\k}&=&  -t\sum_{<l,0>} e^{i\k \vec{R}_l}+\e^c= -t\gamma(\k)+\e^c
\label{eq:ek}
\end{eqnarray}
and obviously related to the hybridization matrix element $V_{\k,\Gamma_1}$ of the trivial
irreducible representation $\Gamma_1$ with $\mat{U}=(1/\sqrt{N_f})$:
\begin{eqnarray}
V_{\k,\Gamma_1} &=&
\frac{V}{\sqrt{ N_c N_f}}
\gamma(\k)=-
\frac{V}{\sqrt{N_c N_f}}
\left(\frac{\e^c_{\k}-\e^c}{t}\right)
\label{eq:Vk1}
\end{eqnarray}
This holds for all point groups since $\chi^{\Gamma_1}(g)$ for all $g\in P$
and $\hat O(g)$ mapped each $f$-orbital onto each other orbital of the lattice.
Independent of the point group, the operator $f_{\Gamma_1,\sigma}$ always has the form
\footnote{Since the trivial irrep of any point group is one dimensional we drop the index $\alpha$ in this case.}
\begin{eqnarray}
f_{\Gamma_1,\sigma} &=& \frac{1}{\sqrt{N_f}} \sum_l f_{l,\sigma}
\label{eq:f_Gamma1}
\end{eqnarray}
to fulfill Eq.~\eqref{eq:anti-commuator-21}.
We also assume that the total Hamiltionian is  invariant under the 
point group operations, i.\ e.\  $\e_l^f=const.$.
Breaking the point group symmetry by the quantum impurity Hamiltonian
would generate hopping terms between  the single particle orbitals of different 
irreducible representations which we exclude in our analysis below.

\begin{center}
\textit{Pseudo-gap in the effective hybridization} 
\end{center}

Using the equation of motion for Green's functions
we can calculate the complex hybridization function $\Delta_{\Gamma_1}(z)$ of the one dimensional irreducible representation $\Gamma_1$, which completely
determines the influence of the bath on the single $f_{\Gamma_1,\sigma}$-orbital \cite{Bulla2008} and
enters the non-interacting Green's function
$\langle\langle f^\prime_{\Gamma_1,\sigma},f^{\dagger\prime}_{\Gamma_1,\sigma}\rangle\rangle^0(z)=\left[z-\Delta_{\Gamma_1}(z)\right]^{-1}$:
\begin{align}
\Delta_{\Gamma_1}(z)&=
\sum_{\vec{k}}\frac{\left|V_{\vec{k},\Gamma_1}\right|^2}{z-\epsilon^c_{\vec{k}}}.
\label{eqn:HybFunction}
\end{align}
Inserting equations \eqref{eq:Vk1} and \eqref{eq:ek}, the imaginary part of the hybridization function, $\Gamma_{\Gamma_1}(\omega)=\Im\Delta_{\Gamma_1}(z)$, reads 
\begin{align}
\Gamma_{\Gamma_1}(\omega)=
\frac{\pi V^2}{N_f N_c}
\sum_{\k}\delta(\w-\e^c_{\k})\left|\frac{\e^c_{\k}-\e^c}{t}\right|^2
\label{eq:hyb-gap}
\end{align}
and, for $\e^c=0$, obviously exhibits a pseudo-gap $\Gamma_{\Gamma_1}\propto|\w|^r$ with $r>1$

Consequently, the coupling of the $f_{\Gamma_1,\sigma}$-orbital to the host is irrelevant for the resulting fixed point 
in the sense of an RG treatment,
and a single localized orbital decouples, such that single occupancy of this orbital leads to a stable local moment.
Since the effective conduction bands of the other irreducible representations in general do not decouple and the $f_{\Gamma_1,\sigma}$-orbital is a uniform mixing of the original $f_{l,\sigma}$-orbitals
with the amplitude $N_f^{-1/2}$, each of the local moments
in real space gets only partially screened by a fraction of $(N_f-1)/N_f$.

The spatial location of this decoupled orbital depends on the relative strength of the coupling $V/t$, just as in the lattice model.
To understand this we need to recall, that the pseudo-gap has some specific width $\delta_{\text{gap}}$, which is proportional to the hopping $t$ and which defines the energy scale at which the conduction band electrons (Wilson sites) gradually decouple.
In the wide band limit, $V/\delta_\text{gap}\to 0$, the itinerant electrons decouple from the $f$-orbital way before the screening sets in and the decoupled orbital has pure $f$-character.
In the other limit, $V/\delta_\text{gap}\to \infty$, its vice verse. Even if the remaining Wilson sites decouple on the energy scale of $\delta_{\text{gap}}$, the screening of the impurity is nearly completed and, consequently, the decoupled orbital has mainly $c$-character.

In case of $V/t\to\infty$ and half filling,
we can understand the decoupling in a purely local picture.
At each impurity site, the $c$- and $f$-orbitals form a binding and anti-binding linear combination which are energetically separated by $V$ and the binding one is doubly occupied.
The hopping of a single electron located in the conduction electron orbital at the Kondo-hole 
to a neighboring lattice site gets suppressed, since 
such a process involves high energy excitation of the order $V$ due to adding of electron into the anti-binding 
orbital. Due to suppression of the local hopping,
the decoupled orbital localizes in the conduction electron orbital of the Kondo-hole site in this limit.
We note that this is exactly the same behavior for large hybridization as  
emerged  from the single-particle super cell discussion of the decoupled orbital $d_k$ defined in Eq.\ \eqref{eq:d-k}.

\subsection{NRG and low-energy Hamiltonian of the MIAM in the wide band limit}  
\label{sec:low-energy-model}

For the Kondo-hole problem we have primarily a single charge-neutral substitution in mind. 
It has already been shown \cite{Clare96} that the spatial extension of the induced bound state is very important for
its physical properties. Within a DMFT treatment \cite{Schlottmann91I,Schlottmann91II} these spatial correlations
and the interactions of the induced local moments with the rest of the lattice are lost. Since an 
exact treatment of a PAM is not possible, we follow a different strategy: We include the spatial correlations
by investigating very large local correlated clusters which captures some of the lattice physics  \cite{MIAM2020} but sacrifice the feedback of the rest of the correlated sites onto the smaller cluster. As long as this feedback does not alter the physics - for instance by additional decoupling of the remaining screening channels as in the metal insulator transition of a Hubbard model - the FP structure is fixed by the geometry of the cluster and the lattice feedback would only change the absolute values of the low-energy scales.

In order to study the complex multi impurity models in the strongly interacting regime we use the NRG \cite{Krishna-murthy1980I,Krishna-murthy1980II} in combination with a wide band approximation \cite{MIAM2020} in the following.

The NRG was developed by Wilson in 1975 \cite{Wilson1975} to accurately solve the SIAM in the featureless wide band limit.
Since then,
the NRG has been extended to include
the energy dependence of the conduction band electrons and a tremendous amount of
papers have been devoted to quantum impurity problems in various incarnations  
addressed with tailored versions of the NRG. For a detailed review 
and  examples see Ref.\ \cite{Bulla2008}.

The central point of NRG is the construction of the semi infinite Wilson chain, which results from a tridiagonalization of the prior logarithmically discretized conduction band continuum.
For a single impurity at site $l$ and a local $f$-$c$ hybridization, the local conduction electron states $|0_\sigma\rangle=c^\dagger_{l,\sigma}|\text{vac}\rangle$
are used as starting vectors for the iterative construction of the Lanczos vectors, since these are the only states that directly couple to the impurity.
This mapping onto a linear chain problem requires an orthonormal basis set.
Since the expansion of the local Wannier conduction electron orbitals at different impurity sites comprise
linear combinations of energy states that are not orthogonal due to the phase correlations of the underlying plain waves,
such an approach to construct semi infinite Wilson chain for  multi impurity models is not straight forward \cite{MitchellBulla2015}.
Essentially, the hybridization part of the Hamiltonian, Eq.~\eqref{eq:Hhyb}, needs to be rewritten in terms of orthonormal conduction electron states.

In order to keep the minimal Kondo hole model tractable with a relatively large number of correlated
sites surrounding the hole, we employ a 
recently developed mapping \cite{MIAM2020} onto an effective low energy Hamiltonian which we summarize in the following. This mapping 
becomes exact in the wide band limit and enables us to solve the model using the NRG.

The effect of the host conduction band onto the dynamics of the correlated lattice sites
is completely determined by the  hybridization function matrix,
\begin{align}
 \Delta_{lm}(z)=V_l V_m G^0_{c,lm}(z),
 \label{eq:delta-mat-def}
\end{align}
where $G^0_{c,lm}(z)$ is the free conduction band electron Green's function in real space accounting for an electron transfer from site $l$ to site $m$. 

The exact real space multi-impurity Green's function matrix of the dimension $N_f\times N_f$, in the absence of the Coulomb interaction,
$U_l=0$, is given by the matrix
\begin{eqnarray}
\mat{G}_f(z) &=&[ z - \mat{E} - \mat{ \Delta} (z)]^{-1},
\end{eqnarray}
where the matrix $\mat{E}$ contains the single-particle 
energies of the localized $f$-orbitals and the matrix elements of the self-energy matrix $\mat{ \Delta} (z)$ are given in Eq.\ \eqref{eq:delta-mat-def}.

In the wide band limit, $V_i/t\to 0$, the energy dependence of the hybridization function matrix can be neglected, $\mat{\Delta}(\w-i0^+)\approx\mat{\Delta}(-i0^+)$,
and we can absorb the real part into the energy matrix: $\mat{E}\to \mat{E}^\prime=\mat{E}+\Re\mat{\Delta}(-i0^+)$.
Using the unitary transformation $\mat{U}$ that diagonalizes the remaining imaginary part, $\mat{\Gamma}^\text{diag}=\mat{U}\Im\Delta(-i0^+)\mat{U}^*$,
the approximated Green's function reads

\begin{eqnarray}
\mat{G}_f(\w-i0^+)&\approx&\mat{U}\mat{G}_{\tilde{f}}(\w-i0^+)\mat{U}^*
\nonumber \\
&= &\mat{U}[\w-i0^+-\tilde{\mat{E}}^\prime-i\mat{\Gamma}^\text{diag}]^{-1}\mat{U}^*,
\end{eqnarray}
with $\tilde{\mat{E}}^\prime=\mat{U}\mat{E}^\prime\mat{U}^*$.
Consequently, the single particle Green's function matrix 
$\mat{G}_{\tilde{f}}(\w-i0^+)$, in the eigenbase of $\Im\mat{\Delta}(-i0^+)$, can equally be generated by an
effective single particle Hamiltonian $\tilde{H}_\text{sp}$ which has the following form:
\begin{align}
\label{eqn:MIAM_hyb}
\tilde{H}_{\rm sp} = \tilde{H}_{\rm cl} + \tilde{H}_{\rm hyb}.
\end{align}
The cluster part of the mapped  Hamiltonian $\tilde{H}_{\rm sp}$,
\begin{eqnarray}
\label{eq:H-cl}
\tilde{H}_{\rm cl}&=& \sum_{m,l}  \tilde{E}'_{lm} \tilde{f}^\dagger_{l}\tilde{f}_m,
\end{eqnarray}
defines the single-particle Hamiltonian of the correlated orbitals in the new basis that
have acquired additional orbital hopping terms due to $\Re\Delta_{lm}(-i0^+)$, mediated by the conduction band electrons of the host.
Defining the effective coupling constants $\bar V_n$, which result from the $n$ eigenvalues $\Gamma^\text{diag}_n=\pi\bar{V}_n^2\rho^0(0)$ and the conduction band DOS $\rho^0(0)$ for $\e^c=0$, the second part,
\begin{eqnarray}
\label{eq:H-hyb}
 \tilde{H}_{\rm hyb} &=& \sum_{n=1}^{N_f}
 \sum_{\k}
 (\e^c_{\k}-\e^c)c^\dagger_{\k,n}c_{\k,n}
\\
&& \nonumber
 +  
 \sum_{n=1}^{N_f}  \sum_{\k} \left(\frac{\bar V_n}{\sqrt{N_c}} c^\dagger_{\k,n}\tilde{f}_{n} +\text{h.c.}\right),
\end{eqnarray}
includes $N_f$ new effective conduction band channels and the flavor diagonal
coupling to the cluster orbitals for each conduction band flavor $n$.

Note that the $\tilde{f}$-orbitals decouple from the effective conduction band if the corresponding eigenvalue vanishes, $\Gamma^\text{diag}_n=0$, which implys an incomplete 
rank of $\Im\mat{\Delta}(-i0^+)$.
Such a vanishing of the coupling $\bar{V}_n$ indicates a pseudo-gap in the energy dependent hybridization function as it appears in Eq.\ \eqref{eq:rho-eom-disance-R} and Eq.\ \eqref{eq:hyb-gap} in the context of the decoupled $d$-orbital in depleted lattice as well as impurity models. If the width $\delta_\text{gap}$ of the pseud-gap is larger than the coupling $V_n$, as is the case in the wide band limit, $V/\delta_\text{gap}\to 0$, the conduction band channel can be neglected and the decoupling in the hybridization part of the effective Hamiltonian in Eq.\ \eqref{eq:H-hyb} is fully justified.

In addition, the rank of $\Im\mat{\Delta}(-i0^+)$ can be used to distinguish between two types of MIAM's, see Fig. 1 of Ref.\ \cite{MIAM2020}. 
A MIAM of the first kind is defined by rank$[\Im\mat{\Delta}(-i0^+)]=N_f$, whereas a MIAM of the second kind contains decoupled $\tilde{f}$-orbitals and, hence, rank$[\Im\mat{\Delta}(-i0^+)]<N_f$.
Note that the PAM is a representative of a MIAM of the second kind \cite{MIAM2020}.

\section{Screening mechanisms in Kondo hole Hamiltonians}
\label{sec:SreeningMechanism}

So far, using a supercell analysis and the modified Lieb-Mattis theorem, we predicted the 
emerging of LM FPs at PH symmetry, when some correlated $f$-orbitals are removed in a dense lattice and impurity models.
In this section we study the possible Kondo screening mechanisms of these local moments.

In Sec.\ \ref{sec:TypeI_3Imp} and Sec.\ \ref{sec:TypeI_7Imp} we  focus on the Type I Kondo hole models, where the local cluster magnetic moment is stable at PH symmetry, $S^\text{tot}_z=S^{\text{cl}_0}_z\not=0$, and demonstrate that RKKY couplings can induce an effective AF Kondo coupling when (i) the unoccupied hole orbital is considered in the modulation or (ii) the band center $\e^c$ is shifted. In both of these cases PH symmetry is broken such that the Lieb-Mattis theorem is not applicable any longer.

In Sec.\ \ref{sec:TypeII} we maintain PH symmetry but study the Kondo screening of Type II Kondo hole models, where the local cluster magnetic moment is AF coupled to the remaining continuum, $S^\text{tot}_z<S^{\text{cl}_0}_z\not=0$.
The delocalization of the local moments by a finite Coulomb interaction $U>0$
leads to its coupling to another conduction electron channel.  A KT transition is found with a critical coupling $U_c=0$
and exponentially vanishing low energy scale with a counter intuitive Kondo coupling $\propto U^n$, where $n>1$ depends on the precise model.
The supercell analysis is recovered only at $U=0$, while the finite $U$ results are in accordance with the 
modified Lieb-Mattis theorem.

\subsection{Type I Kondo hole model: Single hole surrounded by nearest neighbor correlated orbitals }
\label{sec:TypeI_3Imp}

\begin{figure}[tbp]
\begin{center}
\includegraphics[width=0.42\textwidth,clip]{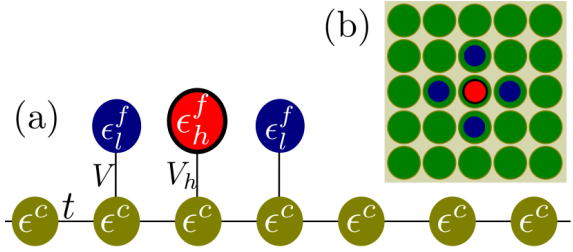}
\caption{Simplest realization of a hole with finite on-site energy $\epsilon_\text{h}$ inside an array of impurities in (a) one and (b) two dimension.}
\label{fig:CN_1d2d_Kh}
\end{center}
\end{figure}

In experiments, the removal of local moments in a dense Kondo lattice is typically realized by the substitution of magnetic 
atoms such as $\mathrm{Ce}$ or $\mathrm{Yb}$ by the nonmagnetic counterparts Th or La.
In the literature \cite{Clare96,GrenzebachAndersCzychollPruschke2008,Potthoff2015} 
and in the previous sections, we modeled that situation by removing the correlated site completely. Since La has excitable
$4f$ states at some large but finite energy that are just not occupied in equilibrium, we include these $f$ states
by demanding that $\e_h^f>\Gamma_0$ rather than removing them completely.
A schematic sketch of the extended model for such a Kondo-hole is depicted in Fig.\ \ref{fig:CN_1d2d_Kh}
which is a realistic generalization of the scenario shown in Fig.\ \ref{fig:CN_1d2d}.

The Hamiltonian $H=H_{\rm MIAM} + H_{\text{hole}}$ 
extends $H_\text{MIAM}$ in Eq.\ \eqref{eqn:H_MIAM}  by taking an high-energy unoccupied f-orbital at the hole location
explicitly into account:
\begin{align}
H_{\rm hole} = \sum_{\sigma}\left(\e^f_h f^\dagger_{h,\sigma}f_{h,\sigma}+\frac{V_h}{\sqrt{N_c}}\sum_{\k}\left[ c^\dagger_{\k,\sigma}f_{h,\sigma}+\text{h.c.}\right]\right).
\end{align}
The index $h$ labels the Kondo hole operators, $V_h$ denotes the coupling 
strength of the conduction electrons to  the Kondo-hole orbital 
placed at $R=0$ with the orbital energy $\e^f_h$.
Since the hole orbital is assumed to be 
nearly unoccupied in a realistic description we can neglect a possible Coulomb repulsion $U_h$ in $H_\text{hole}$ for simplicity.
All nearest neighbor correlated orbitals are set to be equal.

\subsubsection{The non-interacting limit}
\label{sec:nonInteractinLimit}

Before addressing the fully interacting problem using the NRG,
it is helpful to understand the analytically exactly solvable non-interacting limit $U_l=0$.
The matrix $\mat{\Delta}(z)$, defined in Eq.\ \eqref{eq:delta-mat-def},
is block diagonal in the irreducible representations of the local point group.
Since the hole is located at the origin, its orbital transforms according to the trivial representation $\Gamma_1$,
such that the $\Gamma_1$-subspace is now two dimensional.
Therefore, using the definition of $V_{\k,\Gamma_1}$ in Eq.\ \eqref{eq:Vk1} and dropping the spin index $\sigma$ for better readability, the matrix $\mat{\Delta}_{\Gamma_1}(z)$ of this subspace reads for each spin channel
\begin{eqnarray}
\bm{\Delta}_{\Gamma_1}(z)&=&
\begin{pmatrix}
\Delta_{hh}(z) & \Delta_{h+}(z)\\
\Delta_{+h}(z) & \Delta_{++}(z) 
\end{pmatrix}
\nonumber \\
&=&\frac{1}{N_c}\sum_{\k}
 \begin{pmatrix}
\frac{V_\text{h}^2}{z-\e^c_{\k}} &  \frac{V_\text{h}V_{\k,\Gamma_1}}{z-\e^c_{\k}} \\
 \frac{V_\text{h}V_{\k,\Gamma_1}}{z-\e^c_{\k}} &  \frac{V_{\k,\Gamma_1}^2}{z-\e^c_{\k}} \\
\end{pmatrix},
\label{eqn:Hybfunction_threeImp}
\end{eqnarray}
where $f_+=1/\sqrt{N_f}\sum_l f_l$ corresponds to the even combination of the $f$-orbitals at 
the neighboring sites $\vec{R}_l\not=0$.

The hole Green function $G_{hh}(z)$ and the $f_+$ GF 
are given by
\begin{subequations}
\begin{eqnarray}
G_{hh}(z) &=&  \frac{1}{z-\e^f_h -\Delta_{hh}(z)   - \frac{\Delta^2_{h+}(z)}{z-\e^f_+ -\Delta_{++}(z) }} \label{eq:Ghh} \\
G_{++}(z)&=&  \frac{1}{z-\e^f_+ -\Delta_{++}(z)   - \frac{\Delta^2_{h+}(z)}{z-\e^f_h -\Delta_{hh}(z) }}  \label{eq:G++}
\end{eqnarray}
\end{subequations}

For $V_\text{h}=0$ and $V\not=0$, we recover  the Kondo-hole modulation 
presented in Sec.\ \ref{sec:finite_cluster}. 
The even (+) combination decouples for $\e^c=0$ from the conduction band at low temperatures due to $\Im\Delta_{++}(\omega-i0^+)\propto\omega^2$  for a featureless conduction band without van Hove singularity at $\w=0$.
In addition the Kondo-hole orbital is  trivially
disconnected from the problem.

In the opposite limit, $V_\text{h}\not=0$ and $V=0$, only the Kondo-hole couples to the conduction band.
With $\Gamma^h_0=\Im\Delta_{hh}(-i0^+)=\pi V_h^2 \rho_0$ its spectral function is given by
\begin{align}
 \rho^{V=0}_{f_h}(\e^f_h,\omega)=\frac{\Gamma_{0}^{h}}{\pi\left([\omega-\epsilon^f_h]^2+[\Gamma_{0}^{h}]^2\right)}
\label{eqn:Spec_Khole_V0}
\end{align}
in the wide-band limit where $\Re \Delta_{hh}(-i0^+)\to 0$. 

Employing the mapping onto an effective Hamiltonian as laid out in Sec.\ \ref{sec:low-energy-model} requires the
diagonalization of the imaginary part of all sub-matrices in each subspace separately.
However, for $\e^c=0$ the off-diagonal matrix elements in the $\Gamma_1$ subspace are purely real,
\begin{align}
\label{eq:effective-Delta-plus}
 \bm{\Delta}_{\Gamma_1}(-i0^+)=
\begin{pmatrix}
 i \Gamma_{0}^{h} & t_{h+} \\
   t_{h +} & 0              
\end{pmatrix},
\end{align}
and account for an effective
hopping $t_{h+}$ between the two orbitals.
Obviously, the rank of $\bm{\Delta}_{\Gamma_1}$ is one, and only the Kondo hole directly couples to the $\Gamma_1$ conduction band states. The resulting 
effective single particle Hamiltonian 
is extracted to
\begin{align}
 \tilde{H}_{\Gamma_1}&=\sum_{\k}\left[\e^c_{k}c^\dagger_{\k}c_{\k}+ V_\text{h}(c^\dagger_{\k}f_h+f^\dagger_h c_{\k})\right]\nonumber\\
&+\sum_{\nu\in\{h,+\}}\e^f_\nu f^\dagger_\nu f_\nu+ t_{h+}\left(f^\dagger_{h}f_{+}+f^\dagger_{+}f_{h}\right)
\end{align}
and describes the Kondo hole that couples to the $\Gamma_1$ conduction band states and the even combination $f_+$. 
Note that $t_{h+}\propto V$ depends on the coupling $V$ of the original $f_l$-orbitals.

Note that deviations from $\e^c=0$ result in finite imaginary off-diagonal matrix elements in Eq.\ \eqref{eq:effective-Delta-plus} and a precursive diagonalization is necessary in order to obtain 
an effective description with independent conduction band states for each orbital.
Since this additional change of basis mixes the single particle properties of the hole and local moment $f$-orbitals and, consequently, makes the interpretation more difficult, we first concentrate on $\e^c=0$ and discuss the case $\e^c\not=0$ later on.

Substituting \eqref{eq:effective-Delta-plus} into
\eqref{eq:Ghh} and \eqref{eq:G++}, the  spectral functions of the two $\Gamma_1$
orbitals are approximated to
\begin{align}
 \tilde{\rho}_{f_h}(\e_h^f,\omega)&=\frac{\Gamma_{0}^{h}}{\pi}\left(\left\{\omega-\e^f_\text{h}-\frac{t^2_{h+}}{\omega-\epsilon^f_+}\right\}^2+ [\Gamma_{0}^h] ^2\right)^{-1}
\label{eqn:Spec_Khole}
\\
\tilde{\rho}_{f_+}(\e_h^f,\omega)&=\frac{\tilde{\Gamma}_{0}^{+}(\omega)}{\pi\left(\left\{\omega-\e^f_+-\Delta \e_+^f(\omega)
\right\}^2
+\left\{\tilde{\Gamma}_{0}^{+}(\omega) \right\}^2
\right)},
\label{eqn:Spec_f+}
\end{align}
where the energy shift $\Delta\e_+^f(\omega)$ is given by
\begin{equation}
\Delta\e_+^f(\omega) = t^2_{h+} 
\frac{\omega -\e_h^f}{(\omega -\e_h^f)^2 + [\Gamma_0^h]^2},
\end{equation}
and the new effective width of the $f_+$-orbital spectrum reads
\begin{eqnarray}
\tilde{\Gamma}_{0}^{+}(\omega) = t^2_{h+}\frac{\Gamma^h_0}{(\omega -\e_h^f)^2 + [\Gamma_0^h]^2}.
\end{eqnarray}
$\Delta\e_+^f(\omega)$ and $\tilde{\Gamma}_{0}^{+}(\omega)$  
are related to the real and the imaginary part of the $f_+$ GF and vanish in the usually considered limit
$|\e_h^f|\to \infty$.  
In this limit a disconnected $f_+$-orbital that carries a free moment at finite $U$ is recovered. For a finite
$\e_h^f$, however,  with  $|\e_h^f|\gg \Gamma^h_0$, 
$\tilde{\Gamma}_{0}^{+}(\omega)$ can be approximated by 
\begin{eqnarray}
\tilde{\Gamma}_{0}^{+}(\omega) &\approx &\Gamma^h_0  \frac{ t^2_{h+}}{(\e_h^f)^2}.
\label{eq:Gamma0+}
\end{eqnarray}

At finite $U_l$ the  $f_+$ orbital  carries the finite magnetic moment located in the vicinity of the Kondo hole.
We have just proven that this moment does not decouple from the conduction band: $\tilde{\Gamma}_{0}^{+}(\omega)$
allows for another Kondo screening mechanism that has previously been
overlooked in the discussions of the Kondo hole physics.

\begin{figure}[tbp]
\begin{center}
\includegraphics[width=0.45\textwidth,clip]{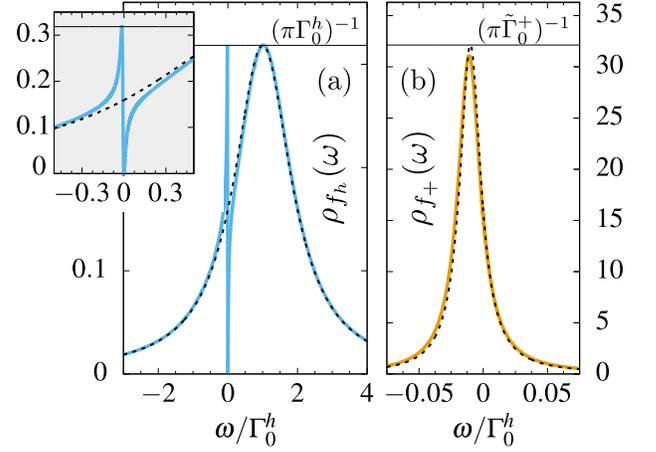}
\caption{Non-interacting spectral functions of the $\Gamma_1$ subspace of the simplified Kondo hole model in 1d. Panel (a) depicts the spectrum of the hole orbital (light blue solid line) obtained from Eq.\ \eqref{eq:Ghh} and the wide band approximation $\rho_{f_\text{h}}^{V=0}(\omega)$ in Eq.\ \eqref{eqn:Spec_Khole_V0}, whereas panel (b) shows the spectral function of the even (+) combination of the outer impurities (orange solid line) obtained from Eq.\ \eqref{eq:G++} in comparison with the wide band approximation in Eq.\ \eqref{eqn:Spec_f+} (black dashed line). The Inset depicts a zoom of panel (a) around $\w=0$.
Parameters are $D/\Gamma^h_0=100$, $\e_h^f/\Gamma^h_0=1$, $\e_+=0$, $\e^c=0$ and $V_\text{h}=10V$.}
\label{fig:Speck_threeImp}
\end{center}
\end{figure}

For any finite orbital hopping $t_{h+}$, $\rho_{f_h}(\omega)$ will be gapped 
close to $\e^f_+$ as can be seen in Eq.\ \eqref{eqn:Spec_Khole}.
In Fig.\ \ref{fig:Speck_threeImp} we illustrate the properties of the non-interacting spectral functions of the $\Gamma_1$
subspace for the 1d Kondo hole model that is schematically diepicted in Fig.\ \ref{fig:CN_1d2d_Kh}(a).
In order to show comparable frequency intervals we chose $\e^f_h$ only very moderately above the chemical potential,
i.\ e.\  $\e^f_h/\Gamma_0^h=1$, and a particle-hole symmetric correlated orbital by $\e^f_+=0$.
To simulate the parameter regime of the interacting case with $T_{\text{K},h}\approx\Gamma_{0,h}\gg T_{\text{K},l}$,
we artificially enlarge the hybridization matrix element to the Kondo-hole orbital, $V_\text{h}=10 V$, and set $V_1=V_2=V$
for the non-interacting model.

Fig.\ \ref{fig:Speck_threeImp}(a) depicts the full spectral function $\rho_{f_h}(\omega)$ of the Kondo hole (light blue solid line)
obtained from Eq.\ \eqref{eq:Ghh} and the wide band approximation $\rho_{f_\text{h}}^{V=0}(\omega)$ in Eq.\ \eqref{eqn:Spec_Khole_V0} 
for $V=0$ (black dashed line).
The inter-orbital coupling $t_{h+}/\Gamma_0^h =0.1\sqrt{2}$ induces a sharp anti-resonance at $\w=\e^f_+=0$
into the spectral function of the Kondo hole. This anti-resonance illustrates the feedback of the free 
$f_+$-orbital onto the Kondo hole spectral function.

Fig.\ \ref{fig:Speck_threeImp}(b) depicts the corresponding $\rho_{f_+}(\omega)$ from Eq.\ \eqref{eq:G++} 
(orange solid line) in comparison with the wide band approximation in Eq.\ \eqref{eqn:Spec_f+} (black dashed line). 
The relevant energy scale, that describes the hybridization of the $f_+$-orbital and, consequently, the height and width of the 
spectral function $\rho_{f_+}(\omega)$, is given by $\tilde{\Gamma}_{0}^{+}$ and depends on the onsite energy $\e_h^f$ of the Kondo hole.

If we consider a decoupling of the correlated orbitals from the conduction band, we can identify a two stage process. In a first step
a Lorentzian resonance curve is generated in the vicinity of the single particle energy of the hole whose width is governed by $\Gamma_h$. Now we switch on
a finite hybridization $V$ that couples indirectly the $f_{\Gamma_1}$ orbital  to the conduction band via the Kondo hole orbital. $\rho_{f_h}(\w)$ serves
as effective density of states and an anti-resonance is generated in  $\rho_{f_h}(\w)$ as in the local conduction electron DOS of the resonant level model \cite{LebanonSchillerAnders2003}.
This physics prevails for correlated orbitals as demonstrated in the next sections.

\subsubsection{The interacting model: $U>0$}

In the following we consider the three impurity model on a 1d tight binding chain as depicted in Fig.\ \ref{fig:CN_1d2d_Kh}(a). We set the parameters to 
$\Gamma^h_0=\Gamma^l_0=\Gamma_0$, $U_l=-2\e^f_l=10\Gamma_0$, $U_h=0$, $D=10\Gamma_0$ and analyze the low energy physics with regard on the remaining two parameters $\e^f_h$ and $\e^c$.
We solve the model by employing the NRG
using the wide band approximation described in Sec.\ \ref{sec:low-energy-model}, a NRG discretization parameter of $\lambda=3$ and kept $N_\text{s}=6000$ NRG states after each iteration.

The low energy scale  $T_0$  governs the crossover from the
last unstable LM FP to the singlet strong coupling FP. We define this temperature 
via the midpoint between the two FP residual entropies:
\begin{align}
S_\text{imp}(T_0)=1/2 k_\text{B}\ln(2).
\label{eq:def_T0}
\end{align}
$T_0$ corresponds to the Kondo temperature in  
the SIAM or Kondo model 
which depends exponentially on the dimensionless coupling constant $g$, $T_K\propto \exp(-1/g)$, where
$g=J\rho$ factorizes in a product of the local Kondo coupling $J$ and the density of states $\rho$ of a featureless conduction band.

\begin{figure}[tbp]
\begin{center}
\includegraphics[width=0.45\textwidth,clip]{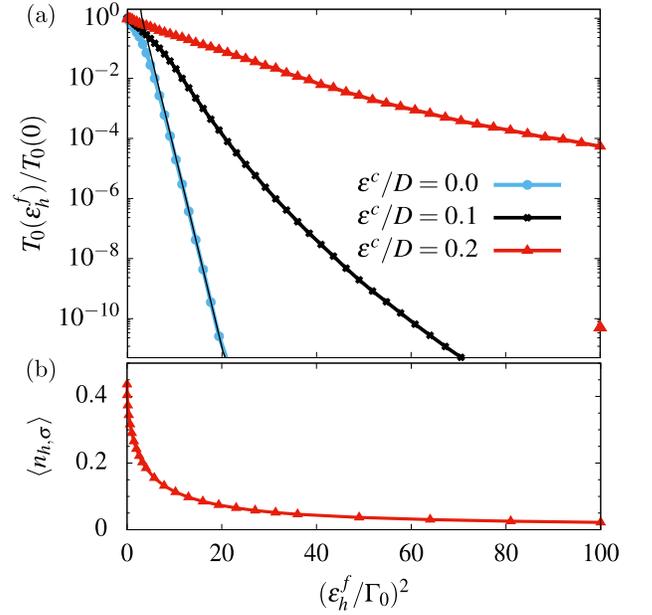}
\caption{(a) normalized low energy scale $T_0(\e^f_h)/T_0(0)$ logarithmically plotted as function of the squared hole orbital on-site energy $(\e^f_h/\Gamma_0)^2$ for three different fillings of the conduction band: $\e^c/D=0$ (light blue line points), $\e^c/D=0.1$ (black line points) and $\e^c/D=0.2$ (red line points with triangles).
The thin black line corresponds to a fit of the $\e^c/D=0$ curve to a function of the form $\exp[-\alpha_\text{hole}(\e^f_h/\Gamma_0)^2]$, which results in $\alpha_\text{hole}=1.4282$, and the single triangle on the right-hand side of the y-axis indicates the value of $T_0(\e^f_h=\infty)/T_0(0)$ for $\e^c/D=0.2$.
(b) expectation value of the hole occupation number $\langle n_{h,\sigma}\rangle$ for $\e^c/D=0.2$ on the same interval for $(\e^f_h/\Gamma_0)^2$ as in (a).}

\label{fig:3Imp_T0_mu}
\end{center}
\end{figure}

In the previous section, we predict a Kondo screening of the hole induced local moment  mediated by
the Kondo hole orbital. Since the effective density of states seen by the momentum carrying orbital 
is proportional to $(\e^f_h)^{-2}$
according to Eq.\ \eqref{eq:Gamma0+}, we logarithmically plotted the normalized low energy scale $T_0(\e^f_h)/T_0(0)$ as function of the on-site energy of the hole orbital, $(\e^f_h/\Gamma_0)^2$, for three different fillings of the conduction band, adjusted by $\e^c/D=0$, $0.1$ and $0.2$ to test its exponential form.

\begin{figure}[tbp]
\begin{center}
\includegraphics[width=0.45\textwidth,clip]{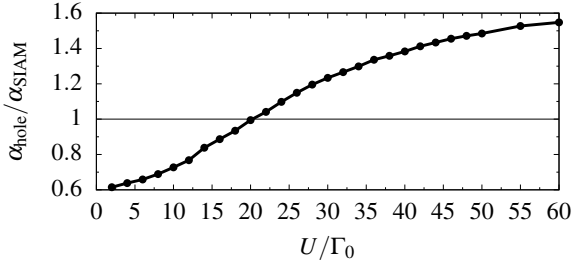}
\caption{U dependency of the ratio $\alpha_\text{hole}/\alpha_\text{SIAM}$. For each value of $U$, $\alpha_\text{hole}$ results from a fit of the $T_0(\e^f_h)$ NRG data to the function in Eq.\ \eqref{eq:T0_ec-0} and $\alpha_\text{SIAM}$ is given by Eq.\ \eqref{eq:aplha_SIAM}.}
\label{fig:3Imp_alpha-U}
\end{center}
\end{figure}

We added a fit of the $\e^c=0$ data points (light blue dots)
to a function of the form
\begin{align}
 f(\e^f_h)\propto\exp[-\alpha_\text{hole}(\e^f_h/\Gamma_0)^2]
\label{eq:T0_ec-0}
\end{align}
as thin black line to Fig.\ \ref{fig:3Imp_T0_mu}(a). 
While for $\e^f_h/\Gamma_0\ll 1$, the full energy dependency of the effective $\rho$ influences the absolute value of $T_0$ and deviation to the simplified fit function is visible, 
we clearly see that the data agrees perfectly with the fit function for $\e^f_h/\Gamma_0\gg 1$. 
Within this effective Kondo model, the prefactor $\alpha_\text{hole}/\Gamma_0$ is a measure of its inverse effective Kondo coupling $1/J$.
The $U$ dependency of the ratio $\alpha_\text{hole}/\alpha_\text{SIAM}$ is shown in Fig.\ \ref{fig:3Imp_alpha-U}, where $\alpha_\text{SIAM}$ is the value that we would expect for $\alpha_\text{hole}$ if Eq.\ \eqref{eq:T0_ec-0} corresponds to $T_\text{K}$ of a SIAM with Coulomb interaction $U$ and $\Gamma_0$ replaced by $\tilde{\Gamma}_0^+$ in Eq.\ \eqref{eq:Gamma0+},
\begin{align}
\alpha_\text{SIAM}=\frac{\pi U \Gamma_0}{8 t_{h+}^2}=\frac{U}{\Gamma_0}\frac{\pi}{16}.
\label{eq:aplha_SIAM}
\end{align}

Since $\alpha_\text{hole}/\alpha_\text{SIAM}<1$ for small $U/\Gamma_0$, the low energy scale $T_0$ falls off more slowly as one would expect from a SIAM with the coupling strength given by $U/\tilde{\Gamma}_0^+$ in this limit, whereas for large $U/\Gamma_0\gtrapprox 20$ it is vice verse.

In order to understand this deviation from the SIAM $T_K$
we recall that the decoupled orbital carrying the local moment
is a even mixture of the original local moments in real space. Due to the rotation into the 
parity eigenstate basis, the Coulomb interaction matrix is rotated as well coupling
charge and spin in different parity subspaces. The delocalized Coulomb interaction in the parity orbital space yields a modification of the prefactor $\alpha_\text{hole}$ but does not change the screening mechanism.

For small $U/\Gamma_0\to 0$ the Kondo temperature in the odd subspace defines the largest energy scale, such that the influence of the non-local Coulomb matrix elements on the screening in the even subspace can be assumed to be small.
Moreover, the local Coulomb interaction in the parity eigenstate basis is given by $U^\prime = U/2$ and, consequently, $\alpha_\text{hole}$ should be reduced by a factor of $0.5$ compared to $\alpha_\text{SIAM}$. This is in qualitative agreement with our result of $\alpha_\text{hole}/\alpha_\text{SIAM}\approx 0.6 < 1$.

In the opposite limit of strong Coulomb interaction $U/\Gamma_0\gg 1$ the FM RKKY interaction $J^\text{FM}_\text{RKKY}$ dominates over the Kondo effect: $J_\text{RKKY}/T_\text{K}\gg 1$. Hence, at intermediate temperature the local moments couple to a triplet state and the system flows to an unstable LM fixed point with $S_\text{imp}=k_\text{B}\ln(3)$.
This triplet state is than screened in a two stage process by the odd and even electron continuum.
If we assume these even and odd conduction band channels to be identical and consider the limit $J_\text{RKKY}/T_\text{K}\to \infty$, the model can be mapped onto a $k=2$-channel spin $S=k/2$ Kondo problem with an reduced Kondo coupling $J^\prime_\text{K}= J_\text{K}/k$ \cite{Nevidomskyy2009,Schrieffer1967}, where $J_\text{K}$ denotes the original Kondo coupling of the individual spin $S=1/2$ local moments.
Consequently, this implys $\alpha_\text{hole}/\alpha_\text{SIAM}= 2 > 1$ which, again, is in qualitative agreement with our result of $\alpha_\text{hole}/\alpha_\text{SIAM}\approx 1.55>1$ for $U/\Gamma_0=60$. Note that $J_\text{K}$ of the even and odd channel are quite different such that the large spin $S=1$ is screened in a two stage process and, therefore, the ratio of $\alpha_\text{hole}/\alpha_\text{SIAM}= 2$ that are reported in the Refs.\ \cite{Nevidomskyy2009,Schrieffer1967} is only a rough estimate.

Now we extend the investigation to 
the more general situation which includes deviations from a symmetric conduction band, $\e^c\not=0$.
Then, the
imaginary part of $\bm{\Delta}_{\Gamma_1}(-i0^+)$ in Eq.\ \eqref{eqn:Hybfunction_threeImp} is given by
\begin{align}
\bm{\Gamma}_{\Gamma_1}=\Im\left[\bm{\Delta}_{\Gamma_1}(-i0^+)\right]= \Gamma_0
\begin{pmatrix}
 1 & \sqrt{2}\e^c/D \\
   \sqrt{2}\e^c/D & 2(\e^c/D)^2          
\end{pmatrix}.
\end{align}
Consequently, the hybridization matrix becomes non-diagonal in case of $\e^c\not=0$ and a precursive diagonalization of $\bm{\Gamma}_{\Gamma_1}$ is necessary in order to obtain a independent bath description as discussed in Sec.\ \ref{sec:low-energy-model}.
Since the $\text{rank}(\bm{\Gamma}_{\Gamma_1})=1$ in the even subspace
is not affected by introducing a finite $\e^c$,
still only one of the two orbitals in the new basis couples to a conduction band channel.
Nevertheless, the two stage screening of the local moments prevails: the orbital which is coupled to the conduction band forms a local Fermi-liquid and the remaining hopping matrix element, stemming from
$\Re\left[\bm{\Delta}_{\Gamma_1}(-i0^+)\right]$ replaces the  hybridization term in the second effective SIAM.

In contrast to the $\e^c=0$ case, however, the decoupled orbital 
developing a local moment at intermediate temperature is a mixture of the hole orbital and the even combination of the interacting impurities.
In this case, $\e^f_h$ does influence both, the effective $\rho$ as well as the effective Kondo coupling, 
such that the scaling $g\propto (\e^f_h)^{-2}$ is modified. The  dependence of the low energy scale $T_0$ on the hole orbital on-site energy $\e^f_h$  becomes more complex for $\e^c\not=0$ 
as seen by the black and red linepoints in Fig.\ \ref{fig:3Imp_T0_mu}(a).

With increasing $\e^f_h$ the low energy scale $T_0$ still decreases, however, for large values of the on-site hole orbital energy, $T_0(\e^f_h)$  saturates and approaches a constant value for $\e^f_h\to\infty$.
In this limit we can completely neglect the hole orbital such that the problem 
is reduced to those of a two impurity model, where the hybridization $\Gamma_e$ of the even combination of the two orbitals with the even parity states of the conduction band is given by $\Gamma_e=2\Gamma_0(\e^c/D)^2$.
Hence, small values of $\e^c/D$ result in an exponentially small low energy scale $T_0$ at which the local moment in the even subspace gets screened by conduction band electrons. For $\e^c/D=0.2$ we added the value of $T_0(\e^f_h\to\infty)/T_0(0)$ as single red triangle on the right-hand side of the y-axis of Fig.\ \ref{fig:3Imp_T0_mu}(a) to illustrate its asymptotic value after removing the Kondo hole orbital.

The hole orbital occupation $\langle n_{h,\sigma}\rangle$ is shown in Fig.\ \ref{fig:3Imp_T0_mu}(b)
for $\e^c/D=0.2$.
Even if the hole orbital is nearly unoccupied for $\e^f_h/\Gamma_0=10$,
its influence on the low energy scale $T_0$ is immense since $T_0(\e^f_h/\Gamma_0=10)/T_0(\e^f_h\to\infty)$ is of the order of $10^{6}$.
This demonstrates that the screening of the local moment in the even subspace is still driven by the RKKY coupling to the hole orbital for $\e^f_h/\Gamma_0=10$:
In general one can \textit{not} neglect the hole orbital solely for the reason that it is nearly unoccupied.

Recently, we demonstrated \cite{MIAM2020} that the screening in 
MIAMs 
of the second kind is a collective effect where the $f$-orbitals 
also contribute via the dynamically generated f-orbital hopping.
From this point of view it is not surprising that the influence of hole orbitals in such models can be quite stronger than one would expect from the slightly misleading Doniach picture \cite{Doniach77}, where each local moment in a lattice can be screened independently.
Pruschke et al.~\cite{PruschkeBullaJarrell2000} showed that in a Kondo insulator 
the effective medium of the DMFT 
comprises a pseudo-gap conduction band coupling function as well as a coupling  to a non-interacting localized f-orbital. This demonstrates that the singlet formation in the PAM is dominated by the effective f-f orbital interactions and less by the conduction band reflecting the exhaustion of conduction electron screening channels. 
The qualitative difference to an onsite Kondo effect becomes apparent in the local correlation function which opens up a hybridization gap.

The 1d three impurity problem is the simplest model to understand and study the effect of RKKY driven Kondo screening, where the density of states of an appropriate $f$-orbital degree of freedom serves as effective electron continuum to screen the local moment of some other $f$-orbital.
In 1d the rank of the hybridization matrix of a MIAM is given by rank$[\bm{\Gamma}]\leq 2$
\footnote{We have shown in Ref.~\cite{MIAM2020} that the rank$[\bm{\Gamma}]$ is limited by
the number of Fermi wave vectors hence by 2 in 1d.}.
 Therefore, the effective model consists of maximum two conduction band channels in the wide band limit and 1d, such that a MIAM with $N_f\geq 3$ is always a MIAM of the second kind (more $f$-orbitals than screening channels).

If we introduce a hole orbital and continuously shift the on-site energy $\e^f_h$ to infinity, we end up with a model where the number of $f$-orbitals is reduced by one, $\tilde{N}_f=N_f-1$.
Consequently, in case of $N_f=3$ we start with a MIAM of the second kind ($N_f>\text{rank}[\bm{\Gamma}]$) but end up with a MIAM of the first kind ($\tilde{N}_f=\text{rank}[\bm{\Gamma}]$) in 1d.
Since the local moment fixed point can only be stable in a MIAM of the second kind, this is the reason why the low energy scale $T_0(\e^f_h)$
in Fig.\ \ref{fig:3Imp_T0_mu}(a) does not vanish in general (except for $\e^c=0$ and $\e^f_h=\infty$ where rank$[\bm{\Gamma}]=1<\tilde{N}_f$):
There are always enough conduction band screening channels to completely compensate the local moments of the $f$-orbitals.

However, even if the mechanism investigated in this section is still relevant, the phase diagram of a more general Kondo hole model in which the correlated orbitals are not only placed at the nearest neighbors of the hole site might be slightly different due to the class change of the MIAM.
For this reason we extend the simplest Kondo hole model by adding additional correlated orbitals in the next section to ensure $\tilde{N}_f>\text{rank}[\bm{\Gamma}]$ for the full parameter range
to make connection with  the full lattice model.

\subsection{Type I Kondo hole model: Single hole surrounded by three nearest neighbor correlated orbitals}
\label{sec:TypeI_7Imp}

\begin{figure}[t]
\begin{center}
\includegraphics[width=0.47\textwidth,clip]{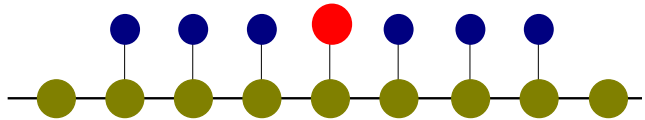}
\caption{Schematic sketch of the model comprising an array of $N_f=7$ $f$-orbitals on a 1d tight binding chain (green). In the center of the $f$-orbital array we replace the correlated impurity (blue) by a non-interacting hole orbital (red).}
\label{fig:7Imp_Model}
\end{center}
\end{figure}

In the wide band limit we can solve any  MIAM in 1d using a two channel NRG due to rank$[\bm{\Gamma}]\leq2$: We are only restricted by the dimension of the
impurity Hilbert-space that grows exponentially with the number of $f$-orbitals $N_f$.
Making use of the parity, total $S_z$ component and total particle number as conserved quantum numbers to divide the Hamiltonian in block-diagonal subspaces, we are able to handle up to
$N_f=7$ $f$-orbitals in such a way that we do not need to truncate before the first Wilson site of each channel has been added.
A schematic sketch of the model discussed in this section is depicted in Fig.\ \ref{fig:7Imp_Model}
where the hole position is indicated by the red dot. 
For the first 10 iterations we kept $N_s=15000$ states after each iteration and reduce to
$N_s=10000$ for the remaining ones. Further we set the NRG discretization to $\lambda=4$ and the bandwidth of the conduction electrons to $D/\Gamma_0=10$.

\subsubsection{Infinite on-site energy of the hole orbital}
\label{sec:7Imp_eh_to_infty}

\begin{figure}[tbp]
\begin{center}
\includegraphics[width=0.45\textwidth,clip]{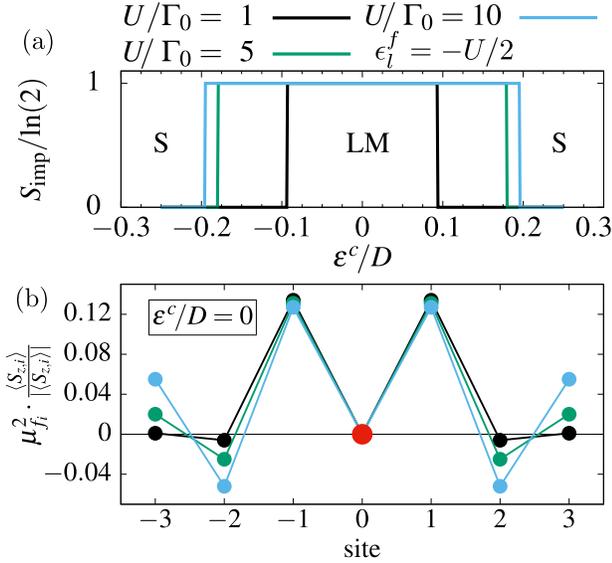}
\caption{NRG calculations for the model depicted in Fig.\ \ref{fig:7Imp_Model} with the hole $f$-orbital (red) removed: $\e^f_h=\infty$.
(a) entropy phase diagram as function of $\e^c/D$ for three different strengths of Coulomb interaction $U=-2\e^f_l=1$ (black), $5$ (green), and $10$ (lightblue). The local moment (LM) and strong coupling (SC) fixed point are separated by a QCP of KT type.
(b) Site dependent local magnetic moment of the individual $f$-orbitals at temperature $T/\Gamma_0=10^{-8}$ for  $\e^c=0$, the same values for $U=-2\e^f_l$ as in (a) and the hole orbital placed at site zero. The sign of the magnetic moment indicates the local polarization of the $S_{z,i}$ component.}
\label{fig:7Imp_S_Mag}
\end{center}
\end{figure}

We first concentrate on the case of $\e^f_h/\Gamma_0\to\infty$ where we can completely neglect the hole orbital.
In Fig.\ \ref{fig:7Imp_S_Mag}(a) we plotted the entropy 
phase diagram as function of the conduction band center $\e^c$ for three different values $U=-2\e^f_l=1\Gamma_0$ (black), $5\Gamma_0$ (green) and $10\Gamma_0$ (lightblue).
In each case we can differentiate between two phases. A local moment (LM) phase emerges around $\e^c=0$, which turns into a singlet (S) phase once a critical value $|\e^c|>\e^c_c(U)$ is exceeded: The larger the Coulomb interaction $U$, the larger the critical value $\e^c_c(U)$.
Our numerical results, however, show that the critical value approaches an upper bound $\e^c_c/D<\alpha\approx 0.22$. 
These results are in agreement with the analysis of the non-interacting supercell calculations 
in Sec.\ \ref{sec:supercell}: The super-cell operator $d_{\k}$  defined
in Eq.\ \eqref{eq:d-k} only decouples for $\e^c=0$ such that $\e^c_c$ must vanish for $U\to 0$.

Right at $|\e^c|=\e^c_c$, we found a quantum critical point (QCP) of Kosterlitz Thouless (KT) type.
For $|\e^c|=\e^c_c+\delta$ and small $\delta$ the system flows to an unstable local moment fixed point at intermediate temperature, which crosses over to a strong coupling fixed point on an exponentially suppressed low energy scale $T_0$.
Applying a Schrieffer-Wolff transformation \cite{SchriefferWol66} at the intermediate local moment fixed point would result in an effective single impurity Kondo model which flows to the strong coupling fixed point on the scale of $T_\text{K}$.
The effective Kondo coupling vanishes right at $\e^c=\e_c^c$ such that $T_\text{K}$ is exponentially suppressed, typically for a KT type QCP.
For large $\delta$ the antiferromagnetic RKKY interactions between the $\tilde{N}_f=6$ local moments dominate such that they lock into a inter impurity singlet state (IIS) that decouples from the conduction band electrons.
Consequently, the intermediate LM FP disappears and the RG directly crosses over from the high temperature unstable FP that contains $\tilde{N}_f=6$ independent spin-1/2 moments to the IIS FP with singlet ground state.
However, the SC and the ISS FP are adiabatically connected 
such that there is no additional QCP between a Kondo screened singlet and a RKKY driven singlet, just as in the extensively studied TIAM \cite{Affleck1995,Silva1996,Eickhoff2018,Jones1989,Jones1987,Jones1988}.

The interesting question arises how the local moment is distribution in real space around the Kondo hole in the stable LM fixed point  regime around $\e^c=0$. We know from
Eq.\ \eqref{eq:d-k} in Sec.\ \ref{sec:supercell} and the weakly interacting DMFT solution of Solli and Schlottmann \cite{Schlottmann91I,Schlottmann91II} 
that the magnetic moment should be located solely on the correlated orbitals 
nearest to the hole site in the limit of small $U\to0$, the wide band limit $V/D \to 0$ and PH symmetry.
In contrast to that the DMRG calculations of Clare C. Yu \cite{Clare96} demonstrate
for the half filled 1d Kondo lattice, that the spin density induced by a single Kondo hole
extends beyond the nearest neighbors -- see Fig.\ 3 in Ref.\ \cite{Clare96}.

In order to show that the 1d MIAM with $\tilde{N}_f=6$ interacting $f$-orbitals is already sufficient 
to bridge between complementary limits and interpolate between the weakly interacting DMFT solution and the strongly interacting Kondo limit using the wide band approach
we plotted the site dependent magnetic moment of the 
$f$-orbitals, $\mu^2_{f_i} \frac{\langle S_{z,i}\rangle}{|\langle S_{z,i}\rangle|}$,
for three different interaction strengths $U/\Gamma_0$ in Fig.\ \ref{fig:7Imp_S_Mag}(b).
We define this quantity via the local susceptibility,
\begin{eqnarray}
\mu^2_{f_i}  \frac{\langle S_{z,i}\rangle}{|\langle S_{z,i}\rangle|} &=&T\chi_{f_i}.
\end{eqnarray}
The local susceptibility was calculated by applying a very small local magnetic field $H_z/\Gamma_0=10^{-10}$ 
to the $f$-orbitals and expressing
\begin{eqnarray}
\chi_{f_i} &=& \frac{\expect{n^f_{i,\uparrow}}-\expect{n^f_{i,\downarrow}}}{2H_z}
\end{eqnarray}
where
$n^f_{i,\sigma} = f^{\dagger}_{i,\sigma}f_{i, \sigma}$. Note, that we are restricted to temperatures $T\gg H_z$ 
to remain in the linear response regime.

Whereas for $U/\Gamma_0=1$ the $f$-orbitals next to the hole site almost exclusively contribute to the spin density, an increasing strength of interaction $U$ leads to an increasing delocalization of the induced spin density.
The oscillatory behavior in the site dependent polarization of the individual magnetic moments is caused by the Friedel oscillations of the RKKY interaction and agrees perfectly with DMRG calculations \cite{Clare96}.

Alternatively we can start from the decoupled orbital
$d_{\k}$ defined in Eq.\ \eqref{eq:d-k} in the context of the supercell discussion.
Inserting the 1d nearest neighbor tight binding parameters we 
perform the Fourier transformation to obtain the decoupled real space orbital in each supercell $s$, which emerges when a single hole at site $l_h$ is considered:
\begin{align}
d_{s,l_h}=\frac{1}{\sqrt{2 (t/V)^2+1}}\left(\frac{t}{V}f_{s,l_h-1}+\frac{t}{V}f_{s,l_h+1}-c_{s,l_h}\right).
\label{eq:ds}
\end{align}
In the wide band limit, $V/t\to 0$, this yields
\begin{align}
d_{s,l_h}=\frac{1}{\sqrt{2}}\left(f_{s,l_h-1}+f_{s,l_h+1}\right).
\label{eq:ds_Vto0}
\end{align}
Note that this orbital is independent of the supercell size in particular for large supercells where the
single Kondo holes are so far apart  that they can be considered as independent impurities.
We take the  model depicted in Fig.\ \ref{fig:7Imp_Model} as a representation of the impuirty physics
with a large real space supercell
and add to $N_f=6$ sites a finite $U$. Within this model, we
calculate the effective magnetic moment of the $d$-orbital, 
$\mu^2_{d} =T|\chi_{d}|$, in the same way as above.

\begin{figure}[tbp]
\begin{center}
\includegraphics[width=0.5\textwidth,clip]{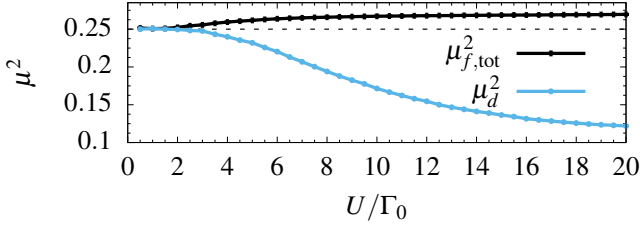}
\caption{NRG calculations for the model depicted in Fig.\ \ref{fig:7Imp_Model} for $\e^c=0$ and the hole $f$-orbital (red) removed: $\e^f_h=\infty$. Local magnetic moment as function of the Coulomb interaction $U=-2\e^f_l$ at temperature $T/\Gamma_0=10^{-8}$.
Comparison between the magnetic moment $\mu^2_d$ of the $d$-orbital defined in Eq. \eqref{eq:ds_Vto0} and the sum of the local magnetic moment of the individual $f$-orbitals, $\mu^2_{f,\text{tot}}$.}
\label{fig:7Imp_MagMomHole}
\end{center}
\end{figure}

In Fig.\ \ref{fig:7Imp_MagMomHole} we compare $\mu^2_{d}$ with the total magnetic moment 
contributions from all $f$-orbitals,
$\mu^2_{f,\text{tot}}=T|\sum_i \chi_{f_i}|$, as function of $U/\Gamma_0$. 
In compliance with the supercell prediction for $U=0$,
the total magnetic moment is located on the $d$-orbital for small interaction strengths, however,  increasing $U/\Gamma_0$  leads to a decreasing $\mu^2_d$, whereas the total magnetic moment stays nearly constant.
This implies that the local magnetic moment is transferred out of the decoupled $d$-orbital 
to the other correlated $f$-sites.
The small increase of $\mu^2_{f,\text{tot}}$ originates from the finite temperature $T/\Gamma_0=10^{-8}$. Due to the small magnetic field $H_z/\Gamma_0=10^{-10}$ we are restricted to temperatures $T\gg H_z$.

The spatial spreading of the magnetic moment upon increasing of $U$
can lead to an overlap between magnetic moments originating from neighboring Kondo holes
and, therefore, to interaction between these different bound states. We study this effect in Sec.\ \ref{sec:HHInteraction}.

Note that only the $f$-orbitals contribute to the total magnetic moment in the wide band limit 
employed here. Consequently, $\mu^2_{f,\text{tot}}$ is identical to Wilson's definition \cite{Wilson1975,Krishna-murthy1980I} of $\mu^2$ that is calculated by the difference of the total $\langle [S_z^{tot}]^2 \rangle$ of the system with and without the impurities present.
Nevertheless, for finite $V/D$ some of the non-interacting $c$-orbitals will also contribute to the magnetic moment of the ground state. This can be seen in Eq.\ \eqref{eq:ds} and has been shown for the 1d Kondo lattice by Clare C. Yu \cite{Clare96}.
Due to the mapping onto an effective low energy Hamiltonian required to solve the 1d MIAM via NRG, however, we can only access the wide band limit -- see Sec.\ \ref{sec:low-energy-model}
and the extensive discussion in Ref.\ \cite{MIAM2020}.

\subsubsection{Finite on-site energy of the hole orbital}
\label{sec:finite-hole-orbital}

For the rest of this section we now explicitly include the hole orbital with finite $\e^f_h/\Gamma_0$ in our NRG calculations and set the Coulomb interaction to $U=-2\e^f_l=10\Gamma_0$
for all other six correlated orbitals depicted in Fig.\ \ref{fig:7Imp_Model}.

\begin{figure}[tbp]
\begin{center}
\includegraphics[width=0.46\textwidth,clip]{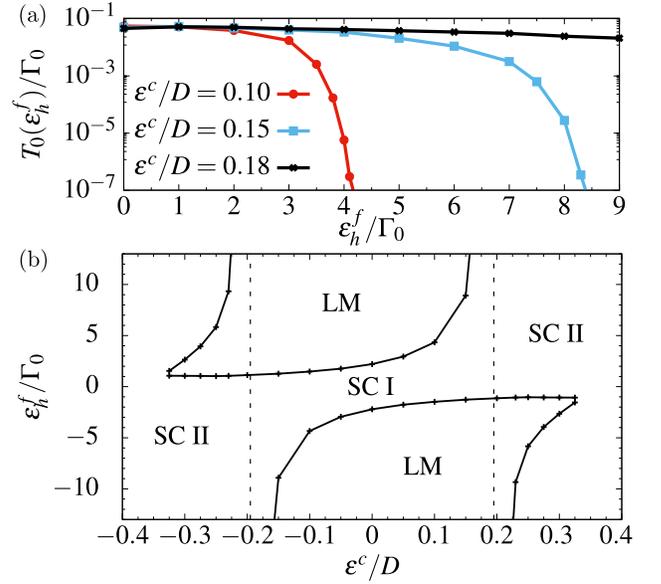}
\caption{NRG calculations for the 1d MIAM with $N_f=7$ $f$-orbitals as depicted in Fig.\ \ref{fig:7Imp_Model} and $U=-2\e^f_l=10\Gamma_0$.
(a) low energy scale $T_0$ as function of the hole orbital on-site energy $\e^f_h$ for three different fillings of the conduction band, $\e^c/D=0.1$ (red linepoints), $\e^c/D=0.15$ (lightblue linepoints) and $\e^c/D=0.18$ (black linepoints).
(b) $T/\Gamma_0=10^{-15}$ phase diagram as function of $\e^c$ (x-axis) and $\e^f_h$ (y-axis). The black linepoints separate a local moment (LM) phase from a singlet (SC) phase and the dashed vertical lines indicate the phase boundary at $|\e^f_h|=\infty$. 
The phase SC I (II) denotes points in the phase space that change to the LM phase (stay in the SC phase) for $|\e^f_h|\to\infty$.  }
\label{fig:7Imp_Phasendiag}
\end{center}
\end{figure}

We plotted the low energy scale $T_0$ as function of the hole orbital on-site energy $\e^f_h$ on the interval $[0,9]$ for three different  conduction band centers: $\e^c/D=0.1$ (red linepoints), $\e^c/D=0.15$ (lightblue linepoints) and $\e^c/D=0.18$ (black linepoints)
in Fig.\ \ref{fig:7Imp_Phasendiag}(a).
Since $\e^c/D<\e^c_c/D\approx 0.195$ is always below the critical value, 
the LM fixed point is stable for $\e^f_h\to \infty$.
The low energy scale $T_0$ at which the LM fixed point crosses over to the singlet fixed point vanishes exponentially at a critical $\e^f_{h,c}(\e^c)$ (not shown for $\e^c/\Gamma_0=0.18$ since here $\e^f_{h,c}/\Gamma_0>9$): The larger $\e^c$ the larger the critical $\e^f_{h,c}$, 
at which the LM FP becomes the stable FP
and $\e^f_{h,c}\to\infty$ for $\e^c=\lim\limits_{\delta \rightarrow 0}{\e^c_c-\delta}$.

Fig.\ \ref{fig:7Imp_Phasendiag}(b) depicts the 1d phase diagram at a fixed temperature $T/\Gamma_0=10^{-15}$ as function of $\e^c$ (x-axis) and $\e^f_h$ (y-axis). The black linepoints separate a local moment (LM) phase from a singlet (SC) phase, 
and the dashed vertical lines indicate the phase boundary at $|\e^f_h|=\infty$.
The point symmetry of the phase diagram with regard to $\e^f_h=\e^c=0$ reflects the fact that the parameters of the correlated $f$-orbitals are chosen to be PH symmetric.
We can differentiate between two SC phases.
Whereas the SC I phase refers to the singlet phase for which $|\e^c| < \e^c_c$ holds, the SC II phase indicates a singlet phase with $|\e^c| > \e^c_c$.
Independent of the hole orbital filling, the orbital is responsible for the screening of the induced
local magnetic moment  in the SC I phase. Removing this orbital because of being
unoccupied would remove virtual excitations mediated by the orbitals and
immediately stabilizes  the LM moment  which might be unphysical in real world situations.
Note that the influence of the hole orbital on the low energy scale $T_0$ in the SC II phase near the phase boundary is enormous as well, however, the degeneracy of the ground state does not change when neglecting the hole orbital in this case.

\subsection{Type II Kondo hole model: Interaction induced Kondo coupling}
\label{sec:TypeII}

\begin{figure}[tbp]
\begin{center}
\includegraphics[width=0.46\textwidth,clip]{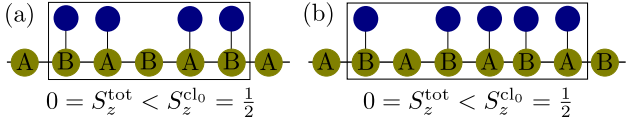}
\caption{Schematic of two type II Kondo hole models in 1d.
The model in (a) comprises an even number of $N_f=4$ $f$-orbitals, whereas the model in (b) comprises $N_f=5$ $f$-orbitals.
For $U>0$ both finite size cluster (black solid rectangle), CL$_0$, carry a local magnetic moment with $S^{\text{cl}_0}_z=1/2$, according to Eq.\ \eqref{eq:LiebMattis}.
Aplying Eq.\ \eqref{eq:JeffAB}, CL$_0$ of the model in panel (a) is AF coupled to both sides, CL$_0$ of the model in (b) is FM coupled to the left and AF to the right continuum. However, in both cases the local cluster magnetic moment gets screened on a low energy scale by electron continuum.
}
\label{fig:Model_TypeII}
\end{center}
\end{figure}

In this section we study two type II Kondo hole models which are schematically depicted in Fig.\ \ref{fig:Model_TypeII}.
For both models in Fig.\ \ref{fig:Model_TypeII}(a) and \ref{fig:Model_TypeII}(b) the modified Lieb-Mattis theorem for $U>0$ in Eq.\ \eqref{eq:LiebMattis} predicts a finite magnetic moment of the finite size cluster CL$_0$ (solid black rectangle) at PH symmetry, $S^{\text{cl}_0}_z=1/2$, that is AF coupled ($J^\text{eff}_\text{K}<0$) to at least one of the remaining electron continuum such that the ground state of the full models is a singlet, $S^\text{tot}_z=S^{\text{cl}_0}_z-1/2=0$.
However, even after successfully predicting
 the sign of $J^\text{eff}_\text{K,A/B}$ using Eq.~\eqref{eq:JeffAB}, 
its order of magnitude and, therefore,
the corresponding Kondo temperature, remains unknown.
Since the supercell analysis  in Sec.\ \ref{sec:supercell} predicts
a complete decoupling of the momentum carrying orbital,
$J^\text{eff}_\text{K,A/B}(U=0)=0$  must hold, setting a lower bound to the non-interacting limit, $U\to 0$.
Consequently, $J^\text{eff}_\text{K,A/B}$ seems to increase with increasing Coulomb interaction $U$. 
Note that $J^\text{eff}_\text{K,A/B}(U=0)=0$ 
does not contradict
the modified Lieb-Mattis theorem since this is only applicable in the strongly interacting Kondo limit $U/\Gamma_0\gg0$ where well defined local moments exist.

\begin{figure}[tbp]
\begin{center}
\includegraphics[width=0.45\textwidth,clip]{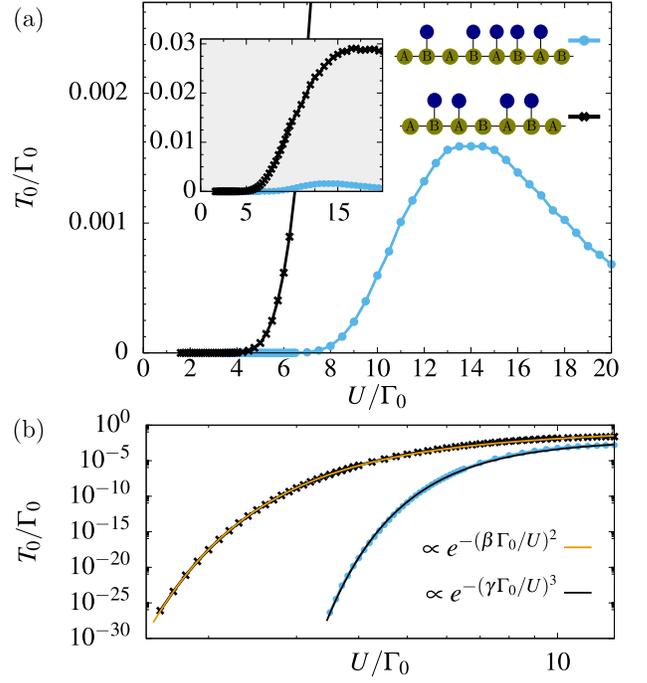}
\caption{Low energy scale $T_0$ as function of the Coulomb interaction $U/\Gamma_0$ for the two models which are schematically depicted in Fig.\ \ref{fig:Model_LiebMattis}(a) (black linepoints) and (b) (light blue linepoints). The inset in panel (a) depicts the same data as the main plot but on a larger scale for $T_0$.
In panel (b) the same data are plotted on a double logarithmic scale and we added a fit to a function of the form $T_0\propto\exp[-(\beta U/\Gamma_0)^2]$ (orange line) for the model comprising $N_f=4$ $f$-orbitals and $T_0\propto\exp[-(\gamma U/\Gamma_0)^3]$ (black line) for the model comprising $N_f=5$ $f$-orbitals. The parameters of the model are chosen to be PH symmetric, i.e. $\e^c=0$ and $U=-2\e^f_l$.}
\label{fig:T0_LiebMattis}
\end{center}
\end{figure}

To quantify the dependence of the effective coupling $J^\text{eff}_\text{K,A/B}$ on the strength of interaction $U$ we solved both models in Fig.\ \ref{fig:Model_TypeII} 
in the wide band limit using the NRG.
We enforced 
PH symmetry by setting $\e^c=0$, $U=-2\e^f$ and calculated the low energy scale $T_0$ 
defined in Eq.\ \eqref{eq:def_T0}.
Fig.\ \ref{fig:T0_LiebMattis} depicts the results of $T_0$ for the $N_f=5$ model (light blue linepoints) and the $N_f=4$ model (black linepoints) as function of $U/\Gamma_0$.
As expected from the supercell analysis, $T_0$ vanishes for both models in the limit $U\to 0$, which corresponds to $\lim\limits_{U \rightarrow 0}J^\text{eff}_\text{K,A/B}(U)\to0$.

$T_0$ decreases exponentially for the $N_f=5$ model 
 in the strongly interacting limit, $U/\Gamma_0\gg 1$,
but stays nearly constant for the $N_f=4$ model, as can be seen in the inset of Fig.\ \ref{fig:T0_LiebMattis}(a) where we plotted the same data but on a larger scale for $T_0$. 
The different behavior of $T_0$ for $U/\Gamma_0\gg 1$ reflects the fact, that both models have a different fixed point structure.
In this limit the RKKY interaction defines the largest energy scale and, consequently, the $N_f=4$ $f$-orbitals 
form a ground state singlet
(there are equal $N_{f,\text{A}}$ and $N_{f,\text{B}}$ sites) which decouples from the rest of the system. Since $J_\text{RKKY}\propto 1/U$ in the wide band limit \cite{Eickhoff2018,MIAM2020}, this energy scale depends only weakly on $U$.
In contrast to that the $N_f=5$ 
cluster ground state is a Kramers
$S_z=1/2$ multiplet screened by conduction electrons on the exponentially suppressed Kondo temperature $T_\text{K}\propto \text{exp}(-\alpha U/\Gamma_0)$.

Assuming that a Kondo effect with a yet to determine value for $J^\text{eff}_\text{K}$
governs the low energy scale $T_0$ in
the weak coupling limit, we plotted the data from Fig.\ \ref{fig:T0_LiebMattis}(a)  on a double logarithmic scale 
in Fig.\ \ref{fig:T0_LiebMattis}(b)
and added a fit (orange and black solid lines) to the functional form:
\begin{align}
\label{eq:T0_fit1}
&T_0\propto \exp[- (\beta \Gamma_0/U)^2]\quad \text{for } N_f=4,\\
&T_0\propto \exp[- (\gamma \Gamma_0/U)^3]\quad \text{for } N_f=5.
\label{eq:T0_fit2}
\end{align}
The fit demonstrates a perfect agreement between the NRG results and Eqs.\ \eqref{eq:T0_fit1} and \eqref{eq:T0_fit2}
which have the typical form of a Kondo scale: In this regime the Kondo screening dominates over the RKKY interaction.
Comparing this result with the textbook expression for the Kondo temperature of the SIKM, $T_\text{K}\propto\exp(-1/J_\text{K}\rho_0)$, we can extract the effective Kondo couplings:
\begin{align}
\label{eq:J_fit1}
&J^\text{eff}_\text{K}\propto (U/\Gamma_0)^2 \quad \text{for } N_f=4,\\
&J^\text{eff}_\text{K}\propto (U/\Gamma_0)^3 \quad \text{for } N_f=5.
\label{eq:J_fit2}
\end{align}

Applying a Schrieffer-Wolff transformation \cite{SchriefferWol66} to the SIAM in the strongly interacting limit, $U/\Gamma_0\gg1$, results in a Kondo coupling 
$J_\text{K}^\text{SIAM}\propto \Gamma_0/U$ and, therefore, our results in Eqs.\ \eqref{eq:J_fit1} and \eqref{eq:J_fit2} are very counter intuitive:
In depleted multi impurity models vanishingly small strengths of interactions can lead to strong correlation effects 
not accessible to perturbative approaches.

The supercell analysis predicts a single orbital disconnected from a free electron gas representing an effective electron continuum.
In the wide band limit, $V/D\to0$, this is equivalent to the appearance of a completely decoupled $\tilde{f}$-orbital in the eigenbase of $\Im\bm{\Delta}(-i0^+)$ as discussed in Sec\ \ref{sec:low-energy-model}.
Whereas a weak interaction $U/\Gamma_0$ is a small perturbation to a free electron gas or a resonant level model resulting in an effective Fermi liquid, the electrons in the disconnected bound $\tilde{f}$-orbital are strongly correlated for $\beta U>1$
enforcing single occupancy and a local magnetic moment. Finding a singlet phase by the NRG for $U>0$  suggests an $U$ induced effective RKKY coupling
$J^\text{RKKY}(U)\vec{S} \vec{S}_{d}$ between the electron spin in the bound state, $\vec{S}_{d}$, and spin $\vec{S} $
in the orbital that couples to the conduction band.
Then, this cluster RKKY interaction can be identified with the effective Kondo coupling
$J^\text{RKKY}(U)\propto J^\text{eff}_\text{K}(U)$.

We quantify this hypothesis in the $N_f=4$ model depicted  in Fig.\ \ref{fig:Model_TypeII}(a)
by  solving  the decoupled cluster using exact diagonalization first, i.\ e.\ setting $ \Im\bm{\Delta}(-i0^+)=0$ 
and include the effect of $ \Im\bm{\Delta}(-i0^+)$ in a second step by the NRG.
While for $U=0$ the decoupled $d$-orbitals result in a finite degeneracy of the cluster ground state, 
we find a unique cluster singlet ground state for $U>0$. Since the cluster energy spectrum is discrete,
and $J^\text{RKKY}(U)$ is  the smallest energy scale for $U/\Gamma_0\ll 1$,
we determine this scale via the cluster crossover energy scale $T^{\rm cluster}_0$ using the definition in Eq.\ \eqref{eq:def_T0} and setting $J^\text{RKKY}(U)=T^{\rm cluster}_0$ which is correct up to an universal but unknown prefactor of $O(1)$.

To remain as general as possible but still maintain parity symmetry we differentiate between the Coulomb interaction $U_\text{A}$ and $U_\text{B}$ for $f$-orbitals placed on $A$- and $B$-sites respectively.
We also include an additional inter-orbital Coulomb interaction $U^\prime_\text{A}$ between the $f$-orbitals on the $A$-sites neighboring to the hole site whose purpose comes apparent below.

\begin{figure}[tbp]
\begin{center}
\includegraphics[width=0.45\textwidth,clip]{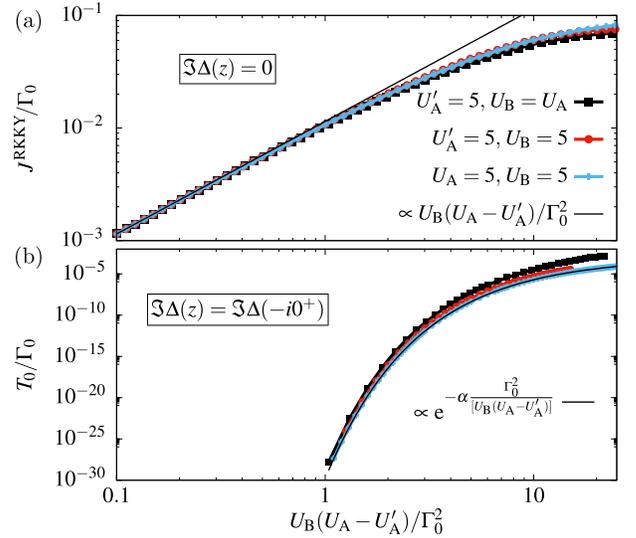}
\caption{Low energy scales for the $N_f=4$ model depicted in Fig.\ \ref{fig:Model_TypeII}(a). Panel (a) depicts the RKKY interaction $\tilde{J}^\text{RKKY}_{ij}(U)$, calculated for the decoupled finite size cluster in Eq.\ \eqref{eq:H-cl}, plotted against the dimensionless parameter $x=U_\text{B}(U_\text{A}-U_\text{A}^\prime)/\Gamma_0^2$. The different linepoints indicate different combinations for the intra orbital Coulomb interactions $U_\text{B}$ and $U_\text{A}$ for the $f$-orbitals on the $B$- and $A$-sites and the inter orbital Coulomb interaction $U_\text{A}^\prime$ between the $f$-orbitals on the $A$-sites. The solid line is fit to a function of the form $f(x)\propto x$ for small $x$.
(b) low energy scale $T_0$ for the same parameters as used in panel (a) but with the hybridization in Eq.\ \eqref{eq:H-hyb} turned on.
The black line shows the fit of one of the curves to a function of the form $f(x)\propto\exp(-\alpha/x)$.
}
\label{fig:T0_LiebMattis_U1-U2-Up}
\end{center}
\end{figure}

$J^\text{RKKY}(U)$ extracted from the entropy crossover scale $T^{\rm cluster}_0$ of the decoupled finite size cluster 
is depicted in Fig.\ \ref{fig:T0_LiebMattis_U1-U2-Up}(a) 
as function of the dimensionless parameter $x=U_\text{B}(U_\text{A}-U_\text{A}^\prime)/\Gamma_0^2$
for  three different cases. We fixed two parameters or linear combinations and altered one of them.
We find universality for $J^\text{RKKY}(U)=f(x)$ and 
a linear relation $J^\text{RKKY}(U)\propto x$ for small values of $x$.
This proves that the cluster singlet formation is driven by a complex interaction patterns that
involve all Coulomb interactions. Analyzing the single-particle cluster orbitals indicates that there is no direct
Coulomb interaction between the decoupled local moment and the spin that couples to one of the effective conduction bands. The interaction must be mediated by a higher order perturbation process by orbitals with different symmetry: All four orbitals contribute and the resulting effective spin-spin coupling in the low-energy subspace of the cluster 
which is very weak as demonstrated in Fig.\ \ref{fig:T0_LiebMattis_U1-U2-Up}(a).
Therefore, it is not surprising to find $J_K^{\rm eff}\propto U^3$ for $N_f=5$.

In Fig.\ \ref{fig:T0_LiebMattis_U1-U2-Up}(b)
the corresponding low energy scale $T_0$ of the full MIAM is plotted, i.\ e.\ 
with the hybridization in Eq.\ \eqref{eq:H-hyb} turned on.
The fit of one of the curves in Fig.~\ref{fig:T0_LiebMattis_U1-U2-Up}(b) to a function of the form $f(x)=\exp(-\alpha/x)$ (black line) demonstrates that the effective Kondo coupling corresponds to $J^\text{RKKY}(U)$ of the decoupled finite size cluster in Fig.\ \ref{fig:T0_LiebMattis_U1-U2-Up}(a).

Now the purpose of the additional interaction $U_\text{A}^\prime$ becomes apparent: 
It controls a KT type quantum phase transition induced by a sign change of $x$.
The sign of the effective Kondo coupling is determined by the sign of $U_\text{A}-U_\text{A}^\prime$:
For $U_\text{A}<U_\text{A}^\prime$, the coupling becomes ferromagnetic
such that the LM FP is stable, as proven by the NRG.
This case  can be easily understood in real space:
For $U_\text{A}<U_\text{A}^\prime$ the singly occupancy of the $f$-orbitals neighboring the hole site gets suppressed, resulting in a singlet state $\propto(|0,2\rangle + |2,0\rangle)$ where each of the orbitals is either empty or doubly occupied, which decouples from the remaining system.
Removing these $f$-orbitals results in an effective two impurity problem for which the Lieb-Mattis theorem in Eq.\ \eqref{eq:LiebMattis} predicts a stable LM FP: $S^\text{tot}_z=1/2$.

A similar analysis of the cluster eigenspectrum can be performed for the $N_f=5$ cluster, leading to $J^\text{eff}_K\propto U^3$.

The low temperature properties of Type II Kondo hole models are very similar to the two stage Kondo effect in T-shaped double quantum dot systems (DQDs) \cite{Cornaglia2005,Tanaka2012,Wojcik2015,Zitko2010}, where only one of two quantum dots is directly coupled to a conducting lead.
If the Kondo temperature of the quantum dot that couples to the lead is larger than the coupling to the second dot, the latter is Kondo screened by the heavy quasi particles of the prior formed Fermi liquid, corresponding to the small $U/\Gamma_0$ limit in the Type II Kondo hole models.
The structure of the stable FP at zero temperature corresponds to that of a free electron gas since each screening process removes one electron from the respective continuum resulting in a Fermi liquid where an even number of electrons have been removed.
Hence, the FP doesn't change in the opposite limit of large inter dot coupling in the DQDs and large $U/\Gamma_0$ (large RKKY interaction) in the Type II Kondo hole model respectively, where a inter impurity singlet is formed that decouples from the continuum.

\section{Hole-hole interactions: Type I and Type III Kondo hole models}
\label{sec:HHInteraction}

Local moments coupled to an environment comprising delocalized electrons typically interact with each other via the indirect RKKY interaction, which is mediated by these delocalized electrons. Since electrons
of all energy scales contribute to the RKKY interaction, the environment not necessarily needs to be metallic, even if the
 leading contributions in case of an metallic environment stems from Fermi surface electrons.

In case of local moments induced by Kondo holes, this RKKY mechanism may break down since the moment carrying orbitals can decouple from the environment, as shown by the supercell analysis in Sec.\ \ref{sec:supercell}. 
On the other hand, the on-site Coulomb interaction as well as single-particle hopping between 
 the $f$-orbitals lead to a spreading of the hole induced magnetic moment as illustrated in Fig.\ \ref{fig:7Imp_S_Mag}(b). The magnetic moments originating from different Kondo holes may overlap and, consequently, interact directly with each other.

For the weakly interacting limit and PH symmetry Schlottmann has shown \cite{Schlottmann1995} using a nearest neighbor tight binding dispersion for the conduction band electrons, that two Kondo holes do only interact if they are placed on adjacent sites.
This rigorous result can already be understood within our non-interacting supercell analysis.
According to the argumentation in Sec.\ \ref{sec:supercell}, introducing a second hole per supercell leads to another flat band and a corresponding decoupled orbital $d_{\k,2}$, unless the second hole is placed on a site that contributes to $d_{\k,1}$.
However, in case of a nearest neighbor tight binding description this does only happen if the two holes are placed on neighboring sites.

\begin{figure}[tbp]
\begin{center}
\includegraphics[width=0.45\textwidth,clip]{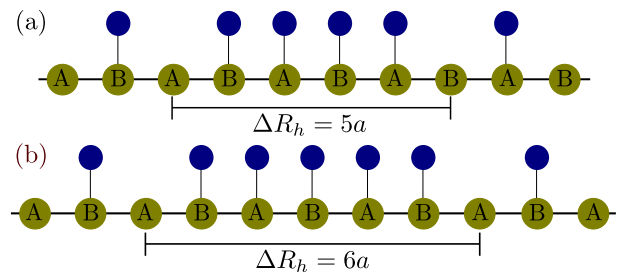}
\caption{Schematic of two 1d MIAMs with two Kondo holes placed inside the impurity array.
(a) if the holes are separated by an odd number of lattice spacings, $\Delta R_h=(2n+1)a$, the ground state for $U>0$ is always a singlet, which indicates AF interactions between the hole induced magnetic moments. (b) if the holes are separated by an even number of lattice spacings, $\Delta R_h=(2n)a$, the ground state for $U>0$ is a triplet if the couplings $J^\text{eff}_{\text{K,A/B}}$ are FM, which demonstrates FM interactions between the hole induced magnetic moments.}
\label{fig:Model_2Holes}
\end{center}
\end{figure}

On the other hand, the question concerning interactions between different Kondo holes in the strongly interacting limit, $U/\Gamma_0\gg 1$, is closely related to the modified Lieb-Mattis theorem.
In case of PH symmetry, two hole orbitals which are placed on different sites of a bipartite sublattice, as shown in Fig.\ \ref{fig:Model_2Holes}(a), always lead to a singlet ground state for $U>0$ ($S^\text{tot}_z=S^{\text{cl}_0}_z=0$), whereas the supercell analysis predicts two decoupled orbitals $d_{1}$ and $d_{2}$ per unit cell in the non-interacting limit.
If the holes are placed on the same sublattice, the local cluster ground state is given by a triplet with $S^{\text{cl}_0}_z=1$ for $U>0$, which is stable in case of FM couplings $J^\text{eff}_{\text{K,A/B}}$ ($S^\text{tot}_z=S^{\text{cl}_0}_z=1$).
An example for such an situation is depicted in Fig.\ \ref{fig:Model_2Holes}(b).
Consequently, we expect that an effective AF exchange interaction $K_{hh}^\text{ex}<0$ between holes 
is generated by the RG in the low-energy regime when the holes are separated by an odd number of lattice spacings, $\Delta R_h=(2n+1)a$, whereas holes separated by an even number, $\Delta R_h=(2n)a$, should be FM coupled, $K_{hh}^\text{ex}>0$
in a 1d lattice.

In Sec.\ \ref{sec:7Imp_eh_to_infty} we have demonstrated that the magnetic moment, which is induced by a single hole in a finite size impurity array, is located on the $d$-orbital defined in Eq.\ \eqref{eq:ds_Vto0} in the limit $U/\Gamma_0\to 0$, which corresponds to the Fourier back transformation of the supercell $d_{\k,l_h}$-orbital in Eq.\ \eqref{eq:d-k} with one hole per unit cell.
Inserting another hole in the finite size impurity array, the same procedure results in two decoupled $d_i$-orbitals in this case, each given by Eq.\ \eqref{eq:ds_Vto0} in the wide band limit.

\begin{figure}[tbp]
\begin{center}
\includegraphics[width=0.5\textwidth,clip]{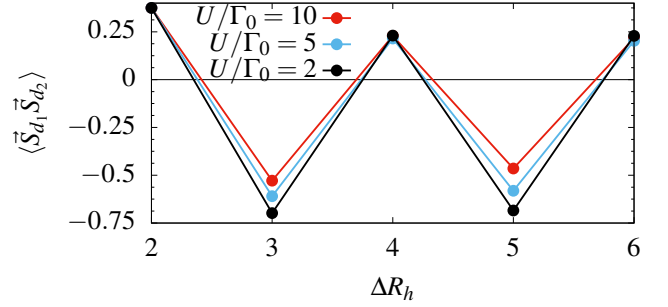}
\caption{Zero temperature spin correlation $\langle \vec{S}_{d_1}\vec{S}_{d_2}\rangle$ of the $d$-orbitals defined in Eq.\ \eqref{eq:ds_Vto0} as function of the distance $\Delta R_h$ between the hole sites for three different strengths of Coulomb interactions, $U/\Gamma_0=2$ (black linepoints), $5$ (light blue linepoints) and $10$ (red linepoints). The parameters of the model are chosen to be PH symmetric, i.e. $\e^c=0$ and $U=-2\e^f_l$.}
\label{fig:SSKorr_2Holes}
\end{center}
\end{figure}

To verify the predictions made by the modified Lieb-Mattis theorem, we calculated the zero temperature spin correlation $\langle \vec{S}_{d_1}\vec{S}_{d_2}\rangle$ as function of the distance $\Delta R_h$ and plotted the results in Fig.\ \ref{fig:SSKorr_2Holes} for three different Coulomb interaction strengths $U/\Gamma_0=2$ (black linepoints), $5$ (light blue linepoints) and $10$ (red linepoints).
In order to ensure that $S^\text{tot}_z=S^{\text{cl}_0}_z$ is always fulfilled, such that the influence of couplings $J^\text{eff}_{\text{K,A/B}}$ can assumed to be small, we always placed the holes at the left and right end of the finite size impurity array and change the distance $\Delta R_h$ by varying the total number $N_f$ of $f$-orbitals, as shown in Fig.\ \ref{fig:Model_2Holes}(a)  for $N_f=6$ and Fig.\ \ref{fig:Model_2Holes}(a) $N_f=7$.
The components $m$ of the spin operators $\vec{S}_{d_i}$ are defined as usual, $S^m_{d_i}=\frac{1}{2}\sum_{\alpha,\beta} d^\dagger_{i,\alpha}\sigma^m_{\alpha,\beta}d_{i,\beta}$, where $\sigma^m$ is the $m$-th Pauli matrix.

For all three values of $U/\Gamma_0$, the spin correlation oscillates between FM ($>0$) for even $\Delta R_h$ and AF ($<0$) correlations for odd $\Delta R_h$ and, consequently, matches the predictions made by the Lieb-Mattis theorem.

The FM correlations are nearly $U$ independent and, except for $\Delta R_h=2a$, from the distance between the hole sites.
For $\Delta R_h=2a$ both $d$-orbitals share the same $f$-orbital located  between the hole sites such that they are not orthogonal in this case, $\{d^\dagger_1,d_2\} \not=0$ causing the large correlation $\langle \vec{S}_{d_1}\vec{S}_{d_2}\rangle=0.35$. 
We orthogonalization of the two linear independent orbital using the parity,
$\tilde{d}_{\pm}=1/N_{\pm}(d_1\pm d_2)$, where $N_+=\sqrt{3}$ and $N_-=1$ ensure normalization, we obtain $\langle \vec{S}_{\tilde{d}_1}\vec{S}_{\tilde{d}_2}\rangle=0.25$ (not shown).
For all other distances $\{d^\dagger_1,d_2\} =0$
holds and the correlation reaches the maximal value of $\langle \vec{S}_{d_1}\vec{S}_{d_2}\rangle=0.25$ for free
FM aligned  local moments.

In contrast to this the AF correlations slightly decrease with increasing $U/\Gamma_0$ and increasing distance $\Delta R_h$.
There are two effects which contribute to this reduction of $\langle \vec{S}_{d_1}\vec{S}_{d_2}\rangle$ for odd $\Delta R_h$: 
(i) The Coulomb interaction $U$ 
leads to a deformation of the effective $d$-orbital that carries the local moment induced by a single hole, as discussed in the context of Fig.\ \ref{fig:7Imp_MagMomHole}. Consequently, the operators $\vec{S}_{d_i}$ are not the exact spin operators of the hole induced magnetic moment for $U>0$.
(ii) Since we are studying holes in a finite size impurity array we need to take into account the couplings $J^\text{eff}_{\text{K,A/B}}$ between the individual hole induced magnetic moments and the continuum. For odd $\Delta R_h$, the magnetic moment induced by the left hole is coupled AF to the continuum on the right site of the finite size cluster and vice verse. Therefore, the Kondo effect can reduce the size of the corresponding magnetic moment and $\langle \vec{S}_{d_1}\vec{S}_{d_2}\rangle$ respectively. However, this effect plays only a minor role. The effective Kondo coupling and the direct exchange interaction are both generated by an finite overlap of the magnetic bound states which falls of exponentially \cite{Clare96} with increasing distance from the hole site. Since the distance between the holes is smaller than the distance between the left (right) hole and the right (left) continuum, the direct exchange interaction is much larger than the effective Kondo coupling, $K_{hh}^\text{ex}\gg J^\text{eff}_\text{K,A/B}$.
 
Note that, due to the Lieb-Mattis theorem, the correlation between the hole induced magnetic moments does not decay in the full lattice problem at zero temperature and oscillates for $\Delta R_h>2a$ between $0.25$ and $-0.75$.
Any finite exchange interaction $K_{hh}^\text{ex}\not=0$ leads to a maximal correlation since the free local moments are infinite susceptible at zero temperature. This is equivalent to the two impurity Kondo model with FM Kondo couplings, where the Kondo effect is absent such that the RKKY interaction defines the only energy scale at $T=0$ - see Fig.\ 5 of Ref.\ \cite{Lechtenberg2018}.
However, a finite temperature $T$ introduces a natural cut off energy scale such that the ratio of $K_{hh}^\text{ex}/T$ determines the strength of the spin correlation.

Since the physical mechanism that induces the exchange interaction $K_{hh}^\text{ex}$ is different from the conventional RKKY mechanism, we still need to analyze the dependence of $K_{hh}^\text{ex}$ on the model parameters, especially with regard to the Coulomb interaction $U$.
In order to calculate the absolute value of the energy scale $K_{hh}^\text{ex}$ via the impurity induced entropy $S_\text{imp}$ in analogy to the low energy scale $T_0$ in Eq.\ \eqref{eq:def_T0}, we need to differentiate between FM and AF couplings.

Assuming the exchange interaction to be larger than the effective couplings $J^\text{eff}_\text{K,A/B}$ but smaller than all other energy scales of the System, the last unstable intermediate fixed point contains two independent local moments and, consequently, $S_\text{imp}=k_\text{B}\ln(4)$.
Reducing the temperature further, the sign of $K_{hh}^\text{ex}$ determines the ground state multiplet:
Whereas AF interactions $K_{hh}^\text{ex}$ lead to a singlet ground state with $S_\text{imp}=k_\text{B}\ln(0)$, FM interactions cause a triplet ground state with $S_\text{imp}=k_\text{B}\ln(3)$.
Therefore, we define the low energy scale of the exchange interaction $K_{hh}^\text{ex}$ by
\begin{align}
\label{eq:T0_AF}
&S_\text{imp}(T_0^\text{AF})=\frac{1}{2}k_\text{B}[\ln(4)+\ln(0)],
\end{align}
where $T_0^\text{AF}$ corresponds to  $K_{hh}^\text{ex}>0$ and 
\begin{align}
&S_\text{imp}(T_0^\text{FM})=\frac{1}{2}k_\text{B}[\ln(4)+\ln(3)],
\label{eq:T0_FM}
\end{align}
where $T_0^\text{FM}$ corresponds to  $K_{hh}^\text{ex}<0$.

\begin{figure}[tbp]
\begin{center}
\includegraphics[width=0.5\textwidth,clip]{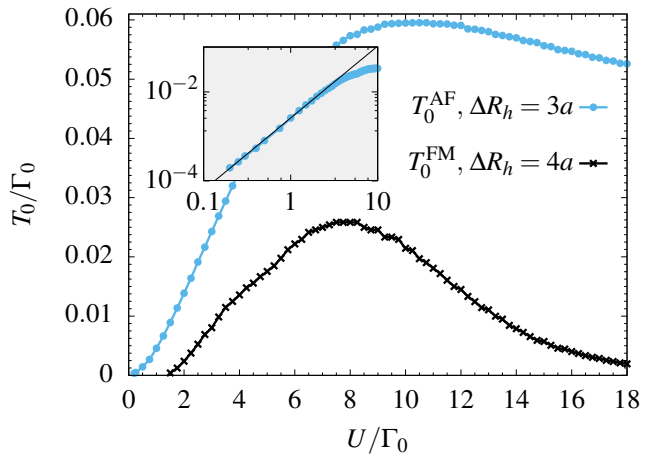}
\caption{Low energy scales for the models depicted in Fig.\ \ref{fig:Model_2Holes} but with $\Delta R_h=3a$ (light blue linepoints) and $\Delta R_h=4a$ (black linepoints). In case of $\Delta R_h=3a$ the ground state is a singlet such that the low energy scale is given by $T^\text{AF}_0$ defined in Eq.\ \eqref{eq:T0_AF}, whereas it is given by $T_0^\text{FM}$ defined in Eq.\ \eqref{eq:T0_FM} for a triplet ground state for $\Delta R_h=4a$.
The inset shows the data for $\Delta R_h=3a$ on a logarithmic scale for $T_0^\text{AF}$ together with the fit function $f(x)=a x^b$ for $a=0.0025$ and $b=1.6169$. The parameters of the model are chosen to be PH symmetric, i.e. $\e^c=0$ and $U=-2\e^f_l$.}
\label{fig:T0_2Holes}
\end{center}
\end{figure}

The results for $T_0^\text{AF}$ and $T_0^\text{FM}$
vs. $U$, calculated for $\Delta R_h=3a$ and $\Delta R_h=4a$ respectively, are depicted in Fig.\ \ref{fig:T0_2Holes} (light blue linepoints for $T_0^\text{AF},\, \Delta R_h=3a$ and black linepoints for $T_0^\text{FM},\, \Delta R_h=4a$).
The overall curves
of $T_0^\text{AF}$ and $T_0^\text{FM}$ matches those of $T_0$ for the $N_f=4$ and $N_f=3$ model in Fig.\ \ref{fig:T0_LiebMattis}(a).
For small $U/\Gamma_0$ the low energy scales $T_0^\text{AF}$ and $T_0^\text{FM}$ increase with increasing Coulomb interaction.
However, in the strongly interacting limit, $U/\Gamma_0\gg 1$, $T_0^\text{AF}$ corresponds to the AF RKKY scale $\propto1/U$ since the model consists of an even number of $N_f=4$ orbitals, whereas $T_0^\text{FM}$ can be linked to the exponentially suppressed Kondo temperature due to an odd number of $N_f=5$ orbitals in the model.
These different mechanisms are also encoded in the different  NRG FP spectra for the two cases (not depicted here.)

In the weakly interacting limit, $U/\Gamma_0\to 0$, the exchange interaction $K_{hh}^\text{ex}$ vanishes as expected from the supercell analysis in Sec.\ \ref{sec:supercell}.
The inset shows the $T_0^\text{AF}$ data on a logarithmic scale and a corresponding fit to a 
power law function of the form $f(x)=a x^b$, with $a=0.0025$ and $b=1.6169$.
In contrast to the conventional RKKY interaction which falls of as $1/U$ in the wide band limit, $K_{hh}^\text{ex}$ increases strongly with increasing Coulomb interaction.  
Note that we can not resolve the small energy scale for $U/\Gamma_0<1.8$ and $\Delta R_h>3$.  
Hence we can not quantify the dependence of $K_{hh}^\text{ex}$ on the distance $\Delta R_h$, however, $|K_{hh}^\text{ex}(\Delta R_h=3)|>|K_{hh}^\text{ex}(\Delta R_h=4)|$ is fulfilled for all $U$. As the wave function of the magnetic bound states falls of exponentially \cite{Clare96} and $K_{hh}^\text{ex}$ is proportional to the overlap of different magnetic bound states, we expect $K_{hh}^\text{ex}$ to fall of exponentially with increasing distance $\Delta R_h$.

\section{Viewpoint onto the Kondo-hole physics puzzle in $\text{Ce}_{1-x}\text{La}_x\text{Pd}_3$}
\label{sec:experiment}

Stable magnetic bound states induced by charge neutral substitution of single Ce by La are expected to exist in systems whose parent compound shows Kondo insulating behavior \cite{Schlottmann91I}.
However, in experiments \cite{lebraski2010,Pietrus2008,Rotundu2007,Malik1995,Adroja1996,Borstel1998} with a finite hole concentration $x$ these local bound states haven't been observed so far since long range interactions between the bound states, which are already present at the lowest experimentally accessible $x$ (for example $x=0.006$ in (Ce$_{1-x}$La$_x$)$_3$Bi$_4$Pt$_3$ \cite{Pietrus2008}), lead to the formation of a very narrow Kondo-hole band in the hybridization gap.

The long-range nature of the interaction between the hole induced bound states 
in Kondo insulators
is in qualitative agreement with our results for Kondo-holes in finite impurity arrays. As discussed in the context of Fig.\ \ref{fig:7Imp_S_Mag}(b), the hole induced magnetic moment is increasingly
delocalized with increasing Coulomb interaction $U$ and spreads over several lattice spacings in the strongly interacting limit $U/\Gamma_0\gg 1$.

In contrary, the insertion of Kondo-holes in heavy-fermion materials, such as Ce$_{1-x}$La$_x$Cu$_6$ \cite{ONUKI1987} are mainly discussed 
as examples for a smooth crossover from a lattice material to single impurity physics \cite{ONUKI1987,Grewe91,GrenzebachAndersCzychollPruschke2008}.

Pristine CePd$_3$ is a heavy fermion metal that is considered to be close to a Kondo insulator but still maintains Fermi liquid properties at low temperatures.
Therefore,  
the electronic degrees of freedom relevant for the low energy physics 
should violate PH symmetry and the filling of the conduction bands must deviate
from half-filling, however, the model parameters are at the brink of becoming a Kondo insulator.

Substituting a small amount of nonmagnetic La for the Ce in $\text{Ce}_{1-x}\text{La}_x\text{Pd}_3$
leads to a  logarithmic increase  of the resistivity  \cite{Lawrence85,Lawrence96} 
with decreasing temperature after the material is cooled  far
below the lattice coherence temperature of CePd$_3$ typical for 
a system with  magnetic impurities in a metal. 
Early on a negative magnetoresistance was taken as another indication  that the second resistively increase
is connected to secondary Kondo effect  \cite{Lawrence96} 
but the microscopic origin of the local moment formation remain a puzzle.

Connecting our investigation of Kondo-hole physics with experiments, the
main message is that Kondo hole insertion generate stable local moments near the PH symmetric point (which corresponds to Kondo insulators in the PAM) that can be screened by two
mechanisms which require PH asymmetry:
(i) hybridizing the localized single particle state with the itinerant electrons due an asymmetric conduction band and (ii) indirect coupling via the unfilled La $4f$-orbitals.
We showed in Fig.\ \ref{fig:7Imp_Phasendiag}
that taking into account a finite orbital energy $\e^f_h<\infty$ of the unoccupied La $4f$-orbitals 
and a band center $\e^c$ away from a half-filled band yields a Kondo screening of the Kondo hole induced local
moment. 

The question of whether exotic physical properties are observable in material is directed linked to
the question of separating the energy or temperature scales of a two-step screening mechanism: First a 
formation of a correlated HF or Kondo insulator phase is required and then, at much lower temperatures, the screening of the Kondo-hole induced local moments can occur.

In real materials, the PH-symmetry  in the conduction bands  is broken
even for half-filling. Furthermore, La substitution changes the band structure of the conduction band slightly 
as shown by LDA calculations \cite{Borstel1998} due to the Lanthanide contraction.
Instead of introducing a complicated tight-binding band structure adequate for only one material, we use the value for the band center $\e^c$ as a control parameter for the degree of PH symmetry breaking.

Although a finite size MIAM does not include the full lattice physics, we have demonstrated \cite{MIAM2020}
that major self-screening mechanism of magnetic moment is correctly captured already in a MIAM of second type.
Pruschke at al.\ \cite{PruschkeBullaJarrell2000} showed already with the DMFT for the PAM that
the singlet-formation in the Kondo-insulators is not due to the Kondo effect since the 
conduction band as well as continuum of effective media shows a pseudo-gap DOS but caused by the effective f-f orbital hybridization that is explicitly included in the matrix $\Re \mat{\Delta(0)}$ in our mapping.

\begin{figure}[tbp]
\begin{center}
\includegraphics[width=0.45\textwidth,clip]{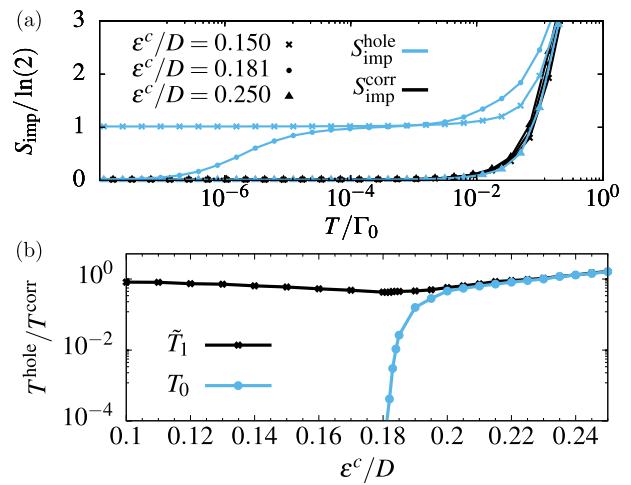}
\caption{NRG calculations for the 1d model depicted in Fig.\ \ref{fig:7Imp_Model} with $N_f=7$ $f$-orbitals.
The properties with index 'hole' 
were calculated for an uncorrelated hole-orbital (red in Fig.\ \ref{fig:7Imp_Model}), $U_h=0$ and $\e^f_h/\Gamma_0=25$, in the center of an otherwise strongly correlated impurity array, $U_l=-2\e^f_l=10\Gamma_0$, whereas all properties indicated by the index 'corr' correspond to the homogenous model where all $N_f=7$ $f$-orbitals are identical with $U=-2\e^f=10\Gamma_0$.
(a) impurity induced entropy $S_\text{imp}$ as function of the dimensionless temperature $T/\Gamma_0$ for the model with a hole orbital (light blue linepoints) and without hole orbital (black linepoints) for three different values $\e^c/D=0.15$, $0.181$ and $0.25$, indicated by different types of points.
(b) ratio $T^\text{hole}/T^\text{corr}$ for the low energy scales $T_0$ (light blue linepoints), defined in Eq.\ \eqref{eq:def_T0}, and $\tilde{T}_1=\kappa T_1$ (black linepoints), defined in Eq.\ \eqref{eq:def_T1}, plotted against the band center $\e^c/D$.
}
\label{fig:Comparison}

\end{center}
\end{figure}

Since we believe that the Kondo-hole physics in the dilute limit is locally driven 
and independent of the details of the conduction band other than PH symmetry breaking,
we used a 1d MIAM representation in order to ensure that the number of screening channels are less than the number of correlated orbitals  in the vicinity of the Kondo hole to mimic the physics of a larger system. 
For our NRG calculation we use the MIAM with $N_f=7$ $f$-orbitals shown in  in Fig.\ \ref{fig:7Imp_Model}.

We need to define the appropriate low energy scales of the MIAM to make connection with the experiment. Within a DMFT(NRG) it has been shown that the lattice coherence temperature is identical to the Kondo temperature of the effective site up to some universal number depending on the definition  \cite{GrenzebachAndersCzychollPruschke2006}. Using that
knowledge, we extracted the crossover scale to the low-energy singlet fixed point in the homogenous model where all $N_f=7$ $f$-orbitals are identical with $U=-2\e^f=10\Gamma_0$ as measure for the  lattice coherence temperature $T^\text{corr}$ in our approach and neglect the DMFT lattice correction.

To connect the low-temperature scale of the fully correlated array of $N_f=7$ 
correlated orbitals with those of the $N_f=6$ Kondo hole MIAM, we plotted
the temperature dependent impurity entropy $S_\text{imp}$
for the model with hole (light blue linepoints) and without hole (black linepoints) for three different values $\e^c/D=0.15$, $0.181$ and $0.25$ in  Fig.\ \ref{fig:Comparison}(a).
In  the full MIAM the  $N_f$ spin-$1/2$ local moments are quenched on a single 
low-temperature crossover scale $T^{\rm corr}$ as expected from a Kondo lattice problem.
This  crossover scale  is nearly independent of the band center $\e^c$ for the full MIAM. 
In the Kondo hole case we, however, observe three different regimes. For large $\e^c$, 
all $N_f-1$ local moments are screened on the same energy scale $T_1$ that qualitatively agrees with that of the full MIAM.
For $\e^c\approx 0.18$, only $N_f-2$ local moments are screened at on a similar scale than the full problem,
and a unstable LM $S=1/2$ FP emerges whose local moment started to disappear on a  
secondary crossover scale $T_0$. This LM FP is stable for $\e^c = 0.15$.

To distinguish between the potential two energy scales in the present of
of the Kondo hole, we
define an additional low energy scale $T_1$,
\begin{align}
S_\text{imp}(T_1) = 1/2 k_\text{B} \left[\ln(2)+\ln(3)\right],
\label{eq:def_T1}
\end{align}
which governs the crossover to the unstable $S=1/2$ LM FP -- see also Fig.\ \ref{fig:Comparison}(a).
$\ln(3)$ has no physical meaning and could be replaced by any other reasonable value, since $T_1$ only defines a crossover scale.
Per definition, the relation $T_0(\e^c)<T_1(\e^c)$ always holds. In the large
$\e^c$ regime, where there is only one crossover scale $T_0(\e^c) \propto T_1(\e^c)$ holds while
for intermediate $\e^c$ the temperatures $T_0$ and $T_1$ refer to physically different scales.
Therefore we introduce the scaling factor $\kappa=T_0(\e^c/D=0.25)/T_1(\e^c/D=0.25)=0.462$ that compensates the mismatch between $T_1$ and $T_0$ at $\e^c/D=0.25$ where both refer to the same energy scale, and introduce
the rescaled $\tilde{T}_1(\e^c)=\kappa T_1(\e^c)$ such that $\tilde{T}_1(\e^c/D=0.25)=T_0(\e^c/D=0.25)$.
For the $N_f=7$ full MIAM, $\tilde{T}_1(\e^c)=T_0(\e^c)$ always holds since there is only one crossover scale
as illustrated in Fig.\ \ref{fig:Comparison}(a), and we set $T^{\rm corr}(\e^c) = T_0(\e^c)$.

In Fig.\ \ref{fig:Comparison}(b) we plotted the ratio $T^\text{hole}/T^\text{corr}$ for the low energy scales $T_0$ (light blue linepoints) and $\tilde{T}_1$ (black linepoints), plotted as function of the band center $\e^c/D$. 
We used an $N_f=7$ MIAM  with a single 
uncorrelated hole (red in Fig.\ \ref{fig:7Imp_Model}), $U_h=0$ and $\e^f_h/\Gamma_0=25$, in the center of an otherwise strongly correlated impurity array, $U_l=-2\e^f_l=10\Gamma_0$.
In contrary to the Kondo hole literature, we included an uncorrelated and
unoccupied $4f$-orbital at an energy of $\e^f_h \approx 2U$ in the calculation.

We note that the ratio $\tilde{T}^\text{hole}_1/T^\text{corr}$ remains almost independent of the band center $\e^c$
and the ratio remains of $O(1)$. The scale $\tilde{T}_1$ takes the role of the lattice coherence scale in the system with and without the Kondo hole. The change is not really significant and in the experiments 
on Ce$_{0.97}$La$_{0.03}$Pd$_3$
the reduction of $T^\text{corr}$ in the presence of Kondo holes has been
interpreted as signature for the lanthanide contraction \cite{Grewe91,Lawrence96,Rotundu2007,lebraski2010}.

We can identify three different regimes. 

(I)
For $\e^c/D\geq 0.2$ we have $T_0\approx \tilde{T}_1$ and $T^\text{hole}\approx T^\text{corr}$, indicating that there is only one relevant low energy scale which is nearly identical for the model with and without a Kondo hole in its center.
We believe that this regime is relevant for Kondo hole substitution in 
Ce$_{1-x}$La$_x$Cu$_6$ where we start from a heavy-fermion compound that is significantly away from half-filling.

(II) $T_0^\text{hole}/T^\text{corr}$ is zero for $\e^c/D\leq 0.179$,
and a stable LM FP is found as exacted to occur in Kondo insulators in the dilute limit.
Free local moments are present for $T\to 0$ and the finite entropy
is probably removed by magnetic polaron formation \cite{lebraski2010}. 
It has been suggested that disorder and the spatially extended
bound states apparently start to overlap in Kondo insulators already at relative low concentrations leading to a very narrow band formation in the Kondo insulator gap \cite{Schlottmann91II,Schlottmann92,Schlottmann96,Adroja1996,lebraski2010}.

(III)  
In between these two regimes at around $\e^c/D\approx 0.179+|\delta|$ and small $\delta$, there is a clear hierarchy, $0<T_0<\tilde{T}_1$, indicating that there is an intermediate unstable LM FP that crosses over to the SC phase on a energy scale clearly below the coherence temperature $T_1$.
We believe that this intermediate regime is relevant for the exotic behavior in doped heavy Fermion materials which are at the brink of being a Kondo insulator such as CePd$_3$.
In this regime the lowest energy scale, $T^\text{hole}_0$, corresponds to the onset of magnetic scattering associated with an increase in the electrical resistivity due to a Kondo effect of the hole induced magnetic moment.

\section{Summary and discussion}
\label{sec:summary}

In this paper we reviewed the effect of Kondo holes in lattice and in impurity models from two complementary perspectives using 
(i) a supercell analysis in the uncorrelated limit and (ii) the Lieb-Mattis theorem in the strongly interacting regime with well defined local moments.

Without any restrictions concerning the spatial dimension and the geometry of the underlying lattice, the supercell analysis predicts the existence of localized decoupled orbitals when Kondo holes are introduced in the PAM or MIAM, without presupposing PH symmetry, \ie a half filled conduction band.
Additionally, for the subset of bipartite lattice structures and PH symmetric models, the Lieb-Mattis theorem makes some rigorous statements about these Kondo hole Hamiltonians, predicting local cluster magnetic moments which can be either stable or subject to further screening processes.

In consequence of these generic statements
 which are in accordance with the established theory about Kondo holes \cite{Schlottmann91I,Schlottmann91II,Schlottmann92,Schlottmann96,Clare96,Figgins2011}, we expect the occurrence of hole induced bound states to be a generic feature for a wide class of HF materials.
Indeed, Hamidian et al \cite{Hamidian2011} observed these bound states inside the hybridization gap in Th-doped
URu$_2$Si$_2$ by comparing the differential conductance far away
from a Th atom with that right at a Th atom site.
We demonstrated that these bound states are the consequence of pseudo-gap formation which can be understood from a local point of view.

Using the NRG in combination with a recently developed wide band approximation for multi impurity models \cite{MIAM2020}, we solved the MIAM with up to $N_f=7$ $f$-orbitals in the strongly interacting regime.
Inserting Kondo holes in such dense impurity arrays our NRG results are in perfect agreement with all exact statements available, ranging from the non-interacting limit (supercell analysis) to the strongly interacting regime (Lieb-Mattis theorem). We identified three different classes at PH symmetry:
The effective Kondo coupling $J^\text{eff}_\text{K}$ between the hole induced bound state and the environment can be (i) FM, leading to a stable LM FP, or (ii) AF, resulting in a low energy scale at which the intermediate LM FP becomes unstable. 
Remarkably, this low energy scale is of a counterintuitive Kondo form due to $J^\text{eff}_\text{K}\propto U^n$, where $n>1$ depends on the number $N_f$ of correlated $f$-orbitals in the model. 
Alternatively, in case of several Kondo holes, the hole induced bound states can (iii) interact directly with each other which may lead to magnetic order in lattice models such that the ordered magnetic moment per unit cell is significantly reduced compared to the number of local moments per unit cell.

Going beyond the PH symmetric limit where the Lieb-Mattis theorem is applicable, by including the nearly unoccupied hole orbital or shifting the band center of the conduction electrons, we further demonstrated that the LM FP can be destabilized.
In this case the local moment phase is replaced by two types of singlet phases that are adiabatically connected.
At a KT type QCP the physics is governed by an exponentially suppressed Kondo scale approaching the strong coupling phase that can be replaced by a singlet formation via antiferromagnetic RKKY interaction for large deviation from the critical values.

In heavy Fermions that are close to a Kondo-insulator such as CePd$_3$
we believe that our analysis can be applied for low concentrations of Kondo-holes.
Breaking particle-hole symmetry in the conduction band as well as including the unoccupied $4f$-orbitals of La for virtual excitation leads to
a coupling of the Kondo-hole induced local moment such that the localized free spin $S=1/2$
is screened. This provides the microscopic mechanism for an Kondo effect as reported in Ce$_{1-x}$La$_x$Pd$_3$.
Deviations of single-ion Kondo behavior is expected when  the Kondo-lattice coherent temperature is
not well separated from the single-impurity Kondo temperature of the residual local moments, as well
as when the concentration is high enough that Kondo-hole induced local moments start to interact with each other.


\begin{thebibliography}{75}%
\makeatletter
\providecommand \@ifxundefined [1]{%
 \@ifx{#1\undefined}
}%
\providecommand \@ifnum [1]{%
 \ifnum #1\expandafter \@firstoftwo
 \else \expandafter \@secondoftwo
 \fi
}%
\providecommand \@ifx [1]{%
 \ifx #1\expandafter \@firstoftwo
 \else \expandafter \@secondoftwo
 \fi
}%
\providecommand \natexlab [1]{#1}%
\providecommand \enquote  [1]{``#1''}%
\providecommand \bibnamefont  [1]{#1}%
\providecommand \bibfnamefont [1]{#1}%
\providecommand \citenamefont [1]{#1}%
\providecommand \href@noop [0]{\@secondoftwo}%
\providecommand \href [0]{\begingroup \@sanitize@url \@href}%
\providecommand \@href[1]{\@@startlink{#1}\@@href}%
\providecommand \@@href[1]{\endgroup#1\@@endlink}%
\providecommand \@sanitize@url [0]{\catcode `\\12\catcode `\$12\catcode
  `\&12\catcode `\#12\catcode `\^12\catcode `\_12\catcode `\%12\relax}%
\providecommand \@@startlink[1]{}%
\providecommand \@@endlink[0]{}%
\providecommand \url  [0]{\begingroup\@sanitize@url \@url }%
\providecommand \@url [1]{\endgroup\@href {#1}{\urlprefix }}%
\providecommand \urlprefix  [0]{URL }%
\providecommand \Eprint [0]{\href }%
\providecommand \doibase [0]{http://dx.doi.org/}%
\providecommand \selectlanguage [0]{\@gobble}%
\providecommand \bibinfo  [0]{\@secondoftwo}%
\providecommand \bibfield  [0]{\@secondoftwo}%
\providecommand \translation [1]{[#1]}%
\providecommand \BibitemOpen [0]{}%
\providecommand \bibitemStop [0]{}%
\providecommand \bibitemNoStop [0]{.\EOS\space}%
\providecommand \EOS [0]{\spacefactor3000\relax}%
\providecommand \BibitemShut  [1]{\csname bibitem#1\endcsname}%
\let\auto@bib@innerbib\@empty
\bibitem [{\citenamefont {Haas}\ and\ \citenamefont {Berg}(1936)}]{DeHaas36}%
  \BibitemOpen
  \bibfield  {author} {\bibinfo {author} {\bibfnamefont {W.~D.}\ \bibnamefont
  {Haas}}\ and\ \bibinfo {author} {\bibfnamefont {G.~V.~D.}\ \bibnamefont
  {Berg}},\ }\href {\doibase https://doi.org/10.1016/S0031-8914(36)80009-3}
  {\bibfield  {journal} {\bibinfo  {journal} {Physica}\ }\textbf {\bibinfo
  {volume} {3}},\ \bibinfo {pages} {440 } (\bibinfo {year} {1936})}\BibitemShut
  {NoStop}%
\bibitem [{\citenamefont {Kondo}(1964)}]{Kondo1964}%
  \BibitemOpen
  \bibfield  {author} {\bibinfo {author} {\bibfnamefont {J.}~\bibnamefont
  {Kondo}},\ }\href {\doibase 10.1143/PTP.32.37} {\bibfield  {journal}
  {\bibinfo  {journal} {Progress of Theoretical Physics}\ }\textbf {\bibinfo
  {volume} {32}},\ \bibinfo {pages} {37} (\bibinfo {year} {1964})}\BibitemShut
  {NoStop}%
\bibitem [{\citenamefont {Hewson}(1993)}]{Hewson1993}%
  \BibitemOpen
  \bibfield  {author} {\bibinfo {author} {\bibfnamefont {A.~C.}\ \bibnamefont
  {Hewson}},\ }\href {\doibase 10.1017/CBO9780511470752} {\emph {\bibinfo
  {title} {{The Kondo Problem to Heavy Fermions}}}},\ Cambridge Studies in
  Magnetism\ (\bibinfo  {publisher} {Cambridge University Press},\ \bibinfo
  {year} {1993})\BibitemShut {NoStop}%
\bibitem [{\citenamefont {Grewe}(1984)}]{Grewe1984-SSC}%
  \BibitemOpen
  \bibfield  {author} {\bibinfo {author} {\bibfnamefont {N.}~\bibnamefont
  {Grewe}},\ }\href {\doibase 10.1016/0038-1098(84)90050-4} {\bibfield
  {journal} {\bibinfo  {journal} {Solid State Communications}\ }\textbf
  {\bibinfo {volume} {50}},\ \bibinfo {pages} {19 } (\bibinfo {year}
  {1984})}\BibitemShut {NoStop}%
\bibitem [{\citenamefont {Grewe}\ and\ \citenamefont
  {Steglich}(1991)}]{Grewe91}%
  \BibitemOpen
  \bibfield  {author} {\bibinfo {author} {\bibfnamefont {N.}~\bibnamefont
  {Grewe}}\ and\ \bibinfo {author} {\bibfnamefont {F.}~\bibnamefont
  {Steglich}},\ }in\ \href@noop {} {\emph {\bibinfo {booktitle} {Handbook on
  the Physics and Chemistry of Rare Earths}}},\ Vol.~\bibinfo {volume} {14},\
  \bibinfo {editor} {edited by\ \bibinfo {editor} {\bibfnamefont {K.~A.}\
  \bibnamefont {{Gschneidner, Jr.}}}\ and\ \bibinfo {editor} {\bibfnamefont
  {L.}~\bibnamefont {Eyring}}}\ (\bibinfo  {publisher} {North-Holland},\
  \bibinfo {address} {Amsterdam},\ \bibinfo {year} {1991})\ p.\ \bibinfo
  {pages} {343}\BibitemShut {NoStop}%
\bibitem [{\citenamefont {Lawrence}\ \emph {et~al.}(1985)\citenamefont
  {Lawrence}, \citenamefont {Thompson},\ and\ \citenamefont
  {Chen}}]{Lawrence85}%
  \BibitemOpen
  \bibfield  {author} {\bibinfo {author} {\bibfnamefont {J.~M.}\ \bibnamefont
  {Lawrence}}, \bibinfo {author} {\bibfnamefont {J.~D.}\ \bibnamefont
  {Thompson}}, \ and\ \bibinfo {author} {\bibfnamefont {Y.~Y.}\ \bibnamefont
  {Chen}},\ }\href {\doibase 10.1103/PhysRevLett.54.2537} {\bibfield  {journal}
  {\bibinfo  {journal} {Phys. Rev. Lett.}\ }\textbf {\bibinfo {volume} {54}},\
  \bibinfo {pages} {2537} (\bibinfo {year} {1985})}\BibitemShut {NoStop}%
\bibitem [{\citenamefont {Lawrence}\ \emph {et~al.}(1996)\citenamefont
  {Lawrence}, \citenamefont {Graf}, \citenamefont {Hundley}, \citenamefont
  {Mandrus}, \citenamefont {Thompson}, \citenamefont {Lacerda}, \citenamefont
  {Torikachvili}, \citenamefont {Sarrao},\ and\ \citenamefont
  {Fisk}}]{Lawrence96}%
  \BibitemOpen
  \bibfield  {author} {\bibinfo {author} {\bibfnamefont {J.~M.}\ \bibnamefont
  {Lawrence}}, \bibinfo {author} {\bibfnamefont {T.}~\bibnamefont {Graf}},
  \bibinfo {author} {\bibfnamefont {M.~F.}\ \bibnamefont {Hundley}}, \bibinfo
  {author} {\bibfnamefont {D.}~\bibnamefont {Mandrus}}, \bibinfo {author}
  {\bibfnamefont {J.~D.}\ \bibnamefont {Thompson}}, \bibinfo {author}
  {\bibfnamefont {A.}~\bibnamefont {Lacerda}}, \bibinfo {author} {\bibfnamefont
  {M.~S.}\ \bibnamefont {Torikachvili}}, \bibinfo {author} {\bibfnamefont
  {J.~L.}\ \bibnamefont {Sarrao}}, \ and\ \bibinfo {author} {\bibfnamefont
  {Z.}~\bibnamefont {Fisk}},\ }\href {\doibase 10.1103/PhysRevB.53.12559}
  {\bibfield  {journal} {\bibinfo  {journal} {Phys. Rev. B}\ }\textbf {\bibinfo
  {volume} {53}},\ \bibinfo {pages} {12559} (\bibinfo {year}
  {1996})}\BibitemShut {NoStop}%
\bibitem [{\citenamefont {Hamidian}\ \emph {et~al.}(2011)\citenamefont
  {Hamidian}, \citenamefont {Schmidt}, \citenamefont {Firmo}, \citenamefont
  {Allan}, \citenamefont {Bradley}, \citenamefont {Garrett}, \citenamefont
  {Williams}, \citenamefont {Luke}, \citenamefont {Dubi}, \citenamefont
  {Balatsky},\ and\ \citenamefont {Davis}}]{Hamidian2011}%
  \BibitemOpen
  \bibfield  {author} {\bibinfo {author} {\bibfnamefont {M.~H.}\ \bibnamefont
  {Hamidian}}, \bibinfo {author} {\bibfnamefont {A.~R.}\ \bibnamefont
  {Schmidt}}, \bibinfo {author} {\bibfnamefont {I.~A.}\ \bibnamefont {Firmo}},
  \bibinfo {author} {\bibfnamefont {M.~P.}\ \bibnamefont {Allan}}, \bibinfo
  {author} {\bibfnamefont {P.}~\bibnamefont {Bradley}}, \bibinfo {author}
  {\bibfnamefont {J.~D.}\ \bibnamefont {Garrett}}, \bibinfo {author}
  {\bibfnamefont {T.~J.}\ \bibnamefont {Williams}}, \bibinfo {author}
  {\bibfnamefont {G.~M.}\ \bibnamefont {Luke}}, \bibinfo {author}
  {\bibfnamefont {Y.}~\bibnamefont {Dubi}}, \bibinfo {author} {\bibfnamefont
  {A.~V.}\ \bibnamefont {Balatsky}}, \ and\ \bibinfo {author} {\bibfnamefont
  {J.~C.}\ \bibnamefont {Davis}},\ }\href {\doibase 10.1073/pnas.1115027108}
  {\bibfield  {journal} {\bibinfo  {journal} {Proceedings of the National
  Academy of Sciences}\ }\textbf {\bibinfo {volume} {108}},\ \bibinfo {pages}
  {18233} (\bibinfo {year} {2011})}\BibitemShut {NoStop}%
\bibitem [{\citenamefont {Rosa}\ \emph {et~al.}(2016)\citenamefont {Rosa},
  \citenamefont {Oostra}, \citenamefont {Thompson}, \citenamefont {Pagliuso},\
  and\ \citenamefont {Fisk}}]{Rosa2016}%
  \BibitemOpen
  \bibfield  {author} {\bibinfo {author} {\bibfnamefont {P.~F.~S.}\
  \bibnamefont {Rosa}}, \bibinfo {author} {\bibfnamefont {A.}~\bibnamefont
  {Oostra}}, \bibinfo {author} {\bibfnamefont {J.~D.}\ \bibnamefont
  {Thompson}}, \bibinfo {author} {\bibfnamefont {P.~G.}\ \bibnamefont
  {Pagliuso}}, \ and\ \bibinfo {author} {\bibfnamefont {Z.}~\bibnamefont
  {Fisk}},\ }\href {\doibase 10.1103/PhysRevB.94.045101} {\bibfield  {journal}
  {\bibinfo  {journal} {Phys. Rev. B}\ }\textbf {\bibinfo {volume} {94}},\
  \bibinfo {pages} {045101} (\bibinfo {year} {2016})}\BibitemShut {NoStop}%
\bibitem [{\citenamefont {Shimozawa}\ \emph {et~al.}(2012)\citenamefont
  {Shimozawa}, \citenamefont {Watashige}, \citenamefont {Yasumoto},
  \citenamefont {Mizukami}, \citenamefont {Nakamura}, \citenamefont {Shishido},
  \citenamefont {Goh}, \citenamefont {Terashima}, \citenamefont {Shibauchi},\
  and\ \citenamefont {Matsuda}}]{shimozawa2012}%
  \BibitemOpen
  \bibfield  {author} {\bibinfo {author} {\bibfnamefont {M.}~\bibnamefont
  {Shimozawa}}, \bibinfo {author} {\bibfnamefont {T.}~\bibnamefont
  {Watashige}}, \bibinfo {author} {\bibfnamefont {S.}~\bibnamefont {Yasumoto}},
  \bibinfo {author} {\bibfnamefont {Y.}~\bibnamefont {Mizukami}}, \bibinfo
  {author} {\bibfnamefont {M.}~\bibnamefont {Nakamura}}, \bibinfo {author}
  {\bibfnamefont {H.}~\bibnamefont {Shishido}}, \bibinfo {author}
  {\bibfnamefont {S.~K.}\ \bibnamefont {Goh}}, \bibinfo {author} {\bibfnamefont
  {T.}~\bibnamefont {Terashima}}, \bibinfo {author} {\bibfnamefont
  {T.}~\bibnamefont {Shibauchi}}, \ and\ \bibinfo {author} {\bibfnamefont
  {Y.}~\bibnamefont {Matsuda}},\ }\href {\doibase 10.1103/PhysRevB.86.144526}
  {\bibfield  {journal} {\bibinfo  {journal} {Phys. Rev. B}\ }\textbf {\bibinfo
  {volume} {86}},\ \bibinfo {pages} {144526} (\bibinfo {year}
  {2012})}\BibitemShut {NoStop}%
\bibitem [{\citenamefont {Pietrus}\ \emph {et~al.}(2008)\citenamefont
  {Pietrus}, \citenamefont {v.~L\"ohneysen},\ and\ \citenamefont
  {Schlottmann}}]{Pietrus2008}%
  \BibitemOpen
  \bibfield  {author} {\bibinfo {author} {\bibfnamefont {T.}~\bibnamefont
  {Pietrus}}, \bibinfo {author} {\bibfnamefont {H.}~\bibnamefont
  {v.~L\"ohneysen}}, \ and\ \bibinfo {author} {\bibfnamefont {P.}~\bibnamefont
  {Schlottmann}},\ }\href {\doibase 10.1103/PhysRevB.77.115134} {\bibfield
  {journal} {\bibinfo  {journal} {Phys. Rev. B}\ }\textbf {\bibinfo {volume}
  {77}},\ \bibinfo {pages} {115134} (\bibinfo {year} {2008})}\BibitemShut
  {NoStop}%
\bibitem [{\citenamefont {Rotundu}\ \emph {et~al.}(2007)\citenamefont
  {Rotundu}, \citenamefont {Andraka},\ and\ \citenamefont
  {Schlottmann}}]{Rotundu2007}%
  \BibitemOpen
  \bibfield  {author} {\bibinfo {author} {\bibfnamefont {C.~R.}\ \bibnamefont
  {Rotundu}}, \bibinfo {author} {\bibfnamefont {B.}~\bibnamefont {Andraka}}, \
  and\ \bibinfo {author} {\bibfnamefont {P.}~\bibnamefont {Schlottmann}},\
  }\href {\doibase 10.1103/PhysRevB.76.054416} {\bibfield  {journal} {\bibinfo
  {journal} {Phys. Rev. B}\ }\textbf {\bibinfo {volume} {76}},\ \bibinfo
  {pages} {054416} (\bibinfo {year} {2007})}\BibitemShut {NoStop}%
\bibitem [{\citenamefont {\ifmmode~\acute{S}\else \'{S}\fi{}lebarski}\ \emph
  {et~al.}(2010)\citenamefont {\ifmmode~\acute{S}\else \'{S}\fi{}lebarski},
  \citenamefont {Spa\l{}ek}, \citenamefont {Fija\l{}kowski}, \citenamefont
  {Goraus}, \citenamefont {Cichorek},\ and\ \citenamefont
  {Bochenek}}]{lebraski2010}%
  \BibitemOpen
  \bibfield  {author} {\bibinfo {author} {\bibfnamefont {A.}~\bibnamefont
  {\ifmmode~\acute{S}\else \'{S}\fi{}lebarski}}, \bibinfo {author}
  {\bibfnamefont {J.}~\bibnamefont {Spa\l{}ek}}, \bibinfo {author}
  {\bibfnamefont {M.}~\bibnamefont {Fija\l{}kowski}}, \bibinfo {author}
  {\bibfnamefont {J.}~\bibnamefont {Goraus}}, \bibinfo {author} {\bibfnamefont
  {T.}~\bibnamefont {Cichorek}}, \ and\ \bibinfo {author} {\bibfnamefont
  {L.}~\bibnamefont {Bochenek}},\ }\href {\doibase 10.1103/PhysRevB.82.235106}
  {\bibfield  {journal} {\bibinfo  {journal} {Phys. Rev. B}\ }\textbf {\bibinfo
  {volume} {82}},\ \bibinfo {pages} {235106} (\bibinfo {year}
  {2010})}\BibitemShut {NoStop}%
\bibitem [{\citenamefont {Onuki}\ and\ \citenamefont
  {Komatsubara}(1987)}]{ONUKI1987}%
  \BibitemOpen
  \bibfield  {author} {\bibinfo {author} {\bibfnamefont {Y.}~\bibnamefont
  {Onuki}}\ and\ \bibinfo {author} {\bibfnamefont {T.}~\bibnamefont
  {Komatsubara}},\ }in\ \href {\doibase
  https://doi.org/10.1016/B978-1-4832-2948-5.50086-1} {\emph {\bibinfo
  {booktitle} {Anomalous Rare Earths and Actinides}}},\ \bibinfo {editor}
  {edited by\ \bibinfo {editor} {\bibfnamefont {J.}~\bibnamefont {Boucherle}},
  \bibinfo {editor} {\bibfnamefont {J.}~\bibnamefont {Flouquet}}, \bibinfo
  {editor} {\bibfnamefont {C.}~\bibnamefont {Lacroix}}, \ and\ \bibinfo
  {editor} {\bibfnamefont {J.}~\bibnamefont {Rossat-Mignod}}}\ (\bibinfo
  {publisher} {Elsevier},\ \bibinfo {year} {1987})\ pp.\ \bibinfo {pages}
  {281--288}\BibitemShut {NoStop}%
\bibitem [{\citenamefont {Malik}\ \emph {et~al.}(1995)\citenamefont {Malik},
  \citenamefont {Menon}, \citenamefont {Ghosh},\ and\ \citenamefont
  {Ramakrishnan}}]{Malik1995}%
  \BibitemOpen
  \bibfield  {author} {\bibinfo {author} {\bibfnamefont {S.~K.}\ \bibnamefont
  {Malik}}, \bibinfo {author} {\bibfnamefont {L.}~\bibnamefont {Menon}},
  \bibinfo {author} {\bibfnamefont {K.}~\bibnamefont {Ghosh}}, \ and\ \bibinfo
  {author} {\bibfnamefont {S.}~\bibnamefont {Ramakrishnan}},\ }\href {\doibase
  10.1103/PhysRevB.51.399} {\bibfield  {journal} {\bibinfo  {journal} {Phys.
  Rev. B}\ }\textbf {\bibinfo {volume} {51}},\ \bibinfo {pages} {399} (\bibinfo
  {year} {1995})}\BibitemShut {NoStop}%
\bibitem [{\citenamefont {Adroja}\ \emph {et~al.}(1996)\citenamefont {Adroja},
  \citenamefont {Rainford}, \citenamefont {Neville}, \citenamefont {Mandal},\
  and\ \citenamefont {Jansen}}]{Adroja1996}%
  \BibitemOpen
  \bibfield  {author} {\bibinfo {author} {\bibfnamefont {D.}~\bibnamefont
  {Adroja}}, \bibinfo {author} {\bibfnamefont {B.}~\bibnamefont {Rainford}},
  \bibinfo {author} {\bibfnamefont {A.}~\bibnamefont {Neville}}, \bibinfo
  {author} {\bibfnamefont {P.}~\bibnamefont {Mandal}}, \ and\ \bibinfo {author}
  {\bibfnamefont {A.}~\bibnamefont {Jansen}},\ }\href {\doibase
  https://doi.org/10.1016/S0304-8853(96)00034-0} {\bibfield  {journal}
  {\bibinfo  {journal} {Journal of Magnetism and Magnetic Materials}\ }\textbf
  {\bibinfo {volume} {161}},\ \bibinfo {pages} {157} (\bibinfo {year}
  {1996})}\BibitemShut {NoStop}%
\bibitem [{\citenamefont {\ifmmode~\acute{S}\else \'{S}\fi{}lebarski}\ \emph
  {et~al.}(1998)\citenamefont {\ifmmode~\acute{S}\else \'{S}\fi{}lebarski},
  \citenamefont {Jezierski}, \citenamefont {M\"ahl}, \citenamefont {Neumann},\
  and\ \citenamefont {Borstel}}]{Borstel1998}%
  \BibitemOpen
  \bibfield  {author} {\bibinfo {author} {\bibfnamefont {A.}~\bibnamefont
  {\ifmmode~\acute{S}\else \'{S}\fi{}lebarski}}, \bibinfo {author}
  {\bibfnamefont {A.}~\bibnamefont {Jezierski}}, \bibinfo {author}
  {\bibfnamefont {S.}~\bibnamefont {M\"ahl}}, \bibinfo {author} {\bibfnamefont
  {M.}~\bibnamefont {Neumann}}, \ and\ \bibinfo {author} {\bibfnamefont
  {G.}~\bibnamefont {Borstel}},\ }\href {\doibase 10.1103/PhysRevB.58.4367}
  {\bibfield  {journal} {\bibinfo  {journal} {Phys. Rev. B}\ }\textbf {\bibinfo
  {volume} {58}},\ \bibinfo {pages} {4367} (\bibinfo {year}
  {1998})}\BibitemShut {NoStop}%
\bibitem [{\citenamefont {Sollie}\ and\ \citenamefont
  {Schlottmann}(1991{\natexlab{a}})}]{Schlottmann91I}%
  \BibitemOpen
  \bibfield  {author} {\bibinfo {author} {\bibfnamefont {R.}~\bibnamefont
  {Sollie}}\ and\ \bibinfo {author} {\bibfnamefont {P.}~\bibnamefont
  {Schlottmann}},\ }\href {\doibase 10.1063/1.347971} {\bibfield  {journal}
  {\bibinfo  {journal} {Journal of Applied Physics}\ }\textbf {\bibinfo
  {volume} {69}},\ \bibinfo {pages} {5478} (\bibinfo {year}
  {1991}{\natexlab{a}})}\BibitemShut {NoStop}%
\bibitem [{\citenamefont {Sollie}\ and\ \citenamefont
  {Schlottmann}(1991{\natexlab{b}})}]{Schlottmann91II}%
  \BibitemOpen
  \bibfield  {author} {\bibinfo {author} {\bibfnamefont {R.}~\bibnamefont
  {Sollie}}\ and\ \bibinfo {author} {\bibfnamefont {P.}~\bibnamefont
  {Schlottmann}},\ }\href@noop {} {\bibfield  {journal} {\bibinfo  {journal}
  {Journal of Applied Physics}\ }\textbf {\bibinfo {volume} {70}},\ \bibinfo
  {pages} {5803} (\bibinfo {year} {1991}{\natexlab{b}})}\BibitemShut {NoStop}%
\bibitem [{\citenamefont {Schlottmann}(1992)}]{Schlottmann92}%
  \BibitemOpen
  \bibfield  {author} {\bibinfo {author} {\bibfnamefont {P.}~\bibnamefont
  {Schlottmann}},\ }\href {\doibase 10.1103/PhysRevB.46.998} {\bibfield
  {journal} {\bibinfo  {journal} {Phys. Rev. B}\ }\textbf {\bibinfo {volume}
  {46}},\ \bibinfo {pages} {998} (\bibinfo {year} {1992})}\BibitemShut
  {NoStop}%
\bibitem [{\citenamefont {Schlottmann}(1996)}]{Schlottmann96}%
  \BibitemOpen
  \bibfield  {author} {\bibinfo {author} {\bibfnamefont {P.}~\bibnamefont
  {Schlottmann}},\ }\href {\doibase 10.1007/BF02570939} {\bibfield  {journal}
  {\bibinfo  {journal} {Czech. J. Phys.}\ }\textbf {\bibinfo {volume} {46}},\
  \bibinfo {pages} {1895} (\bibinfo {year} {1996})}\BibitemShut {NoStop}%
\bibitem [{\citenamefont {Yu}(1996)}]{Clare96}%
  \BibitemOpen
  \bibfield  {author} {\bibinfo {author} {\bibfnamefont {C.~C.}\ \bibnamefont
  {Yu}},\ }\href {\doibase 10.1103/PhysRevB.54.15917} {\bibfield  {journal}
  {\bibinfo  {journal} {Phys. Rev. B}\ }\textbf {\bibinfo {volume} {54}},\
  \bibinfo {pages} {15917} (\bibinfo {year} {1996})}\BibitemShut {NoStop}%
\bibitem [{\citenamefont {Figgins}\ and\ \citenamefont
  {Morr}(2011)}]{Figgins2011}%
  \BibitemOpen
  \bibfield  {author} {\bibinfo {author} {\bibfnamefont {J.}~\bibnamefont
  {Figgins}}\ and\ \bibinfo {author} {\bibfnamefont {D.~K.}\ \bibnamefont
  {Morr}},\ }\href {\doibase 10.1103/PhysRevLett.107.066401} {\bibfield
  {journal} {\bibinfo  {journal} {Phys. Rev. Lett.}\ }\textbf {\bibinfo
  {volume} {107}},\ \bibinfo {pages} {066401} (\bibinfo {year}
  {2011})}\BibitemShut {NoStop}%
\bibitem [{\citenamefont {Baruselli}\ and\ \citenamefont
  {Vojta}(2014)}]{Vojta2014}%
  \BibitemOpen
  \bibfield  {author} {\bibinfo {author} {\bibfnamefont {P.~P.}\ \bibnamefont
  {Baruselli}}\ and\ \bibinfo {author} {\bibfnamefont {M.}~\bibnamefont
  {Vojta}},\ }\href {\doibase 10.1103/PhysRevB.89.205105} {\bibfield  {journal}
  {\bibinfo  {journal} {Phys. Rev. B}\ }\textbf {\bibinfo {volume} {89}},\
  \bibinfo {pages} {205105} (\bibinfo {year} {2014})}\BibitemShut {NoStop}%
\bibitem [{\citenamefont {Xie}\ \emph {et~al.}(2017)\citenamefont {Xie},
  \citenamefont {Hu},\ and\ \citenamefont {Yang}}]{Neng2017}%
  \BibitemOpen
  \bibfield  {author} {\bibinfo {author} {\bibfnamefont {N.}~\bibnamefont
  {Xie}}, \bibinfo {author} {\bibfnamefont {D.}~\bibnamefont {Hu}}, \ and\
  \bibinfo {author} {\bibfnamefont {Y.-f.}\ \bibnamefont {Yang}},\ }\href
  {https://doi.org/10.1038/s41598-017-12240-7} {\bibfield  {journal} {\bibinfo
  {journal} {Scientific Reports}\ }\textbf {\bibinfo {volume} {7}} (\bibinfo
  {year} {2017})}\BibitemShut {NoStop}%
\bibitem [{\citenamefont {Sen}\ \emph {et~al.}(2015)\citenamefont {Sen},
  \citenamefont {Moreno}, \citenamefont {Jarrell},\ and\ \citenamefont
  {Vidhyadhiraja}}]{Sen2015}%
  \BibitemOpen
  \bibfield  {author} {\bibinfo {author} {\bibfnamefont {S.}~\bibnamefont
  {Sen}}, \bibinfo {author} {\bibfnamefont {J.}~\bibnamefont {Moreno}},
  \bibinfo {author} {\bibfnamefont {M.}~\bibnamefont {Jarrell}}, \ and\
  \bibinfo {author} {\bibfnamefont {N.~S.}\ \bibnamefont {Vidhyadhiraja}},\
  }\href {\doibase 10.1103/PhysRevB.91.155146} {\bibfield  {journal} {\bibinfo
  {journal} {Phys. Rev. B}\ }\textbf {\bibinfo {volume} {91}},\ \bibinfo
  {pages} {155146} (\bibinfo {year} {2015})}\BibitemShut {NoStop}%
\bibitem [{\citenamefont {Kumar}\ and\ \citenamefont
  {Vidhyadhiraja}(2014)}]{Kumar2014}%
  \BibitemOpen
  \bibfield  {author} {\bibinfo {author} {\bibfnamefont {P.}~\bibnamefont
  {Kumar}}\ and\ \bibinfo {author} {\bibfnamefont {N.~S.}\ \bibnamefont
  {Vidhyadhiraja}},\ }\href {\doibase 10.1103/PhysRevB.90.235133} {\bibfield
  {journal} {\bibinfo  {journal} {Phys. Rev. B}\ }\textbf {\bibinfo {volume}
  {90}},\ \bibinfo {pages} {235133} (\bibinfo {year} {2014})}\BibitemShut
  {NoStop}%
\bibitem [{\citenamefont {Zhu}\ \emph {et~al.}(2012)\citenamefont {Zhu},
  \citenamefont {Julien}, \citenamefont {Dubi},\ and\ \citenamefont
  {Balatsky}}]{Zhu2012}%
  \BibitemOpen
  \bibfield  {author} {\bibinfo {author} {\bibfnamefont {J.-X.}\ \bibnamefont
  {Zhu}}, \bibinfo {author} {\bibfnamefont {J.-P.}\ \bibnamefont {Julien}},
  \bibinfo {author} {\bibfnamefont {Y.}~\bibnamefont {Dubi}}, \ and\ \bibinfo
  {author} {\bibfnamefont {A.~V.}\ \bibnamefont {Balatsky}},\ }\href {\doibase
  10.1103/PhysRevLett.108.186401} {\bibfield  {journal} {\bibinfo  {journal}
  {Phys. Rev. Lett.}\ }\textbf {\bibinfo {volume} {108}},\ \bibinfo {pages}
  {186401} (\bibinfo {year} {2012})}\BibitemShut {NoStop}%
\bibitem [{\citenamefont {Maruyama}\ \emph {et~al.}(2002)\citenamefont
  {Maruyama}, \citenamefont {Shibata},\ and\ \citenamefont
  {Ueda}}]{Maruyama2002}%
  \BibitemOpen
  \bibfield  {author} {\bibinfo {author} {\bibfnamefont {I.}~\bibnamefont
  {Maruyama}}, \bibinfo {author} {\bibfnamefont {N.}~\bibnamefont {Shibata}}, \
  and\ \bibinfo {author} {\bibfnamefont {K.}~\bibnamefont {Ueda}},\ }\href
  {\doibase 10.1103/PhysRevB.65.174421} {\bibfield  {journal} {\bibinfo
  {journal} {Phys. Rev. B}\ }\textbf {\bibinfo {volume} {65}},\ \bibinfo
  {pages} {174421} (\bibinfo {year} {2002})}\BibitemShut {NoStop}%
\bibitem [{\citenamefont {Wermbter}\ \emph {et~al.}(1996)\citenamefont
  {Wermbter}, \citenamefont {Sabel},\ and\ \citenamefont
  {Czycholl}}]{Wermbter96}%
  \BibitemOpen
  \bibfield  {author} {\bibinfo {author} {\bibfnamefont {S.}~\bibnamefont
  {Wermbter}}, \bibinfo {author} {\bibfnamefont {K.}~\bibnamefont {Sabel}}, \
  and\ \bibinfo {author} {\bibfnamefont {G.}~\bibnamefont {Czycholl}},\ }\href
  {\doibase 10.1103/PhysRevB.53.2528} {\bibfield  {journal} {\bibinfo
  {journal} {Phys. Rev. B}\ }\textbf {\bibinfo {volume} {53}},\ \bibinfo
  {pages} {2528} (\bibinfo {year} {1996})}\BibitemShut {NoStop}%
\bibitem [{\citenamefont {Schweitzer}\ and\ \citenamefont
  {Czycholl}(1990)}]{SchweitzerCzycholl90b}%
  \BibitemOpen
  \bibfield  {author} {\bibinfo {author} {\bibfnamefont {H.}~\bibnamefont
  {Schweitzer}}\ and\ \bibinfo {author} {\bibfnamefont {G.}~\bibnamefont
  {Czycholl}},\ }\href@noop {} {\bibfield  {journal} {\bibinfo  {journal}
  {Solid State Commun.}\ }\textbf {\bibinfo {volume} {74}},\ \bibinfo {pages}
  {735} (\bibinfo {year} {1990})}\BibitemShut {NoStop}%
\bibitem [{\citenamefont {Eickhoff}\ and\ \citenamefont
  {Anders}(2020)}]{MIAM2020}%
  \BibitemOpen
  \bibfield  {author} {\bibinfo {author} {\bibfnamefont {F.}~\bibnamefont
  {Eickhoff}}\ and\ \bibinfo {author} {\bibfnamefont {F.~B.}\ \bibnamefont
  {Anders}},\ }\href {\doibase 10.1103/PhysRevB.102.205132} {\bibfield
  {journal} {\bibinfo  {journal} {Phys. Rev. B}\ }\textbf {\bibinfo {volume}
  {102}},\ \bibinfo {pages} {205132} (\bibinfo {year} {2020})}\BibitemShut
  {NoStop}%
\bibitem [{\citenamefont {Krishna-murthy}\ \emph
  {et~al.}(1980{\natexlab{a}})\citenamefont {Krishna-murthy}, \citenamefont
  {Wilkins},\ and\ \citenamefont {Wilson}}]{Krishna-murthy1980I}%
  \BibitemOpen
  \bibfield  {author} {\bibinfo {author} {\bibfnamefont {H.~R.}\ \bibnamefont
  {Krishna-murthy}}, \bibinfo {author} {\bibfnamefont {J.~W.}\ \bibnamefont
  {Wilkins}}, \ and\ \bibinfo {author} {\bibfnamefont {K.~G.}\ \bibnamefont
  {Wilson}},\ }\href {\doibase 10.1103/PhysRevB.21.1003} {\bibfield  {journal}
  {\bibinfo  {journal} {Phys. Rev. B}\ }\textbf {\bibinfo {volume} {21}},\
  \bibinfo {pages} {1003} (\bibinfo {year} {1980}{\natexlab{a}})}\BibitemShut
  {NoStop}%
\bibitem [{\citenamefont {Krishna-murthy}\ \emph
  {et~al.}(1980{\natexlab{b}})\citenamefont {Krishna-murthy}, \citenamefont
  {Wilkins},\ and\ \citenamefont {Wilson}}]{Krishna-murthy1980II}%
  \BibitemOpen
  \bibfield  {author} {\bibinfo {author} {\bibfnamefont {H.~R.}\ \bibnamefont
  {Krishna-murthy}}, \bibinfo {author} {\bibfnamefont {J.~W.}\ \bibnamefont
  {Wilkins}}, \ and\ \bibinfo {author} {\bibfnamefont {K.~G.}\ \bibnamefont
  {Wilson}},\ }\href {\doibase 10.1103/PhysRevB.21.1044} {\bibfield  {journal}
  {\bibinfo  {journal} {Phys. Rev. B}\ }\textbf {\bibinfo {volume} {21}},\
  \bibinfo {pages} {1044} (\bibinfo {year} {1980}{\natexlab{b}})}\BibitemShut
  {NoStop}%
\bibitem [{\citenamefont {Andrei}\ \emph {et~al.}(1983)\citenamefont {Andrei},
  \citenamefont {Furuya},\ and\ \citenamefont
  {Lowenstein}}]{AndreiFuruyaLowenstein83}%
  \BibitemOpen
  \bibfield  {author} {\bibinfo {author} {\bibfnamefont {N.}~\bibnamefont
  {Andrei}}, \bibinfo {author} {\bibfnamefont {K.}~\bibnamefont {Furuya}}, \
  and\ \bibinfo {author} {\bibfnamefont {J.~H.}\ \bibnamefont {Lowenstein}},\
  }\href {\doibase 10.1103/RevModPhys.55.331} {\bibfield  {journal} {\bibinfo
  {journal} {Rev. Mod. Phys.}\ }\textbf {\bibinfo {volume} {55}},\ \bibinfo
  {pages} {331} (\bibinfo {year} {1983})}\BibitemShut {NoStop}%
\bibitem [{\citenamefont {Schlottmann}(1989)}]{Schlottmann89}%
  \BibitemOpen
  \bibfield  {author} {\bibinfo {author} {\bibfnamefont {P.}~\bibnamefont
  {Schlottmann}},\ }\href@noop {} {\bibfield  {journal} {\bibinfo  {journal}
  {Physics Rep.}\ }\textbf {\bibinfo {volume} {181}},\ \bibinfo {pages} {1}
  (\bibinfo {year} {1989})}\BibitemShut {NoStop}%
\bibitem [{\citenamefont {Jones}\ and\ \citenamefont
  {Varma}(1987)}]{Jones1987}%
  \BibitemOpen
  \bibfield  {author} {\bibinfo {author} {\bibfnamefont {B.~A.}\ \bibnamefont
  {Jones}}\ and\ \bibinfo {author} {\bibfnamefont {C.~M.}\ \bibnamefont
  {Varma}},\ }\href {\doibase 10.1103/PhysRevLett.58.843} {\bibfield  {journal}
  {\bibinfo  {journal} {Phys. Rev. Lett.}\ }\textbf {\bibinfo {volume} {58}},\
  \bibinfo {pages} {843} (\bibinfo {year} {1987})}\BibitemShut {NoStop}%
\bibitem [{\citenamefont {Eickhoff}\ \emph {et~al.}(2018)\citenamefont
  {Eickhoff}, \citenamefont {Lechtenberg},\ and\ \citenamefont
  {Anders}}]{Eickhoff2018}%
  \BibitemOpen
  \bibfield  {author} {\bibinfo {author} {\bibfnamefont {F.}~\bibnamefont
  {Eickhoff}}, \bibinfo {author} {\bibfnamefont {B.}~\bibnamefont
  {Lechtenberg}}, \ and\ \bibinfo {author} {\bibfnamefont {F.~B.}\ \bibnamefont
  {Anders}},\ }\href {\doibase 10.1103/PhysRevB.98.115103} {\bibfield
  {journal} {\bibinfo  {journal} {Phys. Rev. B}\ }\textbf {\bibinfo {volume}
  {98}},\ \bibinfo {pages} {115103} (\bibinfo {year} {2018})}\BibitemShut
  {NoStop}%
\bibitem [{\citenamefont {Schrieffer}\ and\ \citenamefont
  {Wolff}(1966)}]{SchriefferWol66}%
  \BibitemOpen
  \bibfield  {author} {\bibinfo {author} {\bibfnamefont {J.~R.}\ \bibnamefont
  {Schrieffer}}\ and\ \bibinfo {author} {\bibfnamefont {P.~A.}\ \bibnamefont
  {Wolff}},\ }\href {\doibase 10.1103/PhysRev.149.491} {\bibfield  {journal}
  {\bibinfo  {journal} {Phys. Rev.}\ }\textbf {\bibinfo {volume} {149}},\
  \bibinfo {pages} {491} (\bibinfo {year} {1966})}\BibitemShut {NoStop}%
\bibitem [{\citenamefont {Schlottmann}(1995)}]{Schlottmann1995}%
  \BibitemOpen
  \bibfield  {author} {\bibinfo {author} {\bibfnamefont {P.}~\bibnamefont
  {Schlottmann}},\ }\href
  {https://www.sciencedirect.com/science/article/pii/092145269400594L}
  {\bibfield  {journal} {\bibinfo  {journal} {Physica B: Condensed Matter}\
  }\textbf {\bibinfo {volume} {206-207}},\ \bibinfo {pages} {816} (\bibinfo
  {year} {1995})}\BibitemShut {NoStop}%
\bibitem [{\citenamefont {Lieb}\ and\ \citenamefont
  {Mattis}(1962)}]{LiebMattis}%
  \BibitemOpen
  \bibfield  {author} {\bibinfo {author} {\bibfnamefont {E.}~\bibnamefont
  {Lieb}}\ and\ \bibinfo {author} {\bibfnamefont {D.}~\bibnamefont {Mattis}},\
  }\href {\doibase 10.1063/1.1724276} {\bibfield  {journal} {\bibinfo
  {journal} {Journal of Mathematical Physics}\ }\textbf {\bibinfo {volume}
  {3}},\ \bibinfo {pages} {749} (\bibinfo {year} {1962})}\BibitemShut {NoStop}%
\bibitem [{\citenamefont {Shen}(1996)}]{Shen1996}%
  \BibitemOpen
  \bibfield  {author} {\bibinfo {author} {\bibfnamefont {S.-Q.}\ \bibnamefont
  {Shen}},\ }\href {\doibase 10.1103/PhysRevB.53.14252} {\bibfield  {journal}
  {\bibinfo  {journal} {Phys. Rev. B}\ }\textbf {\bibinfo {volume} {53}},\
  \bibinfo {pages} {14252} (\bibinfo {year} {1996})}\BibitemShut {NoStop}%
\bibitem [{\citenamefont {Titvinidze}\ \emph {et~al.}(2014)\citenamefont
  {Titvinidze}, \citenamefont {Schwabe},\ and\ \citenamefont
  {Potthoff}}]{Potthoff2014}%
  \BibitemOpen
  \bibfield  {author} {\bibinfo {author} {\bibfnamefont {I.}~\bibnamefont
  {Titvinidze}}, \bibinfo {author} {\bibfnamefont {A.}~\bibnamefont {Schwabe}},
  \ and\ \bibinfo {author} {\bibfnamefont {M.}~\bibnamefont {Potthoff}},\
  }\href {\doibase 10.1103/PhysRevB.90.045112} {\bibfield  {journal} {\bibinfo
  {journal} {Phys. Rev. B}\ }\textbf {\bibinfo {volume} {90}},\ \bibinfo
  {pages} {045112} (\bibinfo {year} {2014})}\BibitemShut {NoStop}%
\bibitem [{\citenamefont {Wilson}(1975)}]{Wilson1975}%
  \BibitemOpen
  \bibfield  {author} {\bibinfo {author} {\bibfnamefont {K.~G.}\ \bibnamefont
  {Wilson}},\ }\href {\doibase 10.1103/RevModPhys.47.773} {\bibfield  {journal}
  {\bibinfo  {journal} {Rev. Mod. Phys.}\ }\textbf {\bibinfo {volume} {47}},\
  \bibinfo {pages} {773} (\bibinfo {year} {1975})}\BibitemShut {NoStop}%
\bibitem [{\citenamefont {Bulla}\ \emph {et~al.}(2008)\citenamefont {Bulla},
  \citenamefont {Costi},\ and\ \citenamefont {Pruschke}}]{Bulla2008}%
  \BibitemOpen
  \bibfield  {author} {\bibinfo {author} {\bibfnamefont {R.}~\bibnamefont
  {Bulla}}, \bibinfo {author} {\bibfnamefont {T.~A.}\ \bibnamefont {Costi}}, \
  and\ \bibinfo {author} {\bibfnamefont {T.}~\bibnamefont {Pruschke}},\ }\href
  {\doibase 10.1103/RevModPhys.80.395} {\bibfield  {journal} {\bibinfo
  {journal} {Rev. Mod. Phys.}\ }\textbf {\bibinfo {volume} {80}},\ \bibinfo
  {pages} {395} (\bibinfo {year} {2008})}\BibitemShut {NoStop}%
\bibitem [{Note1()}]{Note1}%
  \BibitemOpen
  \bibinfo {note} {Note that we cannot make a statement about the nature of the
  FP: In the TIAM, for instance, two adiabatically connected FP are found, one
  originating from a RKKY interaction the other driven by the Kondo effect
  \cite {Jones1987,Affleck1995}}\BibitemShut {NoStop}%
\bibitem [{\citenamefont {Vojta}(2006)}]{Vojta2006}%
  \BibitemOpen
  \bibfield  {author} {\bibinfo {author} {\bibfnamefont {M.}~\bibnamefont
  {Vojta}},\ }\href@noop {} {\bibfield  {journal} {\bibinfo  {journal}
  {Philosophical Magazine}\ }\textbf {\bibinfo {volume} {86}},\ \bibinfo
  {pages} {1807} (\bibinfo {year} {2006})}\BibitemShut {NoStop}%
\bibitem [{\citenamefont {Pereira}\ \emph {et~al.}(2006)\citenamefont
  {Pereira}, \citenamefont {Guinea}, \citenamefont {dos Santos}, \citenamefont
  {Peres},\ and\ \citenamefont {Neto}}]{Pareira2006}%
  \BibitemOpen
  \bibfield  {author} {\bibinfo {author} {\bibfnamefont {V.~M.}\ \bibnamefont
  {Pereira}}, \bibinfo {author} {\bibfnamefont {F.}~\bibnamefont {Guinea}},
  \bibinfo {author} {\bibfnamefont {J.~M. B.~L.}\ \bibnamefont {dos Santos}},
  \bibinfo {author} {\bibfnamefont {N.~M.~R.}\ \bibnamefont {Peres}}, \ and\
  \bibinfo {author} {\bibfnamefont {A.~H.~C.}\ \bibnamefont {Neto}},\
  }\href@noop {} {\bibfield  {journal} {\bibinfo  {journal} {Phys. Rev. Let.}\
  }\textbf {\bibinfo {volume} {96}},\ \bibinfo {pages} {036801} (\bibinfo
  {year} {2006})}\BibitemShut {NoStop}%
\bibitem [{\citenamefont {Castro~Neto}\ \emph {et~al.}(2009)\citenamefont
  {Castro~Neto}, \citenamefont {Guinea}, \citenamefont {Peres}, \citenamefont
  {Novoselov},\ and\ \citenamefont {Geim}}]{GrapheneRMP2009}%
  \BibitemOpen
  \bibfield  {author} {\bibinfo {author} {\bibfnamefont {A.~H.}\ \bibnamefont
  {Castro~Neto}}, \bibinfo {author} {\bibfnamefont {F.}~\bibnamefont {Guinea}},
  \bibinfo {author} {\bibfnamefont {N.~M.~R.}\ \bibnamefont {Peres}}, \bibinfo
  {author} {\bibfnamefont {K.~S.}\ \bibnamefont {Novoselov}}, \ and\ \bibinfo
  {author} {\bibfnamefont {A.~K.}\ \bibnamefont {Geim}},\ }\href {\doibase
  10.1103/RevModPhys.81.109} {\bibfield  {journal} {\bibinfo  {journal} {Rev.
  Mod. Phys.}\ }\textbf {\bibinfo {volume} {81}},\ \bibinfo {pages} {109}
  (\bibinfo {year} {2009})}\BibitemShut {NoStop}%
\bibitem [{\citenamefont {Nanda}\ \emph {et~al.}(2012)\citenamefont {Nanda},
  \citenamefont {Sherafati}, \citenamefont {Popovi{\'c}},\ and\ \citenamefont
  {Satpathy}}]{Nanda2012}%
  \BibitemOpen
  \bibfield  {author} {\bibinfo {author} {\bibfnamefont {B.~R.~K.}\
  \bibnamefont {Nanda}}, \bibinfo {author} {\bibfnamefont {M.}~\bibnamefont
  {Sherafati}}, \bibinfo {author} {\bibfnamefont {Z.~S.}\ \bibnamefont
  {Popovi{\'c}}}, \ and\ \bibinfo {author} {\bibfnamefont {S.}~\bibnamefont
  {Satpathy}},\ }\href {http://stacks.iop.org/1367-2630/14/i=8/a=083004}
  {\bibfield  {journal} {\bibinfo  {journal} {New Journal of Physics}\ }\textbf
  {\bibinfo {volume} {14}},\ \bibinfo {pages} {083004} (\bibinfo {year}
  {2012})}\BibitemShut {NoStop}%
\bibitem [{\citenamefont {May}\ \emph {et~al.}(2018)\citenamefont {May},
  \citenamefont {Lo}, \citenamefont {Deltenre}, \citenamefont {Henke},
  \citenamefont {Mao}, \citenamefont {Jiang}, \citenamefont {Li}, \citenamefont
  {Andrei}, \citenamefont {Guo},\ and\ \citenamefont
  {Anders}}]{MayGraphen2018}%
  \BibitemOpen
  \bibfield  {author} {\bibinfo {author} {\bibfnamefont {D.}~\bibnamefont
  {May}}, \bibinfo {author} {\bibfnamefont {P.-W.}\ \bibnamefont {Lo}},
  \bibinfo {author} {\bibfnamefont {K.}~\bibnamefont {Deltenre}}, \bibinfo
  {author} {\bibfnamefont {A.}~\bibnamefont {Henke}}, \bibinfo {author}
  {\bibfnamefont {J.}~\bibnamefont {Mao}}, \bibinfo {author} {\bibfnamefont
  {Y.}~\bibnamefont {Jiang}}, \bibinfo {author} {\bibfnamefont
  {G.}~\bibnamefont {Li}}, \bibinfo {author} {\bibfnamefont {E.~Y.}\
  \bibnamefont {Andrei}}, \bibinfo {author} {\bibfnamefont {G.-Y.}\
  \bibnamefont {Guo}}, \ and\ \bibinfo {author} {\bibfnamefont {F.~B.}\
  \bibnamefont {Anders}},\ }\href {\doibase 10.1103/PhysRevB.97.155419}
  {\bibfield  {journal} {\bibinfo  {journal} {Phys. Rev. B}\ }\textbf {\bibinfo
  {volume} {97}},\ \bibinfo {pages} {155419} (\bibinfo {year}
  {2018})}\BibitemShut {NoStop}%
\bibitem [{\citenamefont {Jiang}\ \emph {et~al.}(2018)\citenamefont {Jiang},
  \citenamefont {Lo}, \citenamefont {May}, \citenamefont {Li}, \citenamefont
  {Guo}, \citenamefont {Anders}, \citenamefont {Taniguchi}, \citenamefont
  {Watanabe}, \citenamefont {Mao},\ and\ \citenamefont
  {Andrei}}]{AndreiGraphen2018}%
  \BibitemOpen
  \bibfield  {author} {\bibinfo {author} {\bibfnamefont {Y.}~\bibnamefont
  {Jiang}}, \bibinfo {author} {\bibfnamefont {P.-W.}\ \bibnamefont {Lo}},
  \bibinfo {author} {\bibfnamefont {D.}~\bibnamefont {May}}, \bibinfo {author}
  {\bibfnamefont {G.}~\bibnamefont {Li}}, \bibinfo {author} {\bibfnamefont
  {G.-Y.}\ \bibnamefont {Guo}}, \bibinfo {author} {\bibfnamefont {F.~B.}\
  \bibnamefont {Anders}}, \bibinfo {author} {\bibfnamefont {T.}~\bibnamefont
  {Taniguchi}}, \bibinfo {author} {\bibfnamefont {K.}~\bibnamefont {Watanabe}},
  \bibinfo {author} {\bibfnamefont {J.}~\bibnamefont {Mao}}, \ and\ \bibinfo
  {author} {\bibfnamefont {E.~Y.}\ \bibnamefont {Andrei}},\ }\href@noop {}
  {\bibfield  {journal} {\bibinfo  {journal} {Nature Communications}\ }\textbf
  {\bibinfo {volume} {9}},\ \bibinfo {pages} {2349} (\bibinfo {year}
  {2018})}\BibitemShut {NoStop}%
\bibitem [{\citenamefont {Fritz}\ and\ \citenamefont
  {Vojta}(2004)}]{Vojta2004I}%
  \BibitemOpen
  \bibfield  {author} {\bibinfo {author} {\bibfnamefont {L.}~\bibnamefont
  {Fritz}}\ and\ \bibinfo {author} {\bibfnamefont {M.}~\bibnamefont {Vojta}},\
  }\href {\doibase 10.1103/PhysRevB.70.214427} {\bibfield  {journal} {\bibinfo
  {journal} {Phys. Rev. B}\ }\textbf {\bibinfo {volume} {70}},\ \bibinfo
  {pages} {214427} (\bibinfo {year} {2004})}\BibitemShut {NoStop}%
\bibitem [{\citenamefont {Lechtenberg}\ \emph {et~al.}(2017)\citenamefont
  {Lechtenberg}, \citenamefont {Eickhoff},\ and\ \citenamefont
  {Anders}}]{TIAM2017}%
  \BibitemOpen
  \bibfield  {author} {\bibinfo {author} {\bibfnamefont {B.}~\bibnamefont
  {Lechtenberg}}, \bibinfo {author} {\bibfnamefont {F.}~\bibnamefont
  {Eickhoff}}, \ and\ \bibinfo {author} {\bibfnamefont {F.~B.}\ \bibnamefont
  {Anders}},\ }\href {\doibase 10.1103/PhysRevB.96.041109} {\bibfield
  {journal} {\bibinfo  {journal} {Phys. Rev. B}\ }\textbf {\bibinfo {volume}
  {96}},\ \bibinfo {pages} {041109} (\bibinfo {year} {2017})}\BibitemShut
  {NoStop}%
\bibitem [{\citenamefont {Jones}\ \emph {et~al.}(1988)\citenamefont {Jones},
  \citenamefont {Varma},\ and\ \citenamefont {Wilkins}}]{Jones1988}%
  \BibitemOpen
  \bibfield  {author} {\bibinfo {author} {\bibfnamefont {B.~A.}\ \bibnamefont
  {Jones}}, \bibinfo {author} {\bibfnamefont {C.~M.}\ \bibnamefont {Varma}}, \
  and\ \bibinfo {author} {\bibfnamefont {J.~W.}\ \bibnamefont {Wilkins}},\
  }\href {\doibase 10.1103/PhysRevLett.61.125} {\bibfield  {journal} {\bibinfo
  {journal} {Phys. Rev. Lett.}\ }\textbf {\bibinfo {volume} {61}},\ \bibinfo
  {pages} {125} (\bibinfo {year} {1988})}\BibitemShut {NoStop}%
\bibitem [{\citenamefont {Affleck}\ \emph {et~al.}(1995)\citenamefont
  {Affleck}, \citenamefont {Ludwig},\ and\ \citenamefont
  {Jones}}]{Affleck1995}%
  \BibitemOpen
  \bibfield  {author} {\bibinfo {author} {\bibfnamefont {I.}~\bibnamefont
  {Affleck}}, \bibinfo {author} {\bibfnamefont {A.~W.~W.}\ \bibnamefont
  {Ludwig}}, \ and\ \bibinfo {author} {\bibfnamefont {B.~A.}\ \bibnamefont
  {Jones}},\ }\href {\doibase 10.1103/PhysRevB.52.9528} {\bibfield  {journal}
  {\bibinfo  {journal} {Phys. Rev. B}\ }\textbf {\bibinfo {volume} {52}},\
  \bibinfo {pages} {9528} (\bibinfo {year} {1995})}\BibitemShut {NoStop}%
\bibitem [{\citenamefont {Esat}\ \emph {et~al.}(2016)\citenamefont {Esat},
  \citenamefont {Lechtenberg}, \citenamefont {Deilmann}, \citenamefont
  {ChristianWagner}, \citenamefont {Kr{\"u}ger}, \citenamefont {Temirov},
  \citenamefont {Rohlfing}, \citenamefont {Anders},\ and\ \citenamefont
  {Tautz}}]{AU-PTCDA-dimer}%
  \BibitemOpen
  \bibfield  {author} {\bibinfo {author} {\bibfnamefont {T.}~\bibnamefont
  {Esat}}, \bibinfo {author} {\bibfnamefont {B.}~\bibnamefont {Lechtenberg}},
  \bibinfo {author} {\bibfnamefont {T.}~\bibnamefont {Deilmann}}, \bibinfo
  {author} {\bibnamefont {ChristianWagner}}, \bibinfo {author} {\bibfnamefont
  {P.}~\bibnamefont {Kr{\"u}ger}}, \bibinfo {author} {\bibfnamefont
  {R.}~\bibnamefont {Temirov}}, \bibinfo {author} {\bibfnamefont
  {M.}~\bibnamefont {Rohlfing}}, \bibinfo {author} {\bibfnamefont {F.~B.}\
  \bibnamefont {Anders}}, \ and\ \bibinfo {author} {\bibfnamefont {F.~S.}\
  \bibnamefont {Tautz}},\ }\href@noop {} {\bibfield  {journal} {\bibinfo
  {journal} {Nature Physics}\ }\textbf {\bibinfo {volume} {12}},\ \bibinfo
  {pages} {8} (\bibinfo {year} {2016})}\BibitemShut {NoStop}%
\bibitem [{Note2()}]{Note2}%
  \BibitemOpen
  \bibinfo {note} {Since the trivial irrep of any point group is one
  dimensional we drop the index $\alpha $ in this case.}\BibitemShut {Stop}%
\bibitem [{\citenamefont {Mitchell}\ and\ \citenamefont
  {Bulla}(2015)}]{MitchellBulla2015}%
  \BibitemOpen
  \bibfield  {author} {\bibinfo {author} {\bibfnamefont {A.~K.}\ \bibnamefont
  {Mitchell}}\ and\ \bibinfo {author} {\bibfnamefont {R.}~\bibnamefont
  {Bulla}},\ }\href {\doibase 10.1103/PhysRevB.92.155101} {\bibfield  {journal}
  {\bibinfo  {journal} {Phys. Rev. B}\ }\textbf {\bibinfo {volume} {92}},\
  \bibinfo {pages} {155101} (\bibinfo {year} {2015})}\BibitemShut {NoStop}%
\bibitem [{\citenamefont {Grenzebach}\ \emph {et~al.}(2008)\citenamefont
  {Grenzebach}, \citenamefont {Anders}, \citenamefont {Czycholl},\ and\
  \citenamefont {Pruschke}}]{GrenzebachAndersCzychollPruschke2008}%
  \BibitemOpen
  \bibfield  {author} {\bibinfo {author} {\bibfnamefont {C.}~\bibnamefont
  {Grenzebach}}, \bibinfo {author} {\bibfnamefont {F.~B.}\ \bibnamefont
  {Anders}}, \bibinfo {author} {\bibfnamefont {G.}~\bibnamefont {Czycholl}}, \
  and\ \bibinfo {author} {\bibfnamefont {T.}~\bibnamefont {Pruschke}},\ }\href
  {\doibase 10.1103/PhysRevB.77.115125} {\bibfield  {journal} {\bibinfo
  {journal} {Phys. Rev. B}\ }\textbf {\bibinfo {volume} {77}},\ \bibinfo
  {pages} {115125} (\bibinfo {year} {2008})}\BibitemShut {NoStop}%
\bibitem [{\citenamefont {Aulbach}\ \emph {et~al.}(2015)\citenamefont
  {Aulbach}, \citenamefont {Titvinidze},\ and\ \citenamefont
  {Potthoff}}]{Potthoff2015}%
  \BibitemOpen
  \bibfield  {author} {\bibinfo {author} {\bibfnamefont {M.~W.}\ \bibnamefont
  {Aulbach}}, \bibinfo {author} {\bibfnamefont {I.}~\bibnamefont {Titvinidze}},
  \ and\ \bibinfo {author} {\bibfnamefont {M.}~\bibnamefont {Potthoff}},\
  }\href {\doibase 10.1103/PhysRevB.91.174420} {\bibfield  {journal} {\bibinfo
  {journal} {Phys. Rev. B}\ }\textbf {\bibinfo {volume} {91}},\ \bibinfo
  {pages} {174420} (\bibinfo {year} {2015})}\BibitemShut {NoStop}%
\bibitem [{\citenamefont {Lebanon}\ \emph {et~al.}(2003)\citenamefont
  {Lebanon}, \citenamefont {Schiller},\ and\ \citenamefont
  {Anders}}]{LebanonSchillerAnders2003}%
  \BibitemOpen
  \bibfield  {author} {\bibinfo {author} {\bibfnamefont {E.}~\bibnamefont
  {Lebanon}}, \bibinfo {author} {\bibfnamefont {A.}~\bibnamefont {Schiller}}, \
  and\ \bibinfo {author} {\bibfnamefont {F.~B.}\ \bibnamefont {Anders}},\
  }\href@noop {} {\bibfield  {journal} {\bibinfo  {journal} {Phys. Rev. B}\
  }\textbf {\bibinfo {volume} {68}},\ \bibinfo {pages} {155301} (\bibinfo
  {year} {2003})}\BibitemShut {NoStop}%
\bibitem [{\citenamefont {Nevidomskyy}\ and\ \citenamefont
  {Coleman}(2009)}]{Nevidomskyy2009}%
  \BibitemOpen
  \bibfield  {author} {\bibinfo {author} {\bibfnamefont {A.~H.}\ \bibnamefont
  {Nevidomskyy}}\ and\ \bibinfo {author} {\bibfnamefont {P.}~\bibnamefont
  {Coleman}},\ }\href {\doibase 10.1103/PhysRevLett.103.147205} {\bibfield
  {journal} {\bibinfo  {journal} {Phys. Rev. Lett.}\ }\textbf {\bibinfo
  {volume} {103}},\ \bibinfo {pages} {147205} (\bibinfo {year}
  {2009})}\BibitemShut {NoStop}%
\bibitem [{\citenamefont {Schrieffer}(1967)}]{Schrieffer1967}%
  \BibitemOpen
  \bibfield  {author} {\bibinfo {author} {\bibfnamefont {J.~R.}\ \bibnamefont
  {Schrieffer}},\ }\href {\doibase 10.1063/1.1709517} {\bibfield  {journal}
  {\bibinfo  {journal} {Journal of Applied Physics}\ }\textbf {\bibinfo
  {volume} {38}},\ \bibinfo {pages} {1143} (\bibinfo {year}
  {1967})}\BibitemShut {NoStop}%
\bibitem [{\citenamefont {Doniach}(1977)}]{Doniach77}%
  \BibitemOpen
  \bibfield  {author} {\bibinfo {author} {\bibfnamefont {S.}~\bibnamefont
  {Doniach}},\ }\href@noop {} {\bibfield  {journal} {\bibinfo  {journal}
  {Physica B}\ }\textbf {\bibinfo {volume} {91}},\ \bibinfo {pages} {231}
  (\bibinfo {year} {1977})}\BibitemShut {NoStop}%
\bibitem [{\citenamefont {Pruschke}\ \emph {et~al.}(2000)\citenamefont
  {Pruschke}, \citenamefont {Bulla},\ and\ \citenamefont
  {Jarrell}}]{PruschkeBullaJarrell2000}%
  \BibitemOpen
  \bibfield  {author} {\bibinfo {author} {\bibfnamefont {T.}~\bibnamefont
  {Pruschke}}, \bibinfo {author} {\bibfnamefont {R.}~\bibnamefont {Bulla}}, \
  and\ \bibinfo {author} {\bibfnamefont {M.}~\bibnamefont {Jarrell}},\ }\href
  {\doibase 10.1103/PhysRevB.61.12799} {\bibfield  {journal} {\bibinfo
  {journal} {Phys. Rev. B}\ }\textbf {\bibinfo {volume} {61}},\ \bibinfo
  {pages} {12799} (\bibinfo {year} {2000})}\BibitemShut {NoStop}%
\bibitem [{Note3()}]{Note3}%
  \BibitemOpen
  \bibinfo {note} {We have shown in Ref.~\cite {MIAM2020} that the
  rank$[\protect \bm {\Gamma }]$ is limited by the number of Fermi wave vectors
  hence by 2 in 1d.}\BibitemShut {Stop}%
\bibitem [{\citenamefont {Silva}\ \emph {et~al.}(1996)\citenamefont {Silva},
  \citenamefont {Lima}, \citenamefont {Oliveira}, \citenamefont {Mello},
  \citenamefont {Oliveira},\ and\ \citenamefont {Wilkins}}]{Silva1996}%
  \BibitemOpen
  \bibfield  {author} {\bibinfo {author} {\bibfnamefont {J.~B.}\ \bibnamefont
  {Silva}}, \bibinfo {author} {\bibfnamefont {W.~L.~C.}\ \bibnamefont {Lima}},
  \bibinfo {author} {\bibfnamefont {W.~C.}\ \bibnamefont {Oliveira}}, \bibinfo
  {author} {\bibfnamefont {J.~L.~N.}\ \bibnamefont {Mello}}, \bibinfo {author}
  {\bibfnamefont {L.~N.}\ \bibnamefont {Oliveira}}, \ and\ \bibinfo {author}
  {\bibfnamefont {J.~W.}\ \bibnamefont {Wilkins}},\ }\href {\doibase
  10.1103/PhysRevLett.76.275} {\bibfield  {journal} {\bibinfo  {journal} {Phys.
  Rev. Lett.}\ }\textbf {\bibinfo {volume} {76}},\ \bibinfo {pages} {275}
  (\bibinfo {year} {1996})}\BibitemShut {NoStop}%
\bibitem [{\citenamefont {Jones}\ and\ \citenamefont
  {Varma}(1989)}]{Jones1989}%
  \BibitemOpen
  \bibfield  {author} {\bibinfo {author} {\bibfnamefont {B.~A.}\ \bibnamefont
  {Jones}}\ and\ \bibinfo {author} {\bibfnamefont {C.~M.}\ \bibnamefont
  {Varma}},\ }\href {\doibase 10.1103/PhysRevB.40.324} {\bibfield  {journal}
  {\bibinfo  {journal} {Phys. Rev. B}\ }\textbf {\bibinfo {volume} {40}},\
  \bibinfo {pages} {324} (\bibinfo {year} {1989})}\BibitemShut {NoStop}%
\bibitem [{\citenamefont {Cornaglia}\ and\ \citenamefont
  {Grempel}(2005)}]{Cornaglia2005}%
  \BibitemOpen
  \bibfield  {author} {\bibinfo {author} {\bibfnamefont {P.~S.}\ \bibnamefont
  {Cornaglia}}\ and\ \bibinfo {author} {\bibfnamefont {D.~R.}\ \bibnamefont
  {Grempel}},\ }\href {\doibase 10.1103/PhysRevB.71.075305} {\bibfield
  {journal} {\bibinfo  {journal} {Phys. Rev. B}\ }\textbf {\bibinfo {volume}
  {71}},\ \bibinfo {pages} {075305} (\bibinfo {year} {2005})}\BibitemShut
  {NoStop}%
\bibitem [{\citenamefont {Tanaka}\ \emph {et~al.}(2012)\citenamefont {Tanaka},
  \citenamefont {Kawakami},\ and\ \citenamefont {Oguri}}]{Tanaka2012}%
  \BibitemOpen
  \bibfield  {author} {\bibinfo {author} {\bibfnamefont {Y.}~\bibnamefont
  {Tanaka}}, \bibinfo {author} {\bibfnamefont {N.}~\bibnamefont {Kawakami}}, \
  and\ \bibinfo {author} {\bibfnamefont {A.}~\bibnamefont {Oguri}},\ }\href
  {\doibase 10.1103/PhysRevB.85.155314} {\bibfield  {journal} {\bibinfo
  {journal} {Phys. Rev. B}\ }\textbf {\bibinfo {volume} {85}},\ \bibinfo
  {pages} {155314} (\bibinfo {year} {2012})}\BibitemShut {NoStop}%
\bibitem [{\citenamefont {W\'ojcik}\ and\ \citenamefont
  {Weymann}(2015)}]{Wojcik2015}%
  \BibitemOpen
  \bibfield  {author} {\bibinfo {author} {\bibfnamefont {K.~P.}\ \bibnamefont
  {W\'ojcik}}\ and\ \bibinfo {author} {\bibfnamefont {I.}~\bibnamefont
  {Weymann}},\ }\href {\doibase 10.1103/PhysRevB.91.134422} {\bibfield
  {journal} {\bibinfo  {journal} {Phys. Rev. B}\ }\textbf {\bibinfo {volume}
  {91}},\ \bibinfo {pages} {134422} (\bibinfo {year} {2015})}\BibitemShut
  {NoStop}%
\bibitem [{\citenamefont {\ifmmode~\check{Z}\else
  \v{Z}\fi{}itko}(2010)}]{Zitko2010}%
  \BibitemOpen
  \bibfield  {author} {\bibinfo {author} {\bibfnamefont {R.}~\bibnamefont
  {\ifmmode~\check{Z}\else \v{Z}\fi{}itko}},\ }\href {\doibase
  10.1103/PhysRevB.81.115316} {\bibfield  {journal} {\bibinfo  {journal} {Phys.
  Rev. B}\ }\textbf {\bibinfo {volume} {81}},\ \bibinfo {pages} {115316}
  (\bibinfo {year} {2010})}\BibitemShut {NoStop}%
\bibitem [{\citenamefont {Lechtenberg}\ and\ \citenamefont
  {Anders}(2018)}]{Lechtenberg2018}%
  \BibitemOpen
  \bibfield  {author} {\bibinfo {author} {\bibfnamefont {B.}~\bibnamefont
  {Lechtenberg}}\ and\ \bibinfo {author} {\bibfnamefont {F.~B.}\ \bibnamefont
  {Anders}},\ }\href {\doibase 10.1103/PhysRevB.98.035109} {\bibfield
  {journal} {\bibinfo  {journal} {Phys. Rev. B}\ }\textbf {\bibinfo {volume}
  {98}},\ \bibinfo {pages} {035109} (\bibinfo {year} {2018})}\BibitemShut
  {NoStop}%
\bibitem [{\citenamefont {Grenzebach}\ \emph {et~al.}(2006)\citenamefont
  {Grenzebach}, \citenamefont {Anders}, \citenamefont {Czycholl},\ and\
  \citenamefont {Pruschke}}]{GrenzebachAndersCzychollPruschke2006}%
  \BibitemOpen
  \bibfield  {author} {\bibinfo {author} {\bibfnamefont {C.}~\bibnamefont
  {Grenzebach}}, \bibinfo {author} {\bibfnamefont {F.~B.}\ \bibnamefont
  {Anders}}, \bibinfo {author} {\bibfnamefont {G.}~\bibnamefont {Czycholl}}, \
  and\ \bibinfo {author} {\bibfnamefont {T.}~\bibnamefont {Pruschke}},\
  }\href@noop {} {\bibfield  {journal} {\bibinfo  {journal} {Phys. Rev. B}\
  }\textbf {\bibinfo {volume} {74}},\ \bibinfo {pages} {195119} (\bibinfo
  {year} {2006})}\BibitemShut {NoStop}%
\end{thebibliography}
\end{document}